\documentclass[runningheads]{llncs}

\usepackage{amssymb}
\usepackage{latexsym}

\usepackage{xcolor}
\usepackage{graphicx}
\usepackage{microtype} 
\usepackage{subfigure} 
\usepackage{amsfonts}
\usepackage{hyperref}
\usepackage{color, colortbl}
\usepackage{comment}
\usepackage{wrapfig}
\usepackage{setspace}
\usepackage{times}                     
\usepackage{amsmath}
\usepackage{multirow}

\DeclareGraphicsExtensions{.pdf,.png,.jpg}

\definecolor{newcolor}{rgb}{.8,.349,.1}

\graphicspath{{figs/}{figures/}{pictures/}{images/}{./}} 

\newcommand{\figref}[1]{Fig.~\ref{#1}}

\begin{document}

\title{Topologically-Guided Color Image Enhancement}%

\author{Junyi Tu\inst{1}\orcidID{0000-0001-7026-7454} \and
Paul Rosen\inst{1}\orcidID{0000-0002-0873-9518}}
\authorrunning{J. Tu and P. Rosen}
\institute{University of South Florida, Tampa FL 33620, USA\\
\email{\{junyi,prosen\}@mail.usf.edu}}

\maketitle

\begin{abstract}

Enhancement is an important step in post-processing digital images for personal use, in medical imaging, and for object recognition. Most existing manual techniques rely on region selection, similarity, and/or thresholding for editing, never really considering the topological structure of the image. In this paper, we leverage the contour tree to extract a hierarchical representation of the topology of an image. We propose 4 topology-aware transfer functions for editing features of the image using local topological properties, instead of global image properties. Finally, we evaluate our approach with grayscale and color images.


\end{abstract}

\section{Introduction}
\label{sec:introduction}

Image enhancement techniques aim to provide users maximal control in directing the improvements of the appearance of an image. They are widely used in photography post-processing (i.e., retouching), medical image processing, and object recognition. Many image editing platforms, such as Adobe Photoshop, utilize region selection, similarity, and/or thresholding to determine groups of pixels to edit. Then, editing options (e.g., contrast or brightness) that only consider global image properties are provided, never considering the topological structure of an image. Topological information provides a new perspective on the ``shape'' of the image and new capabilities for manipulating it~\cite{robles2018shape}.

In this paper, we leverage a tool from Topological Data Analysis (TDA)~\cite{EdelsbrunnerLetscherZomorodian2002}, specifically the contour tree~\cite{Boyell63}, to extract a hierarchical representation of key features of an image (i.e., critical points) and the monotonic regions connecting those features (i.e., Morse cells). The contour tree of a scalar function defined on a simply connected domain (i.e., an image) is obtained by encoding the evolution of the connectivity of the level sets induced by the scalar function. The key property of contour tree that makes it a viable tool for image enhancement is its graph-based representation that captures the changes of topology in image data. The contour tree can therefore be searched, modified, and pruned locally, in a quantifiable way, while retaining the global structures in the image. Using the contour tree, topology-aware transfer functions provide new image enhancement functionalities for editing images using local topological properties, instead of using global image properties.

The contributions of this paper are: (1)~we provide a mapping of color images to the \textit{contour tree} for extracting the topological structure of the image; (2)~we describe an interface for selecting topological features for image editing; and (3)~we provide 4 topology-aware transfer function modalities for image editing: brightness enhancement, contrast enhancement, denoising, and gamma correction.

\begin{figure}[!b]
    \centering
    
    \hspace{10pt}
    \begin{minipage}[m]{0.43\linewidth}
    \subfigure[Scalar Func.\label{fig.terrain.iso}]{\includegraphics[width = 0.485\linewidth]{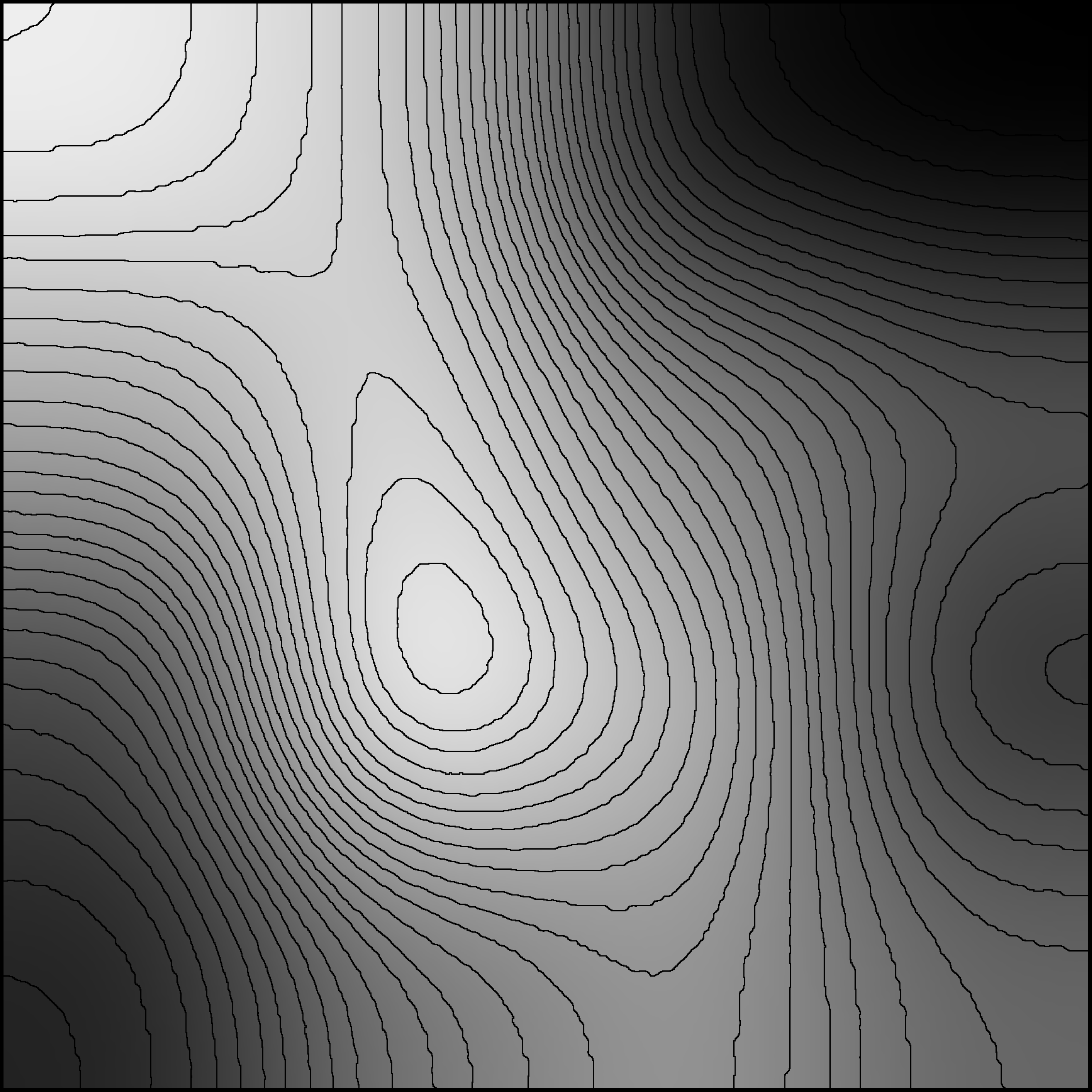}}
    \hfill
    \subfigure[Mono.\ Regions\label{fig.terrain.monotonic}]{\includegraphics[width = 0.485\linewidth]{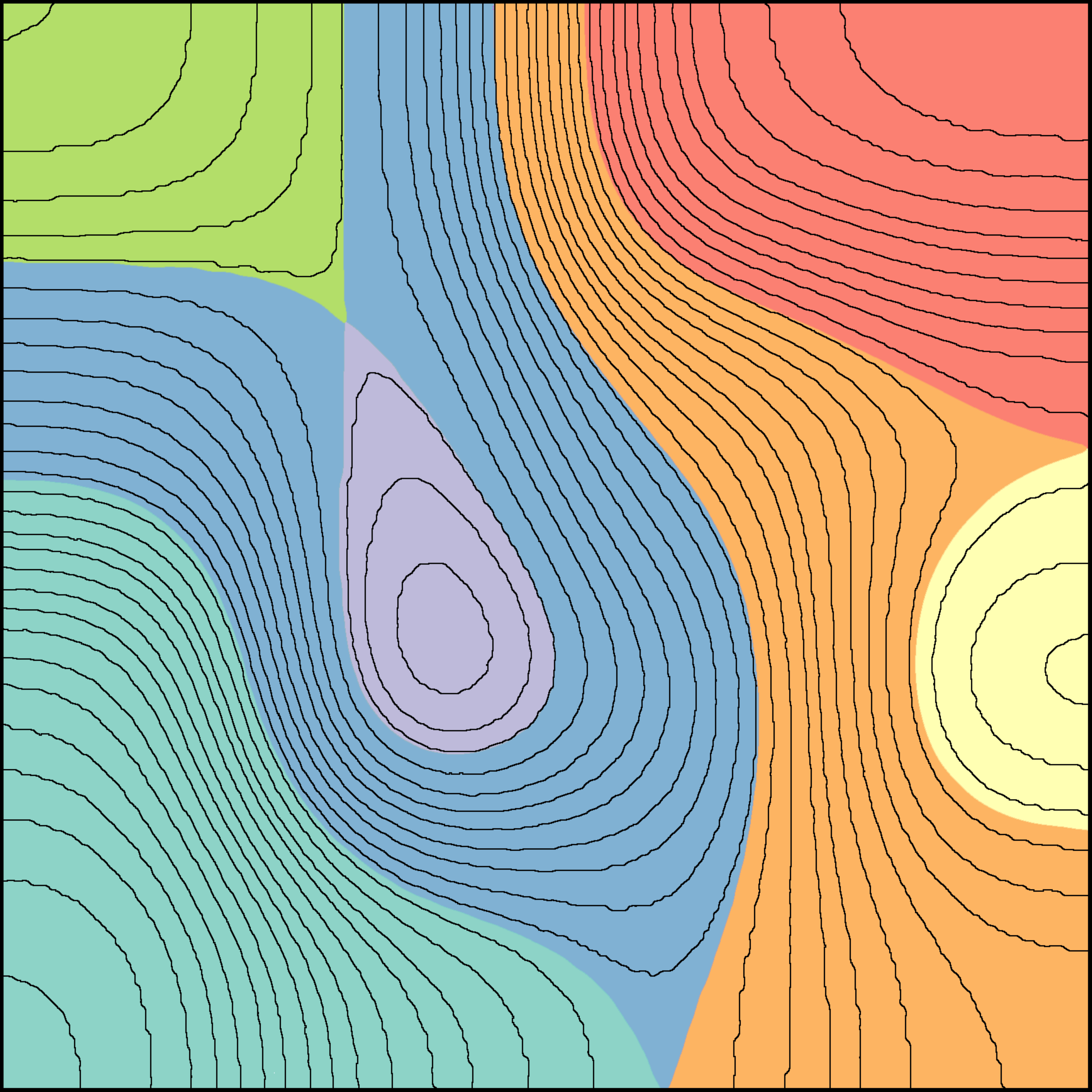}}
    \end{minipage}
    \hfill
    \begin{minipage}[m]{0.49\linewidth}
    \vspace{-0.75cm}\includegraphics[width = 0.975\linewidth]{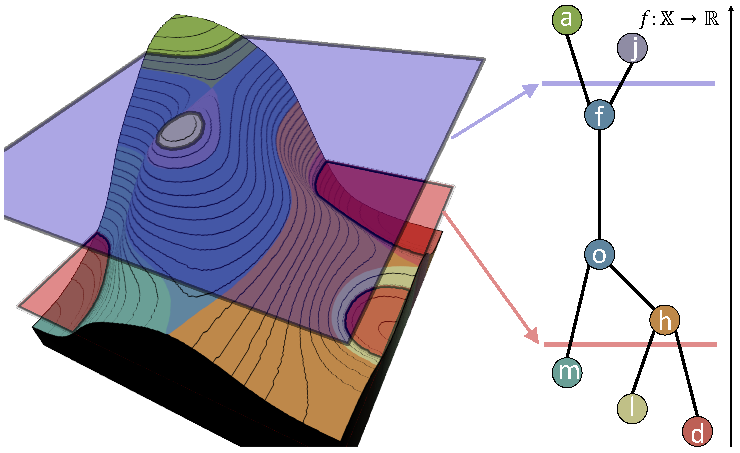}
    \end{minipage}
    \begin{minipage}[m]{0.001\linewidth}
    \hspace{-5.95cm}
    \vspace{-3.1cm}
    \subfigure[Terrain Visualized\label{fig.terrain.3d}]{\hspace{85pt}}
    \end{minipage}
    \begin{minipage}[m]{0.001\linewidth}
    \hspace{-2.75cm}
    \vspace{-3.1cm}
    \subfigure[Contour Tree\label{fig.terrain.ct}]{\hspace{67pt}}
    \end{minipage}

    \vspace{10pt}
    \caption{(a) A scalar function defined on a simple domain (with contours shown for demonstration purposes only) (b) has monotonic regions differentiated by color. The function, (c) visualized as terrain with 2 isovalues highlighted (i.e., the red and blue planes), (d) produces a contour tree, where the nodes represent critical points and edges represent regions of monotonic behavior.}
    \label{fig.terrain.example}
\end{figure}

\section{Capturing the Topology of an Image}
\label{sec:contourtree}

Let $f: \Omega\subset \mathbb{R}^2 \rightarrow \mathbb{R}$ be a continuous function on a simply connected domain~$\Omega$. The level set of a single isovalue $z$ is$f^{-1}(z)=\{(x,y):f(x,y)=z\}$, and a contour is a connected component of a level set. The most familiar context of contours are topographic maps (see \figref{fig.terrain.example}), where $f$ is the elevation, and contours are shown at selected values. The contour tree tracks the creation, merging, splitting, and destruction of contours as a plane is swept across $f$.

Consider the example height map in \figref{fig.terrain.3d} and contour tree in \figref{fig.terrain.ct}. First, a plane $z$ is swept from $-\infty\rightarrow+\infty$. As the plane sweeps up, when it reaches local minima, nodes are created in the contour tree, denoted by labels $m$, $l$, and $d$, since these represent the ``birth'' of a contour. As the plane continues its sweep up, one can observe that at $z=f_{red-plane}$ there are 3 independent contours, each represented by an edge in the contour tree.

At $z=f_h$, a special event occurs, where the contour of $l$ and $d$ merge together. The merge, called a \textit{join event}, represents the ``death'' of the contour that was born more recently, in this case $l$. The event creates the \textit{feature pair} $l/h$. Similarly, at $z=f_o$, the contours of $m$ and $d$ join and $m$ ``dies'', creating the $m/o$ pair.

The birth/death relationships are important, because they segment the space into a hierarchy of regions of uniform (i.e., monotonic) behavior. Furthermore, the difference between the birth and death, $\left|f_{birth}-f_{death}\right|$, of a contour is known as the \textit{persistence} of the feature. Persistence is an important measure in our context, as it captures the amplitude/magnitude of a feature.

Likewise, we also consider a sweep plane $z$ that goes from $+\infty\rightarrow-\infty$. As the plane sweeps downwards, new contours are born at local maxima, such as $a$ and $j$. For downward sweeps, when the contours merge together at $z=f_f$, this is called a \textit{split event}. Similarly to join events, the split represents the ``death'' of the feature born more recently, in this case $j/f$. With splits, $f_{birth}>f_{death}$.

Finally, the global minimum and maximum are paired into a special feature, captured twice as $d/a$ and $a/d$, representing the range of values.

\setlength{\unitlength}{1pt}

\begin{figure}[!b]
	\centering

    \begin{minipage}[t]{0.25\linewidth}
    \vspace{2pt}
    \subfigure[Scalar Field\label{fig.contourexample.sf}]{{\includegraphics[width=1\linewidth]{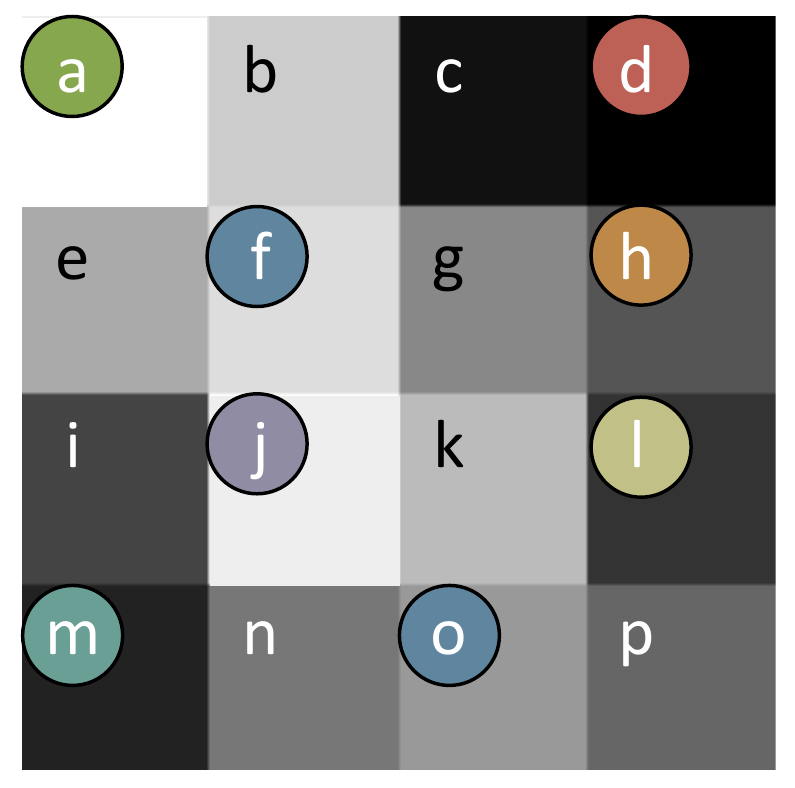}}}	
    \end{minipage}
    \hspace{5pt}
    \begin{minipage}[t]{0.13\linewidth}
    \subfigure[Augmented Join Tree\label{fig.contourexample.ajt}]{{\hspace{7pt}\includegraphics[height=3.1cm]{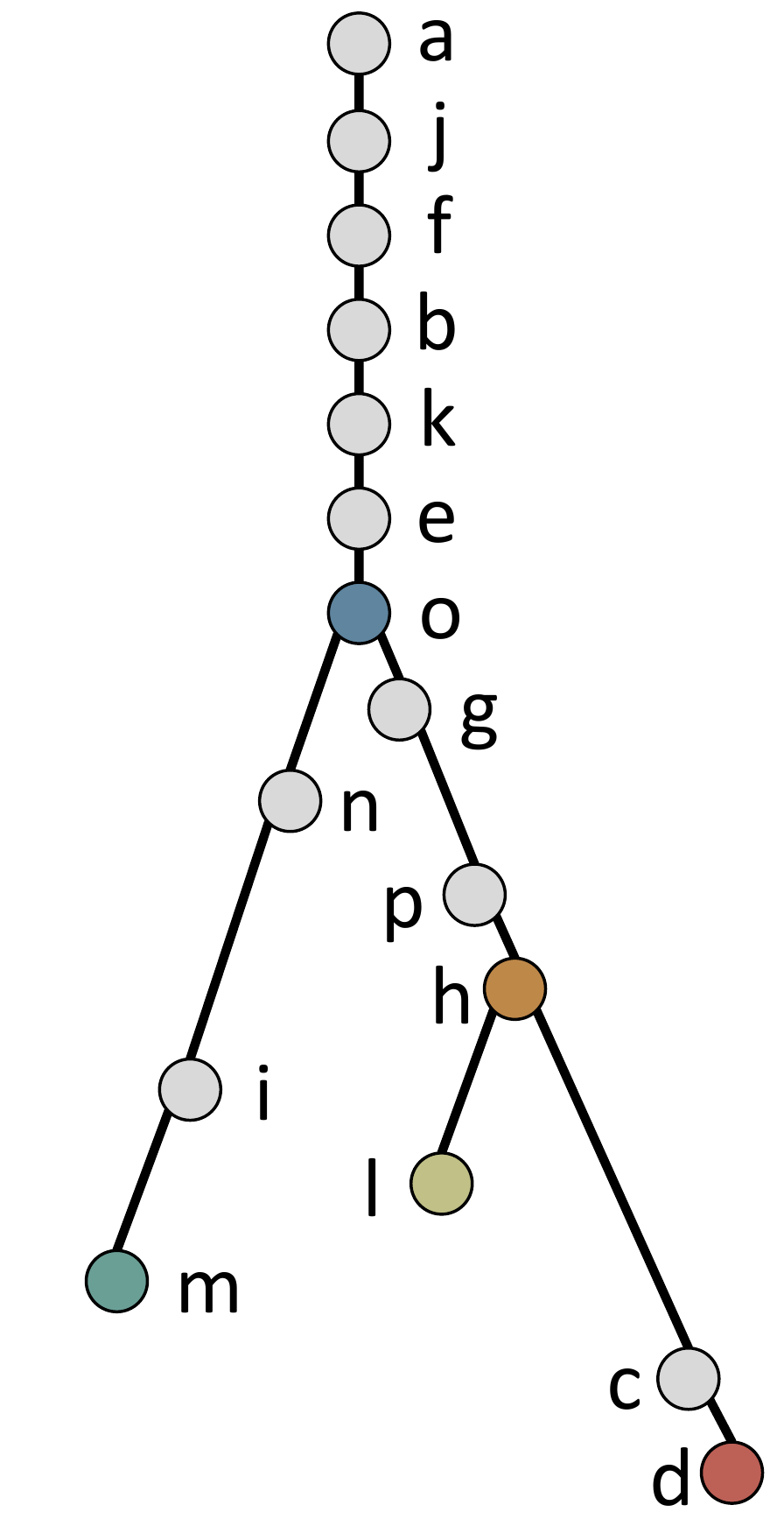}\hspace{7pt}}}
    \end{minipage}
    \begin{minipage}[t]{0.001\linewidth}
        \subfigure[Join Tree\label{fig.contourexample.jt}]{\vspace{-30pt}\hspace{16pt}\includegraphics[height=1.75cm]{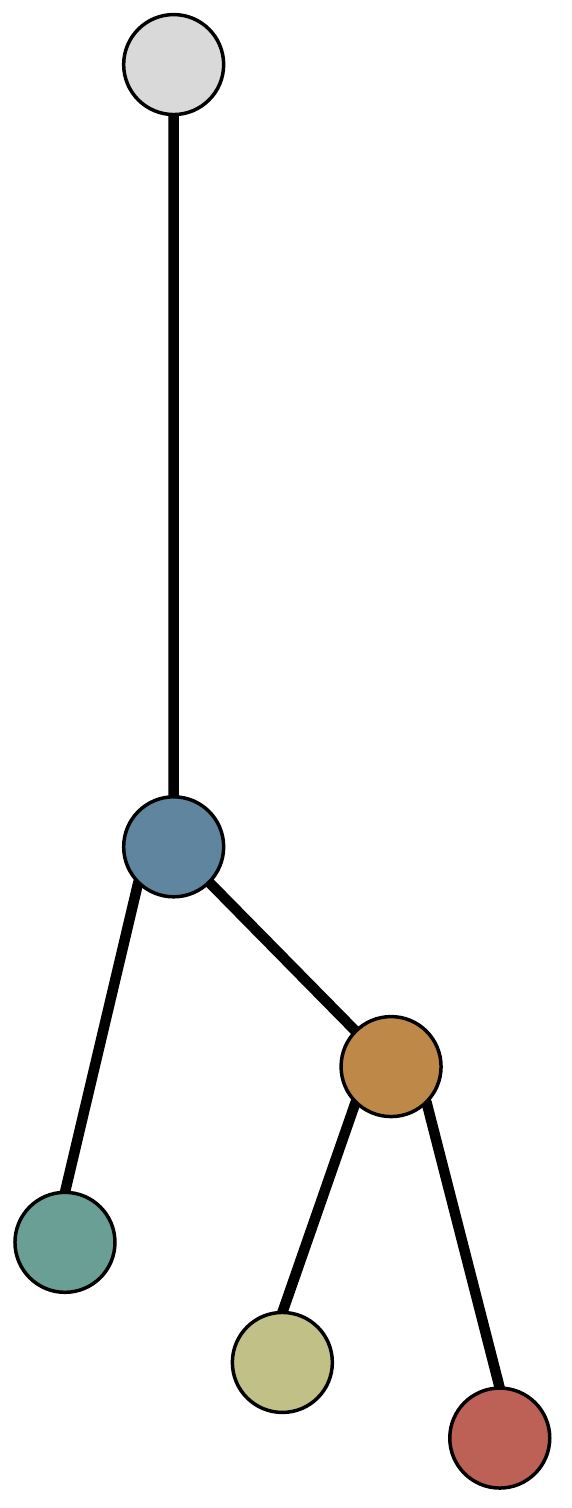}\hspace{16pt}}    
    \end{minipage}
    \hspace{27pt}
    {\begin{minipage}[t]{0.12\linewidth}
    \subfigure[Augmented Split Tree\label{fig.contourexample.ast}]{{\hspace{22pt}\includegraphics[height=3.1cm]{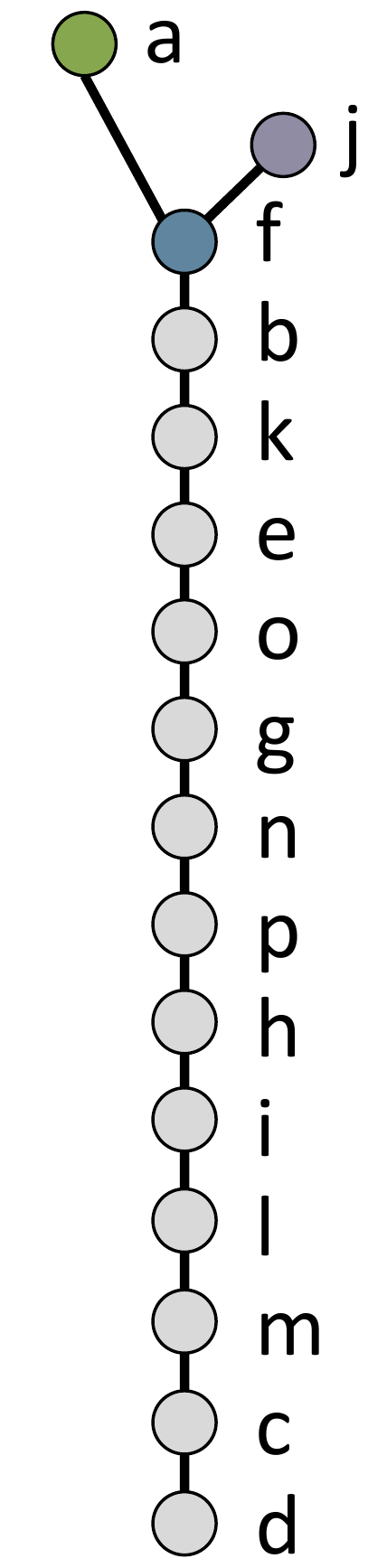}\hspace{22pt}}}
    \end{minipage}}
    \begin{minipage}[t]{0.001\linewidth}
        \subfigure[Split Tree\label{fig.contourexample.st}]{\hspace{20pt}\includegraphics[height=1.75cm]{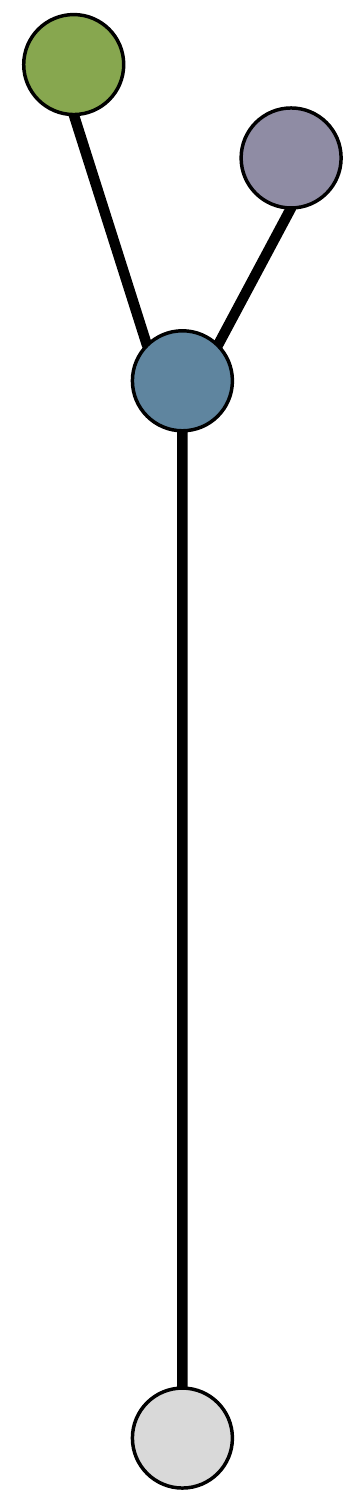}\hspace{20pt}}    
    \end{minipage}
    \hspace{43pt}
    {\begin{minipage}[t]{0.17\linewidth}
    \subfigure[Augmented Contour Tree\label{fig.contourexample.act}]{{\hspace{8pt}\includegraphics[height=3.1cm]{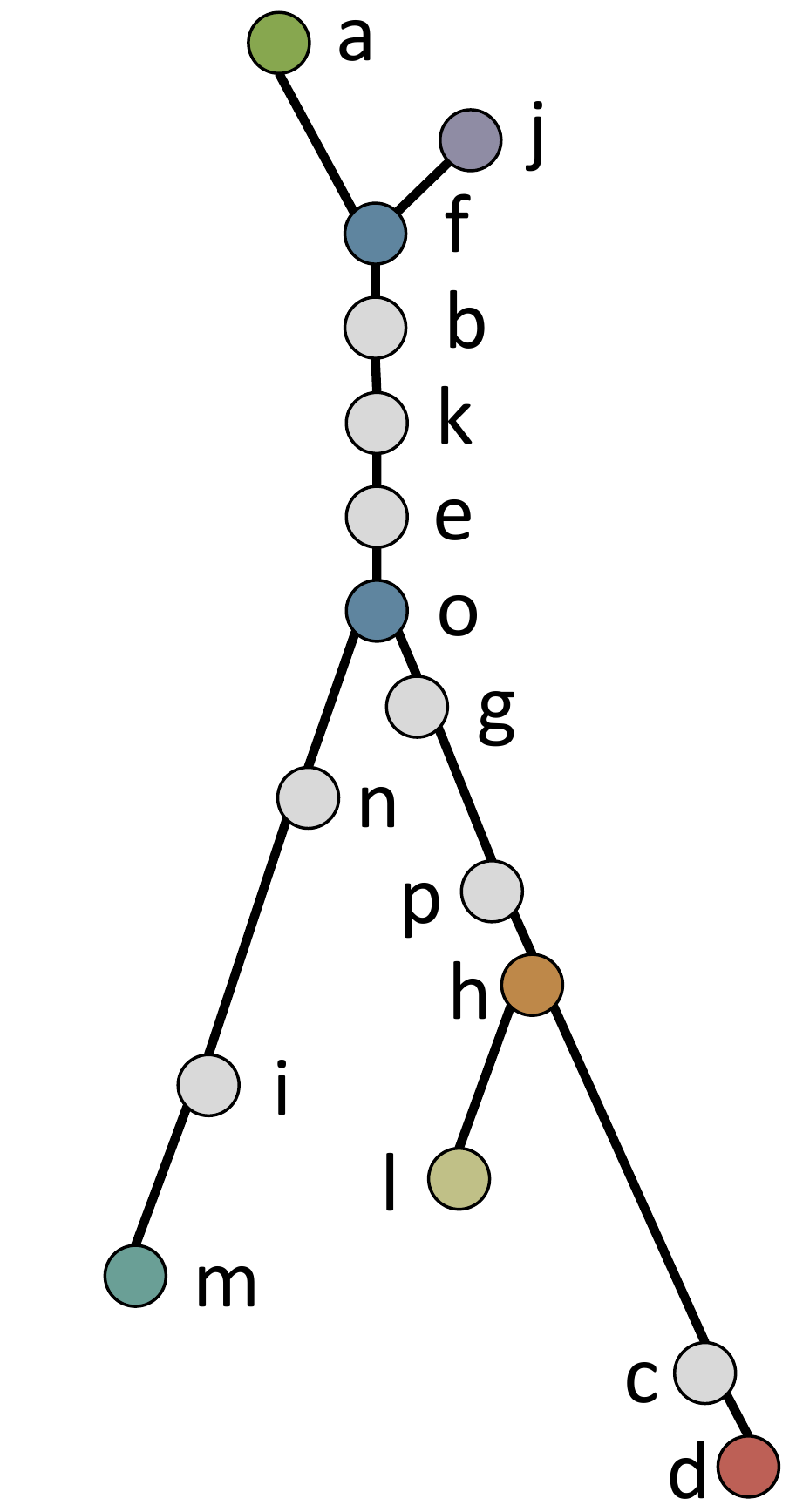}\hspace{8pt}}}
    \end{minipage}}

	\caption{(a) A low resolution version of \figref{fig.terrain.iso} has its (b)~augmented join and (d)~split trees generated. (f) The trees combine into an augmented contour tree. The critical points of the augmented trees generate the (c) join, (e) split, and (\figref{fig.terrain.ct}) contour trees of the scalar field.}
	\label{fig.contourexample}

\end{figure}

\subsection{Computing the Contour Tree}

We briefly describe the computation of the contour tree. For a detailed description and efficient algorithm, see~\cite{carr2003computing,rosen2018hybrid}. 

As the previous section implies, the construction is split into 2 phases: an upward and downward sweep, implemented as join tree (see \figref{fig.contourexample.jt}) and split tree (see \figref{fig.contourexample.st}) construction, respectively. To find the join and split trees, the construction first finds the \textit{augmented join tree} and \textit{augmented split tree}.

Using the scalar field in \figref{fig.contourexample.sf}, which is a downsampled version of \figref{fig.terrain.iso}, the augmented join tree construction is shown in \figref{fig.contourexample.ajt}. First, the pixels are sorted by values, $f_d<f_c<f_m<f_l<...<f_f<f_j<f_a$. Pixels are inserted 1 at a time into the augmented join tree. As they are inserted, connected components are tracked by connecting with neighboring pixels already in the augmented join tree. In our implementation we consider the ring of 8 neighbors surrounding a given pixel. In this example, we only consider 4 (i.e., left/right/up/down) neighbors. If a pixel joins 1 or more existing components, it is connected to the top of those components in the augmented join tree. If it joins no component, it starts a new connected component. For example, when $d$ is added to the tree, no connected components exist, so it creates one. When $c$ is inserted, it joins the $\{d\}$ component, since they are neighbors in the image. Continuing forward, when $h$ is inserted, it links the $\{l\}$ component to the $\{c,d\}$ component.

Augmented split tree construction is performed nearly identically, except it starts at the largest valued pixel first. The augmented contour tree is constructed by pealing leaf nodes off of the join/split trees and adding them to the augmented contour tree, as described in~\cite{carr2003computing}. The join/split/contour trees consist of only critical nodes (i.e., nodes that cause birth and death events). They are calculated by removing ``regular'' nodes, those having only 1 upward and 1 downward edge, from the augmented tree. Finally, using the contour tree, the critical points are paired into birth/death units using the approach in~\cite{tupropagate}.

This construction assumes that all pixels have unique values, which is not the case in real images. There are 2 cases of equal valued pixels we consider. When equal valued pixels are non-adjacent, no special handling is required---the pixels can be processed in arbitrary order, but the output is still deterministic. However, adjacent pixels of equal value are problematic, since changing their insertion order can change the contour tree structure. This is resolved by grouping adjacent equal-valued pixels into ``super-pixel'' units that are processed together.

\subsection{Feature Subtree Extraction}

Extracting the regions of monotonic behavior requires selecting a subtree from the augmented contour tree for a given feature pair. Given a join/split node and local minimum/maximum pair, starting at the join/split, find the 3 subtrees extending from it. The selected subtree for the feature pair is the one containing the local minimum/maximum node. For example, in \figref{fig.contourexample.act}, the $m/o$ feature pair contains join node $o$ and local minimum $m$. Join node $o$ has 3 subtrees: up, down-left, and down-right. The down-left subtree contains the local minimum node $m$, making it the feature subtree, containing nodes $\{m,i,n,o\}$. The selection can be seen in \figref{fig.segementation.subtree}.

\subsection{Using the Contour Tree of Color Images}

The contour tree requires $f : \mathbb{X} \rightarrow \mathbb{R}$ (i.e., a single color channel). However, considering color images in RGB (Red, Green, Blue) colorspace, 3 channels map to each pixel. We consider each channel, red, green, and blue independently, generating 3 contour trees. We also consider HSB (Hue, Saturation, Brightness) colorspace. Saturation and brightness each map to their own contour trees. However, hue maps to $\mathbb{S}^1$ (i.e., circular coordinates), and the contour tree does not work in $\mathbb{S}^1$. Resolving this limitation requires additional theoretical studies. Other colorspaces are possible, when channels map to~$\mathbb{R}$.

\section{Image Processing via Contour Trees}
\label{sec:visualization}

The basic procedure for topology-based enhancement of an image is: (1)~first, the user selects a set of feature pairs of interest; (2)~next, the feature pairs are used to automatically select pixels for editing; (3)~finally, the image is edited by the user, and the contour trees are recalculated so the process can restart.

\subsection{Visualizing the Contour Tree}
\label{sec:visualization:vis}

Direct visualization of the contour tree is generally not advisable, as the size and complexity of the tree is unmanageable for even moderately sized data. To select features to edit, each contour tree is displayed using 2 interactive interfaces---the persistence diagram and persistence-volume diagram.

\begin{figure}[!b]
    \centering
    \subfigure[Persistent diagram (left)\label{fig.vis.pd} and persistent-volume diagram (right)\label{fig.vis.pv} for contour tree in Fig.~\ref{fig.terrain.ct}.]{\hspace{0pt}
    {\begin{minipage}[m][2.95cm][b]{0.205\linewidth}
    \includegraphics[height=2.7cm]{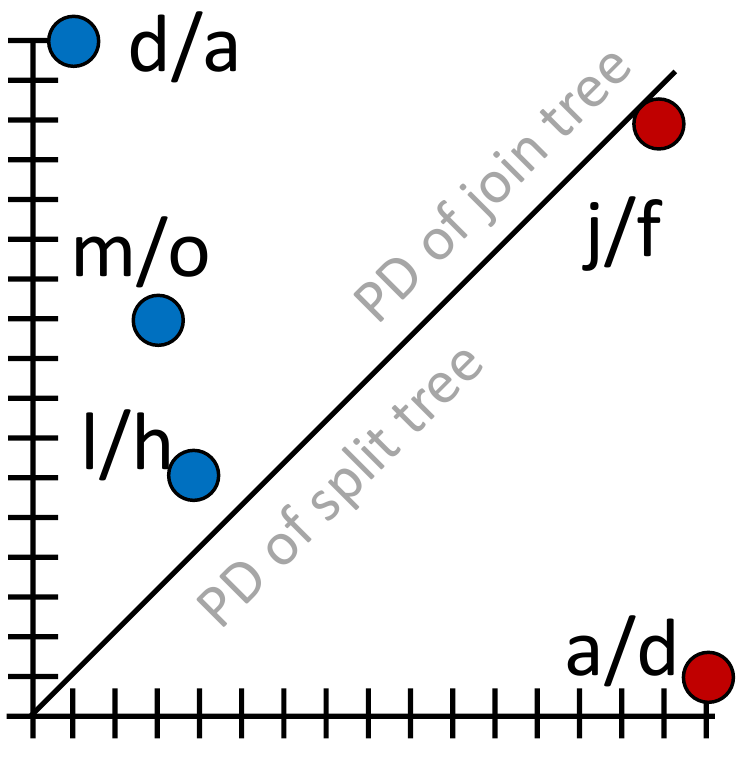}
    \end{minipage}}\hspace{10pt}
    {\begin{minipage}[m][2.95cm][b]{0.205\linewidth}
    \includegraphics[height=2.7cm]{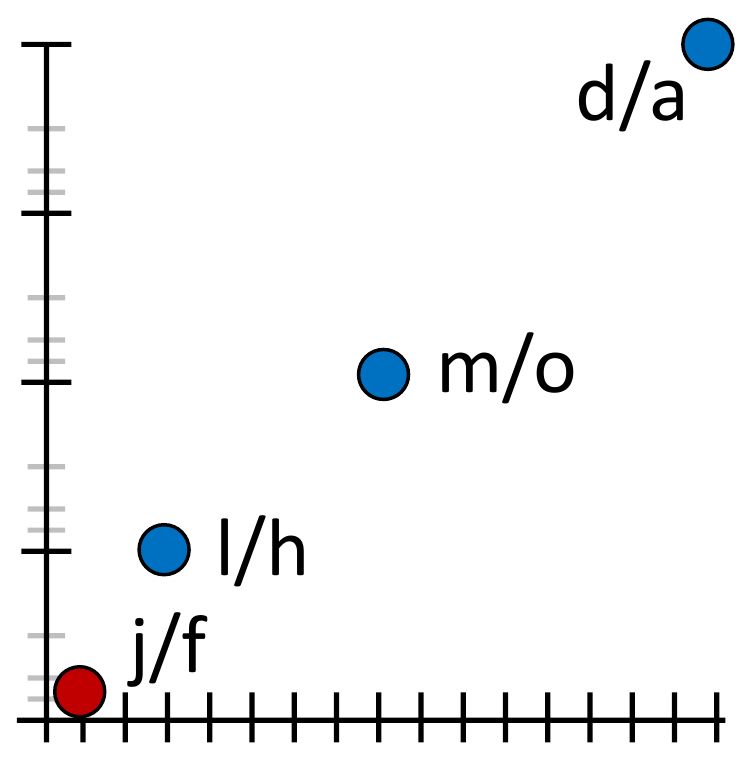}
    \end{minipage}}}
    \hspace{10pt}
    \subfigure[Selection in persistence-volume diagram (left)\label{fig.segementation.selection} precedes edit type (top right) and scale adjustment (bottom right).\label{fig.controls.interface}]{\hspace{8pt}
    {\begin{minipage}[m]{0.18\linewidth}
    \includegraphics[height=2.2cm]{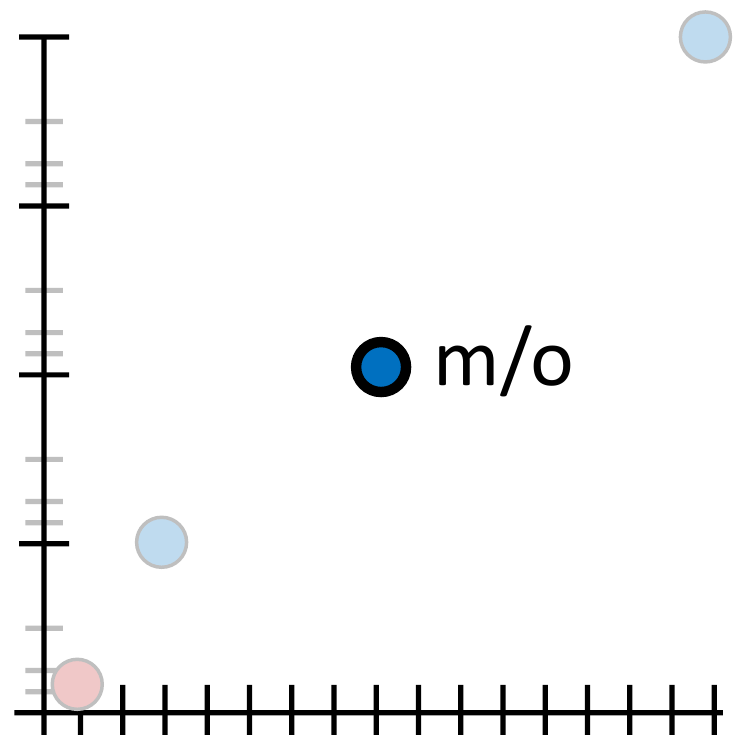}
    \end{minipage}}
    \hspace{5pt}
    {\begin{minipage}[m]{0.205\linewidth}
    \includegraphics[height=2.95cm]{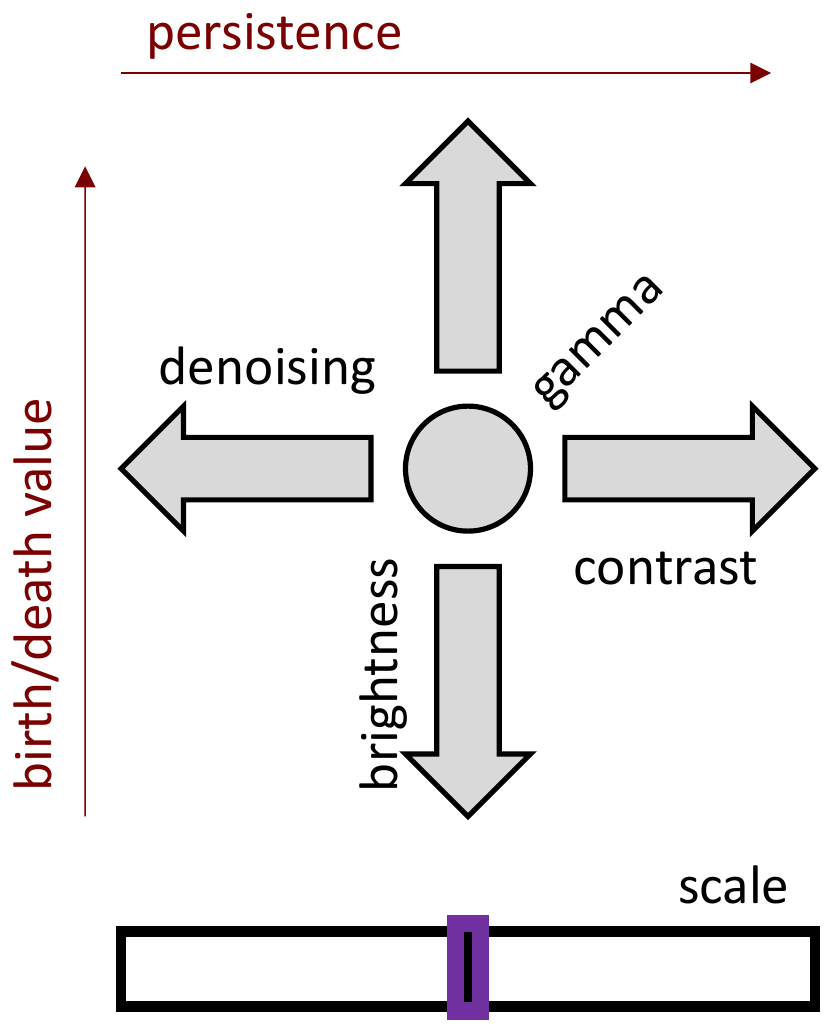}
    \end{minipage}}\hspace{8pt}}
    
    \caption{(a) The main interfaces used for selecting topological features use the persistence diagram (left) and persistence-volume diagram (right). (b) Once features a selected (left) the editing interface enables selecting the type and scale of edits (right).}
\end{figure}

\textbf{Persistence Diagram.} A standard practice in TDA represents the contour tree with birth/death feature pairs in a scatterplot display, called a \emph{persistence diagram}~\cite{Cohen-SteinerEdelsbrunnerHarer2007}. In a persistence diagram, the x-axis is tied to the feature birth value, while the death value is tied to the y-axis. \figref{fig.vis.pd}(left) shows an example for the contour tree in \figref{fig.terrain.ct}. Join features (i.e., $l/h$ and $m/o$) are on the upper left, while split features (i.e., $j/f$) are on the lower right. Features are also colored by their type---join features blue and split features red. The persistence diagram provides 2 powerful selection hints. First, features on the lower left are darker pixels, while those on the upper right are brighter pixels, since the birth and death are parameterized by pixel value. Second, the distance of the point from the diagonal is an analog of the \textit{persistence} of the feature pair. In other words, features with larger magnitude are farther from the diagonal.

\textbf{Persistence-Volume Diagram.} The persistence diagram primarily captures the amplitude of a selected feature. The volume (i.e., the pixel count of a feature) is sometimes important as well. The \textit{persistence-volume diagram} (see \figref{fig.vis.pv}(right)) is an alternative representation that encodes persistence (i.e., $\left|f_{birth}-f_{death}\right|$) on the x-axis using a linear scale and volume on the y-axis using a log scale. Volume is calculated by counting the number of nodes/pixels in the feature subtree. For example, the $m/o$ contains 4 nodes/pixels, $m$, $i$, $n$, and $o$.

\subsection{Subtree Selection}
\label{sec:visualization:seg}

Once the contour trees are generated and visualized, user interaction can proceed. Many topological features come from an image, so while individual selection is possible, selection of multiple features is desirable. We provide a brushing mechanism on both the persistence diagram and persistence-volume diagram for selecting a \textit{set} of features of interest. As the mouse is clicked-and-dragged, the features and feature subtrees are gathered for further processing. For example, brushing across the middle of the persistence-volume diagram in \figref{fig.vis.pv} would select feature $m/o$ and its subtree, as shown in \figref{fig.segementation.selection}. Features of the contour tree are hierarchical, thus, if more than one feature is selected, those features may be inclusions (i.e., one feature may be a subsets of another). In that case, only the larger/outermost feature is processed.

\begin{figure}[!b]
    \centering

    \subfigure[Sel.\ Subtree\label{fig.segementation.subtree}]{\hspace{15pt}\includegraphics[height=2.8cm]{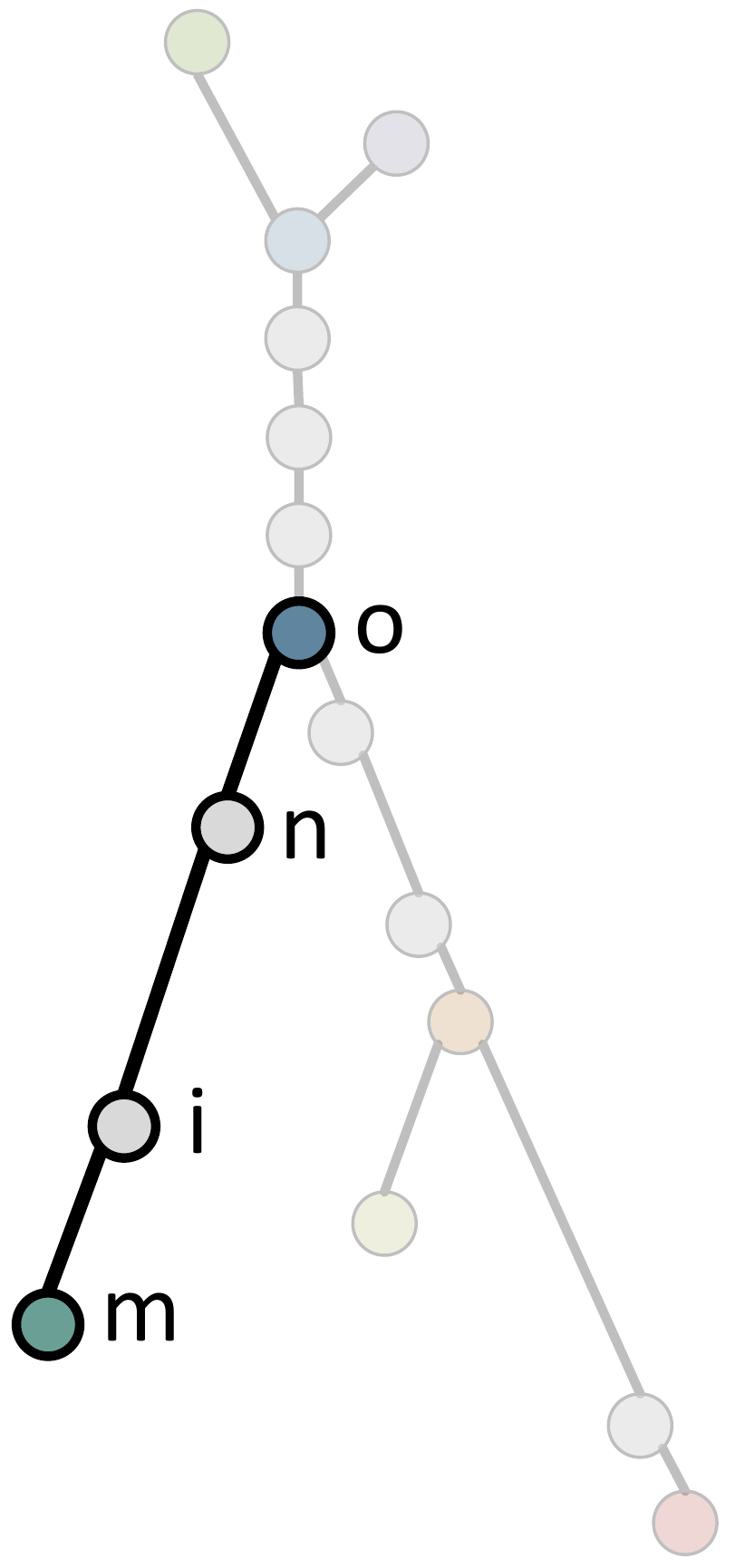}\hspace{15pt}}
    \hfill
    \subfigure[Contrast\label{fig.controls.contrast}]{\includegraphics[height=2.8cm]{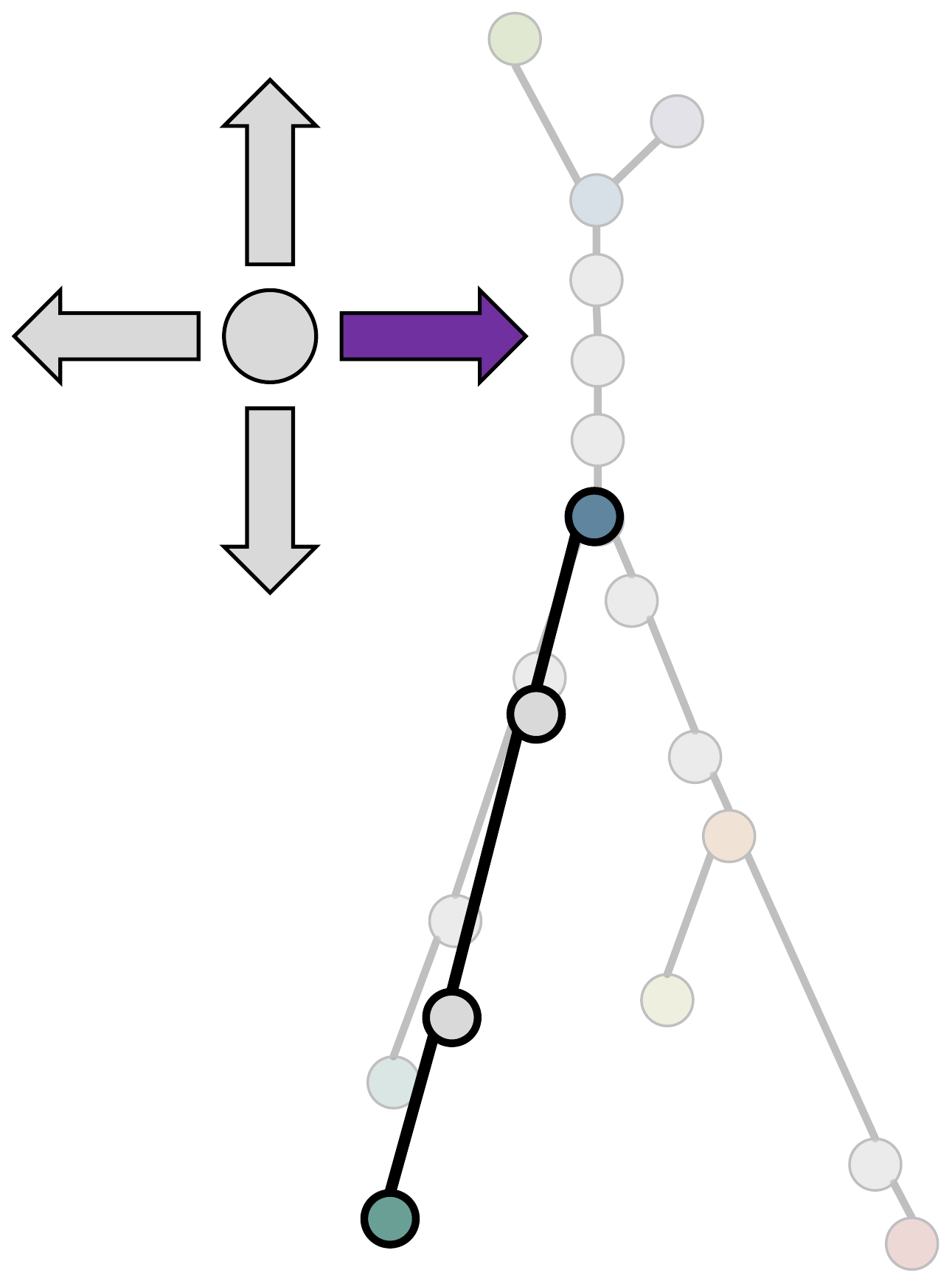}\hspace{15pt}}
    \hfill
    \subfigure[Denoising\label{fig.controls.denoising}]{\hspace{-15pt}\includegraphics[height=2.8cm]{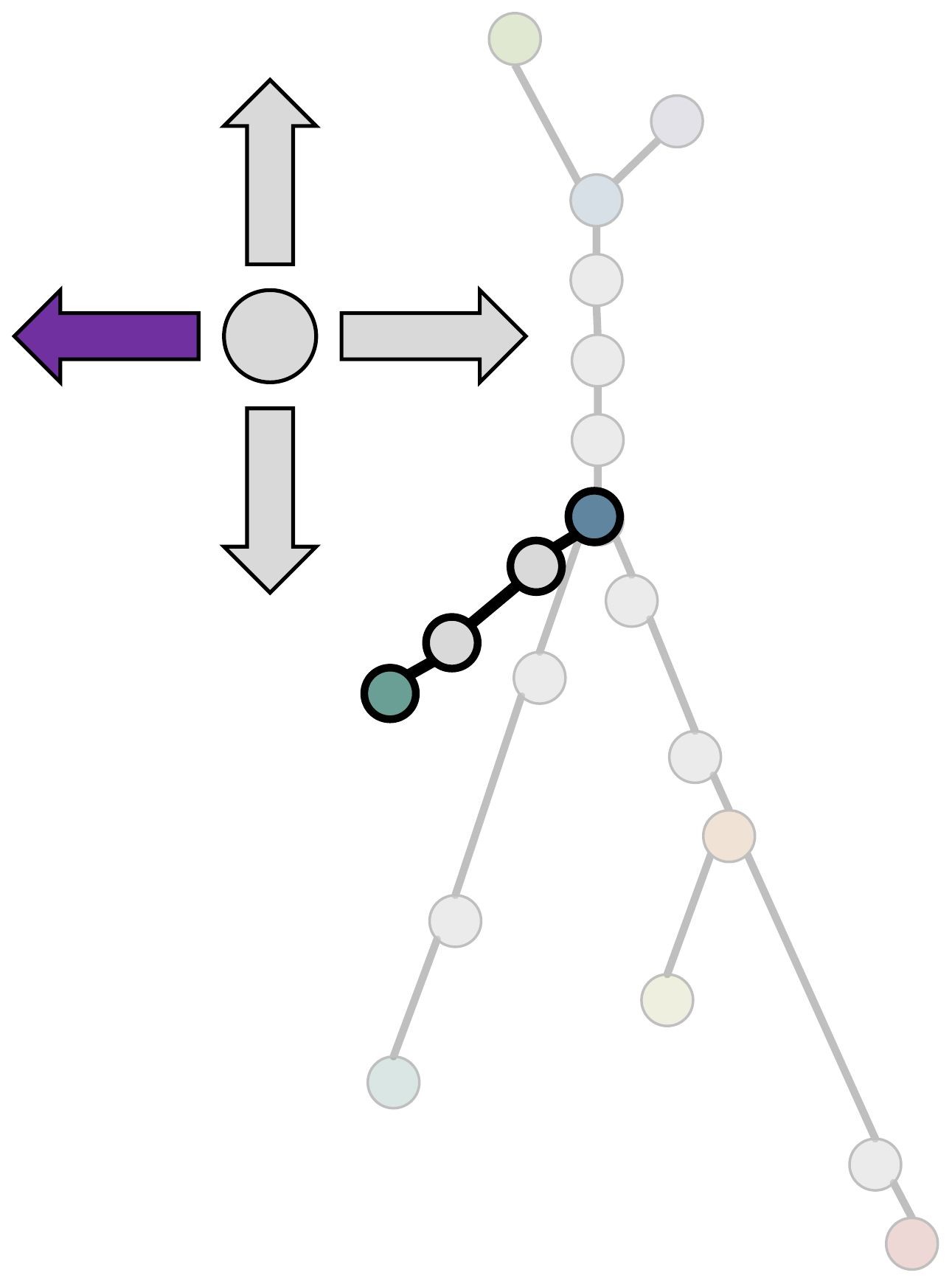}\hspace{15pt}}
    \hfill
    \subfigure[Brightness\label{fig.controls.brightness}]{\hspace{-15pt}\includegraphics[height=2.8cm]{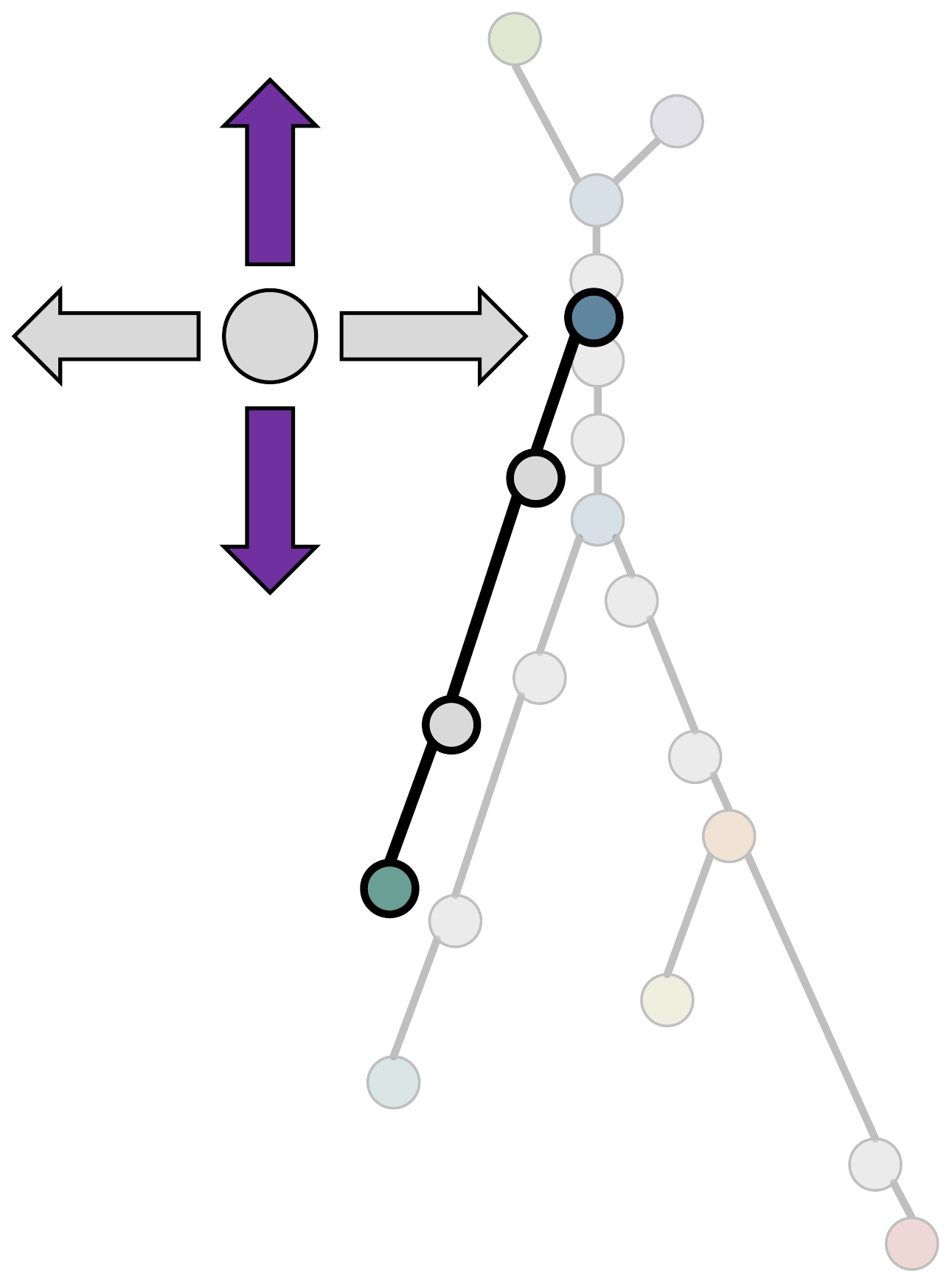}\hspace{15pt}}
    \hfill
    \subfigure[Gamma\label{fig.controls.gamma}]{\hspace{-15pt}\includegraphics[height=2.8cm]{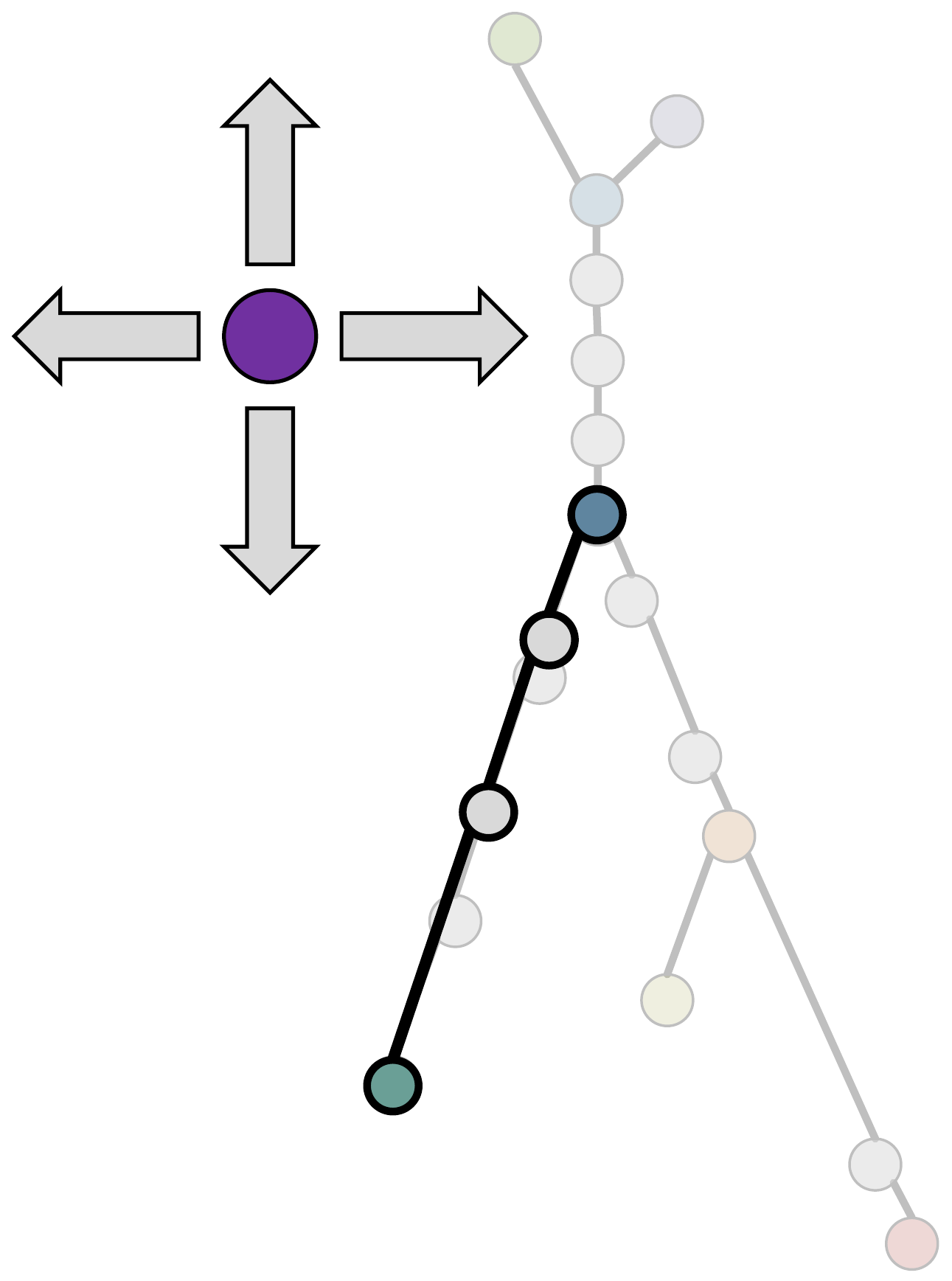}\hspace{15pt}}
    \hfill
    
    \hspace{2pt}
    \subfigure[Selection\label{fig.segementation.segmentation}]{\includegraphics[width=0.17\linewidth]{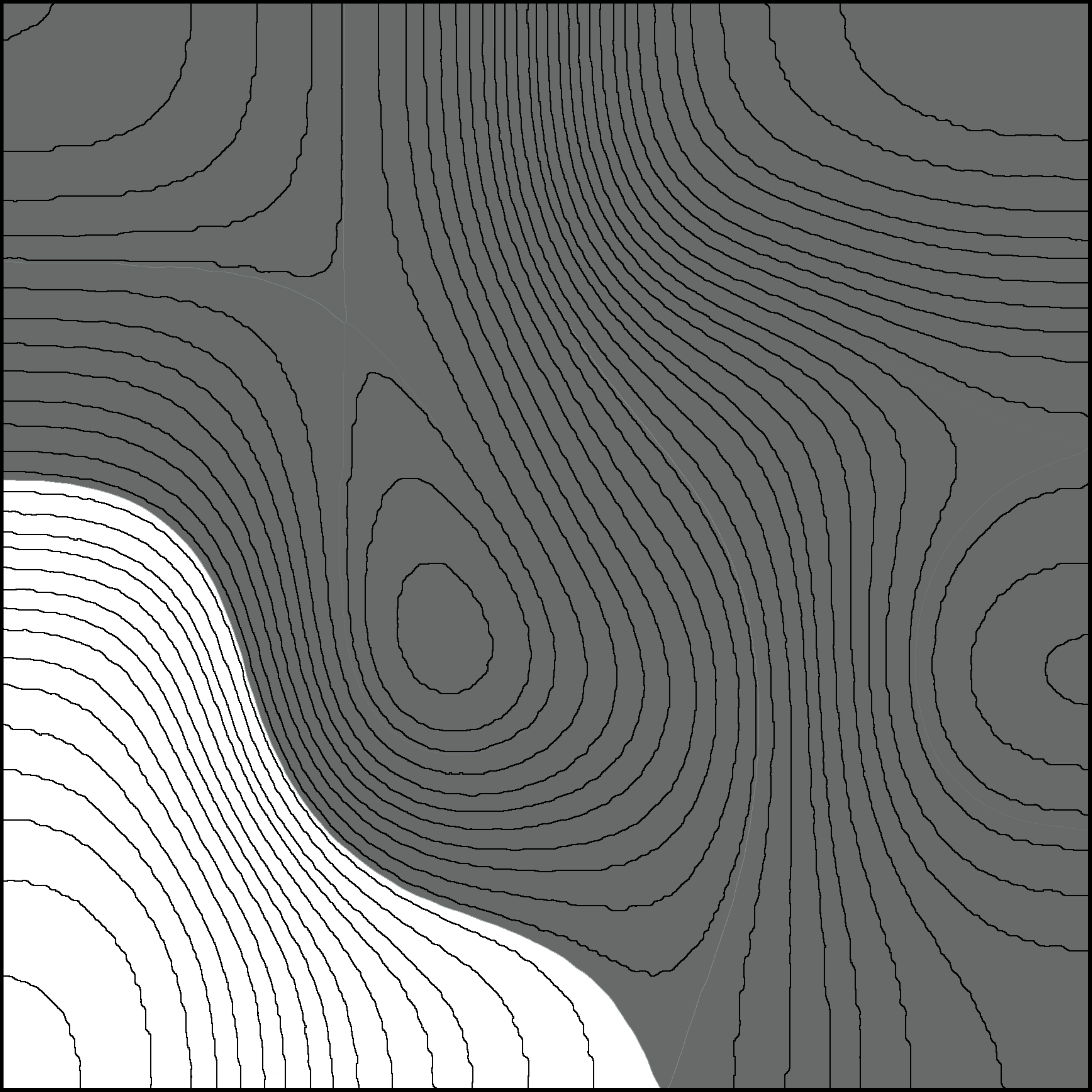}}  
    \hspace{10pt}
    \subfigure[$s$$=$$1.75$\label{fig.synth.contrast}]{\hspace{0pt}\includegraphics[width=0.17\linewidth]{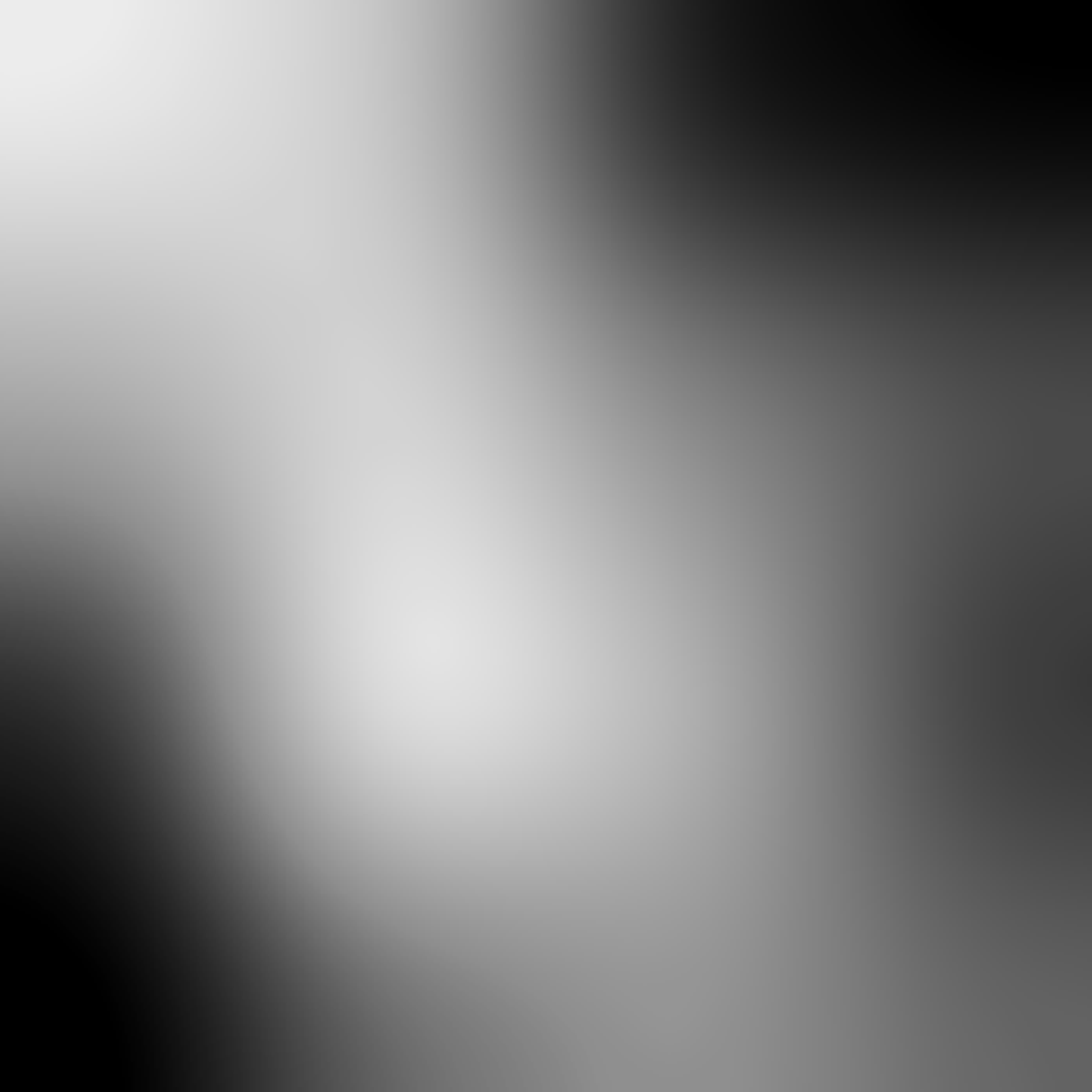}\hspace{0pt}}
    \hspace{2pt}
    \subfigure[$s$$=$$0.3$\label{fig.synth.denoising}]{\hspace{0pt}\includegraphics[width=0.17\linewidth]{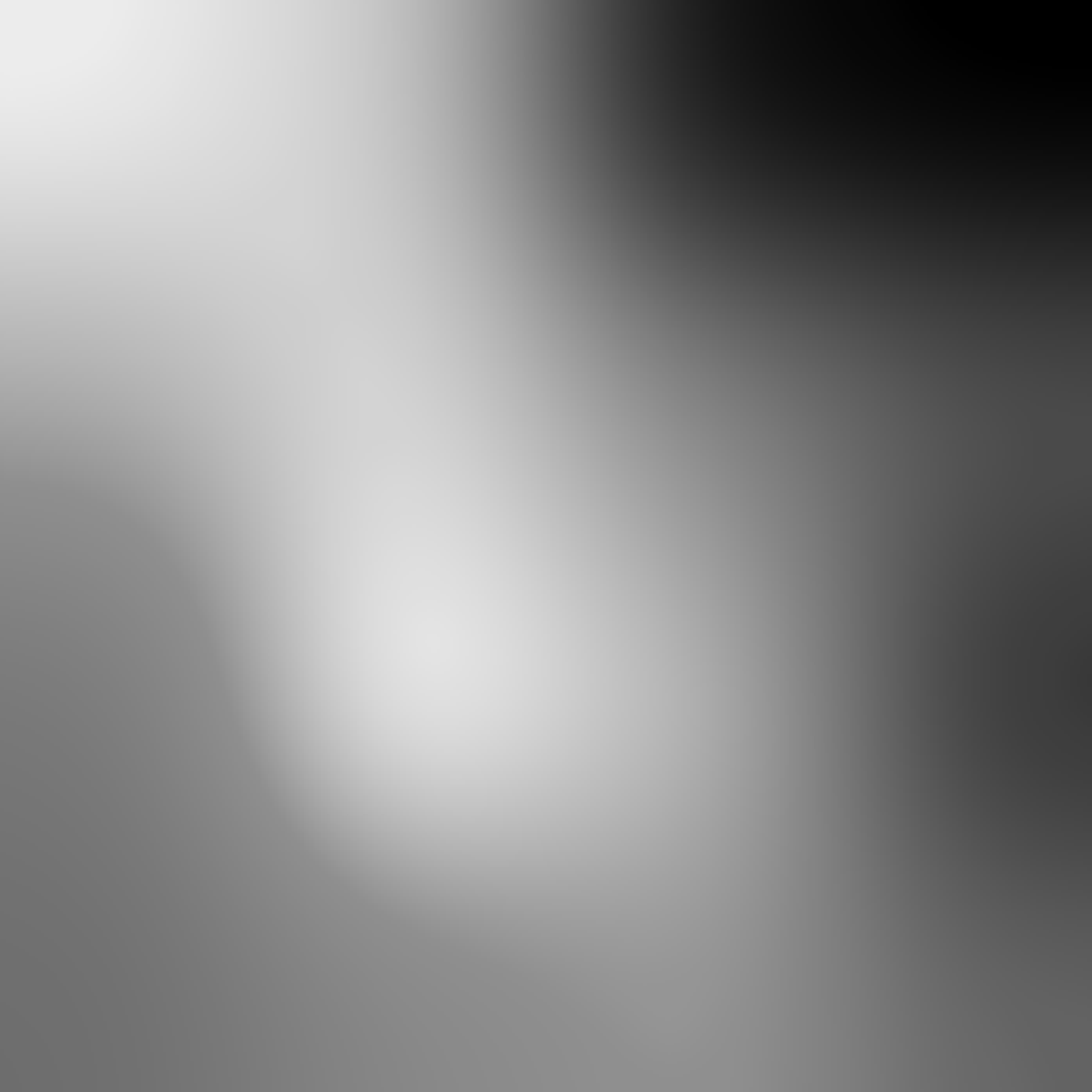}\hspace{0pt}}
    \hspace{2pt}
    \subfigure[$s$$=$$15$\label{fig.synth.brightness}]{\hspace{0pt}\includegraphics[width=0.17\linewidth]{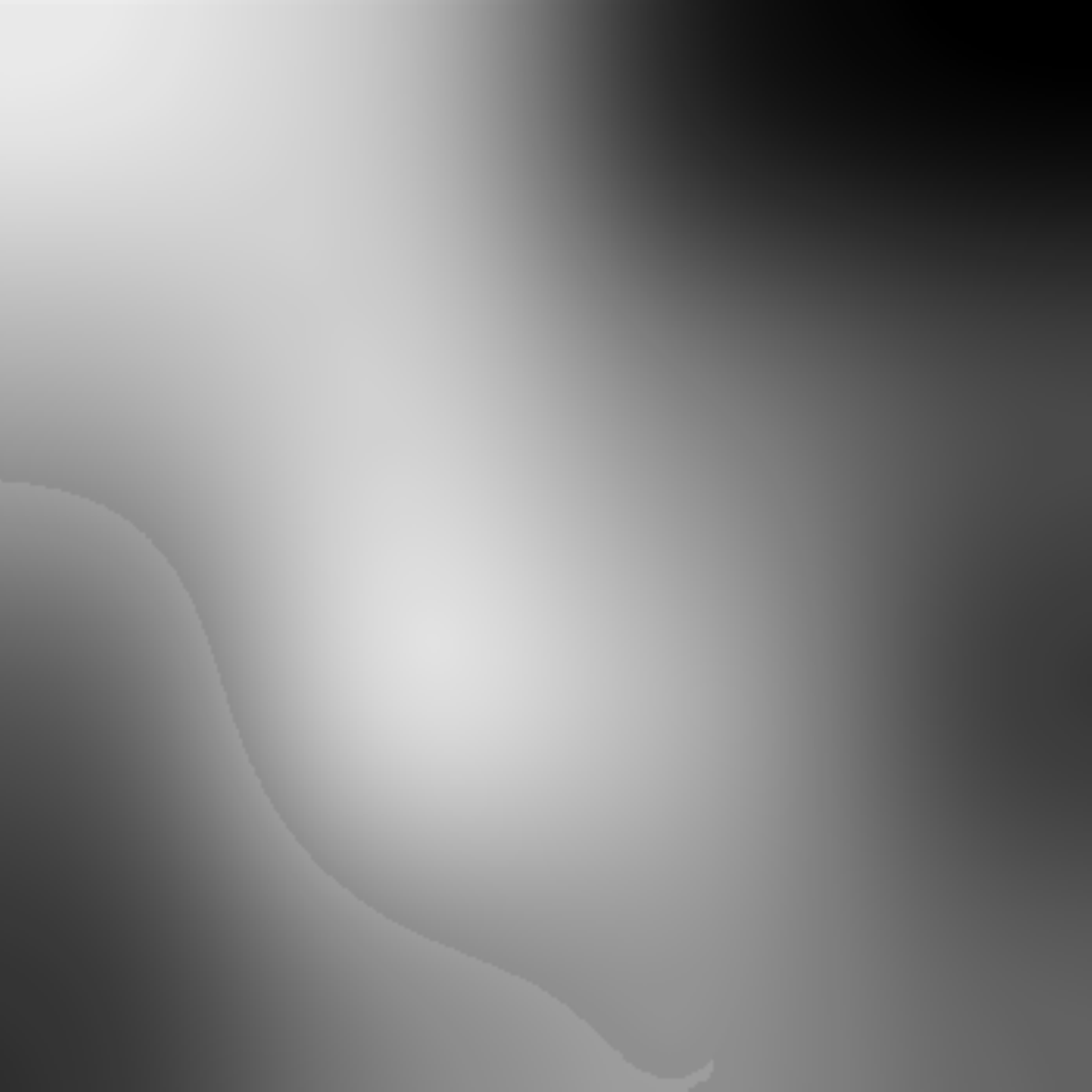}\hspace{0pt}}
    \hspace{2pt}
    \subfigure[$\gamma$$=$$2.5$\label{fig.synth.gamma}]{\hspace{0pt}\includegraphics[width=0.17\linewidth]{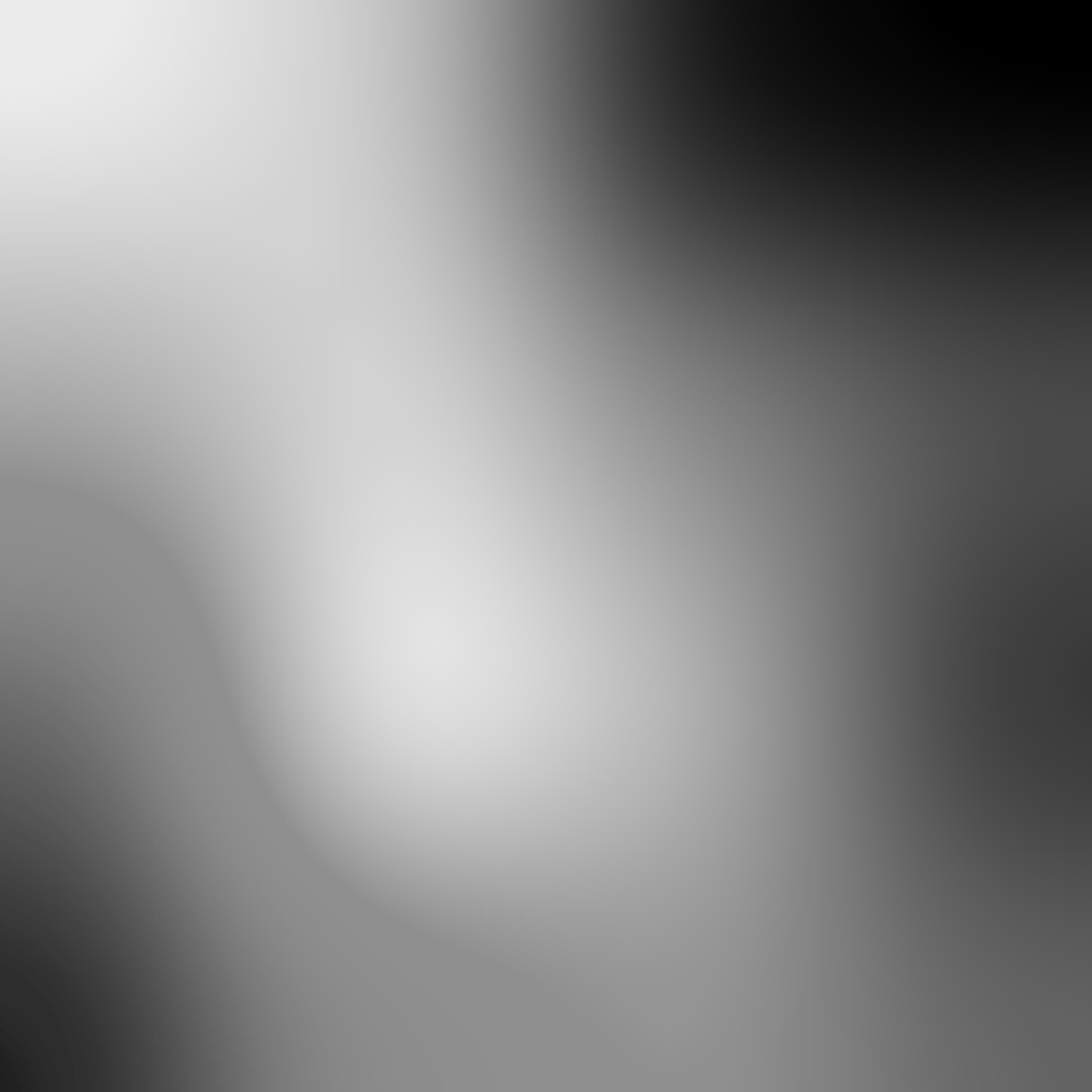}\hspace{0pt}}    
    
    \caption{Using the (a/f) selected subtree, (b/g) contrast enhancement, (c/h) denoising, (d/i) brightness enhancement, and (e/j) gamma correction operations are shown.}
    \label{fig.controls}
\end{figure}

The selected feature pairs and their associated subtrees are relatively easy to use for segmentation. Starting with a full resolution binary mask,  the nodes/pixels of each selected feature is marked in the mask. \figref{fig.segementation.segmentation} shows how this would work given the selection of the $m/o$ feature from \figref{fig.terrain.example}.

\subsection{Subtree Modification as Image Editing}
\label{sec:visualization:edit}

From a segmentation mask, global image editing options (i.e., contrast enhancement, brightness enhancements, etc.) could be easily executed. However, this misses an opportunity for modifications based upon local topology, by using subtree information. For subtree-based modification, we provide 4 transfer function options that utilize the properties of the subtree.

\textbf{Contrast Enhancement.} Contrast enhancement fixes value of the feature pair join/split node and linearly stretches rest of the subtree. For a given node $\sim$ in the subtree and a contrast scale factor $s\geq1$, the value $f'_\sim=f_{death}+(f_\sim-f_{death}) \cdot s$. \figref{fig.controls.contrast} shows an example of the operation, where the local contrast enhancement fixes the death value of a feature, while lowering the birth value.

\textbf{Denoising.} Denoising linearly collapses the subtree, such that all pixels eventually have the same value, $f_{death}$. The calculation of denoising is identical to contrast enhancement---for a given node $\sim$ in the subtree and a denoising scale factor $0\leq s\leq1$, the value $f'_\sim=f_{death}+(f_\sim-f_{death}) \cdot s$. \figref{fig.controls.denoising} shows an example, where the death value is fixed and the values of other nodes/pixels are increased.

\textbf{Brightness Enhancement.} The brightness of the entire subtree can be modified up or down uniformly. For a given node $\sim$ in the subtree and a brightness scale factor $-255\leq s\leq+255$, the value $f'_\sim=f_\sim+s$. \figref{fig.controls.brightness} shows an example.

\textbf{Gamma Correction.}
Gamma correction provides a nonlinear modification, which is usually applied to the luminance of an image. For a given node $\sim$ in the subtree and a gamma correction value $\gamma$, where $\gamma>0$, the value of a pixel $f_\sim$ is first normalized and gamma corrected $\overline{f}_\sim=\left(\frac{f_\sim-f_{death}}{f_{birth}-f_{death}}\right)^\gamma$. The final value of the pixel is then linearly interpolated, $f'_\sim=f_{death} \cdot (1-\overline{f}_\sim) + f_{birth} \cdot (\overline{f}_\sim)$. \figref{fig.controls.gamma} shows an example, where $\gamma=2.5$.

\begin{figure}[!b]
    \centering

    \subfigure[Original Image\label{fig.ALflorala.input}]{
        \begin{minipage}[m]{0.305\linewidth}
            \centering
            \includegraphics[height=1.025in,width=1\linewidth]{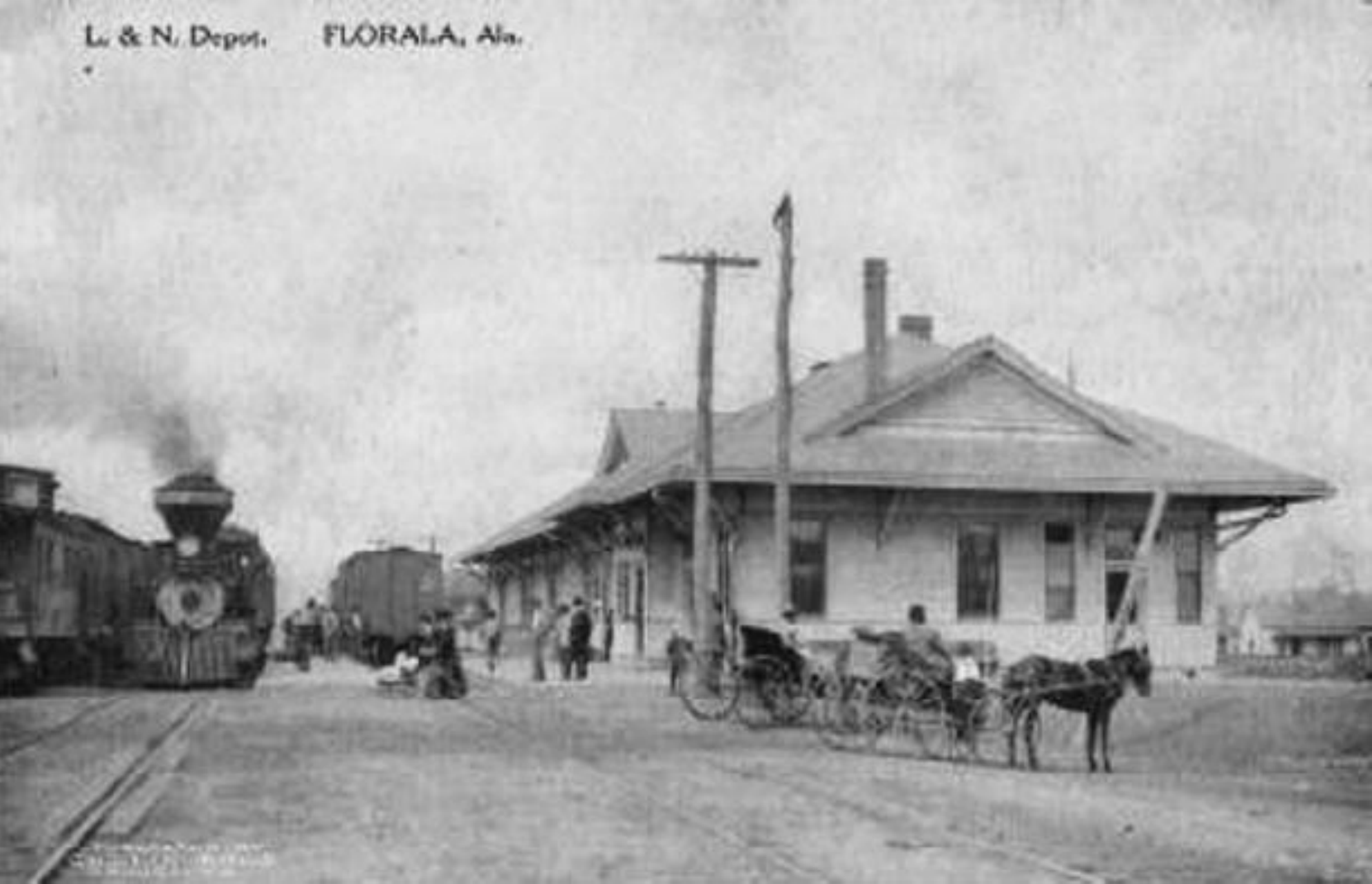}
            
            \vspace{5pt}
            \includegraphics[height=0.60in]{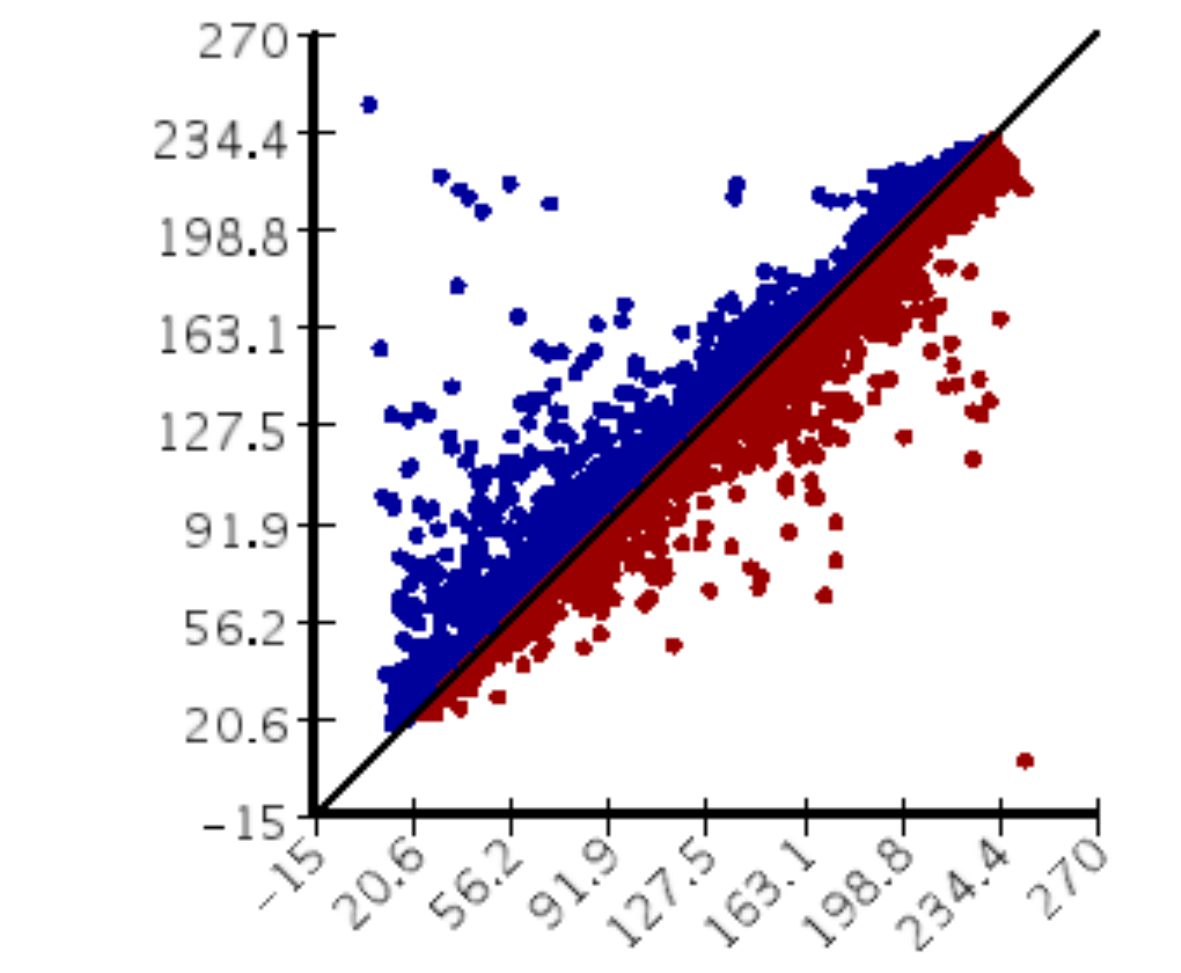}\hfill
            \includegraphics[height=0.60in]{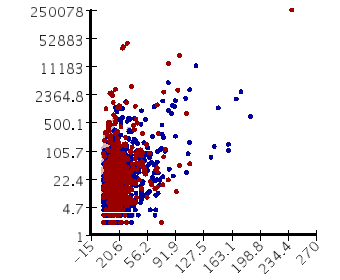}
        \end{minipage}
    }
    \hfill
    \subfigure[Our Approach\label{fig.ALflorala.output}]{
        \begin{minipage}[m]{0.305\linewidth}
        \centering
        \includegraphics[height=1.025in,width=1\linewidth]{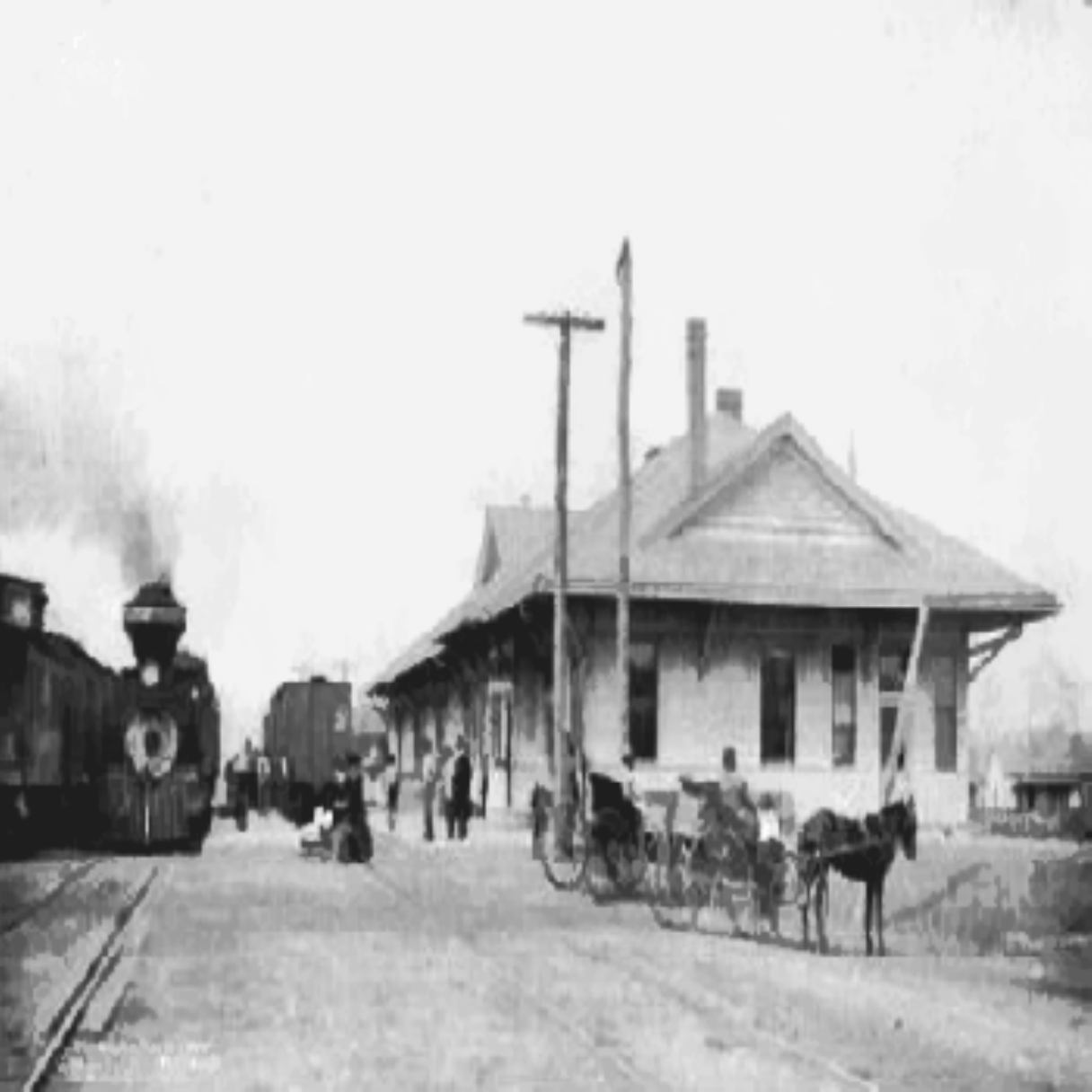}

        \vspace{5pt}
        \includegraphics[height=0.60in]{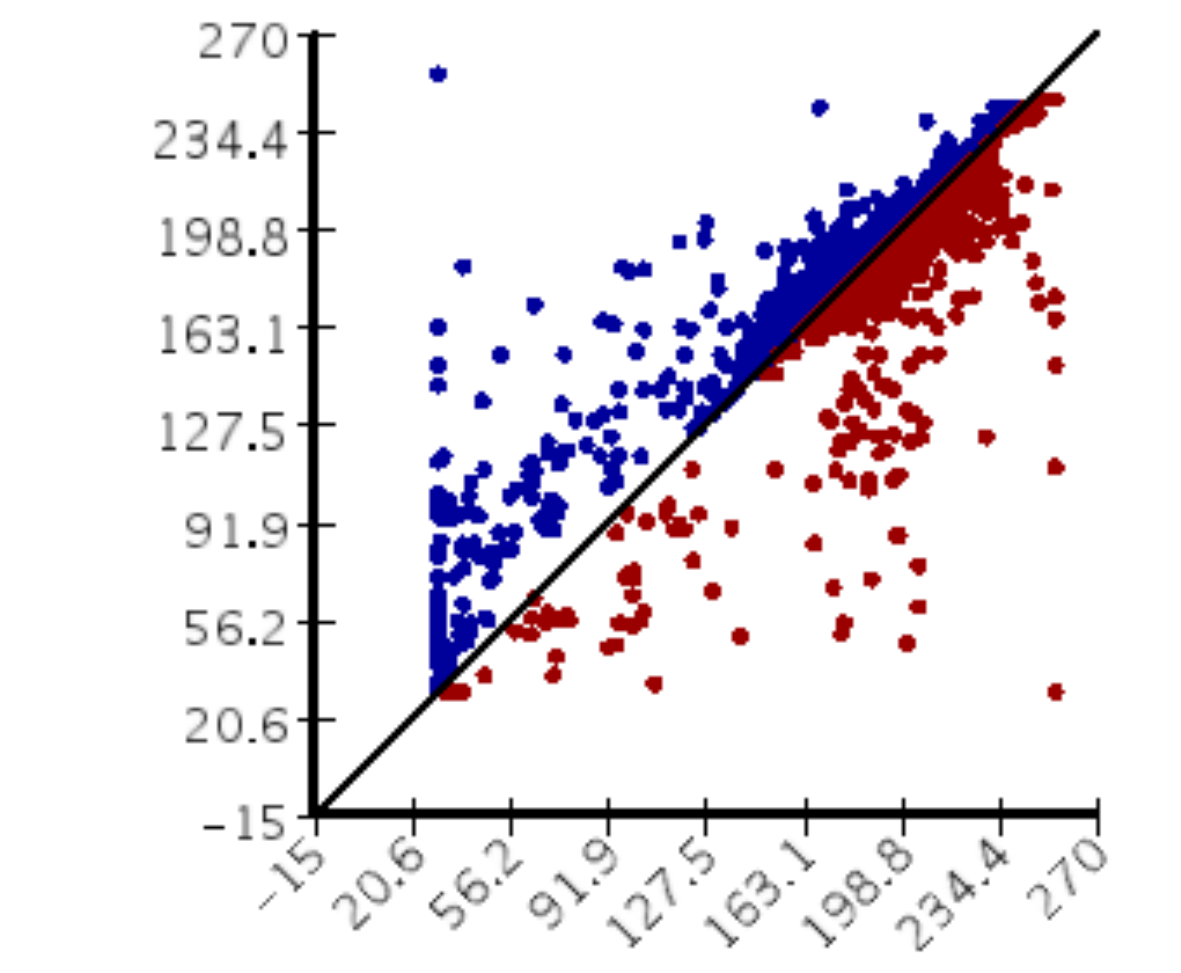}\hfill
        \includegraphics[height=0.60in]{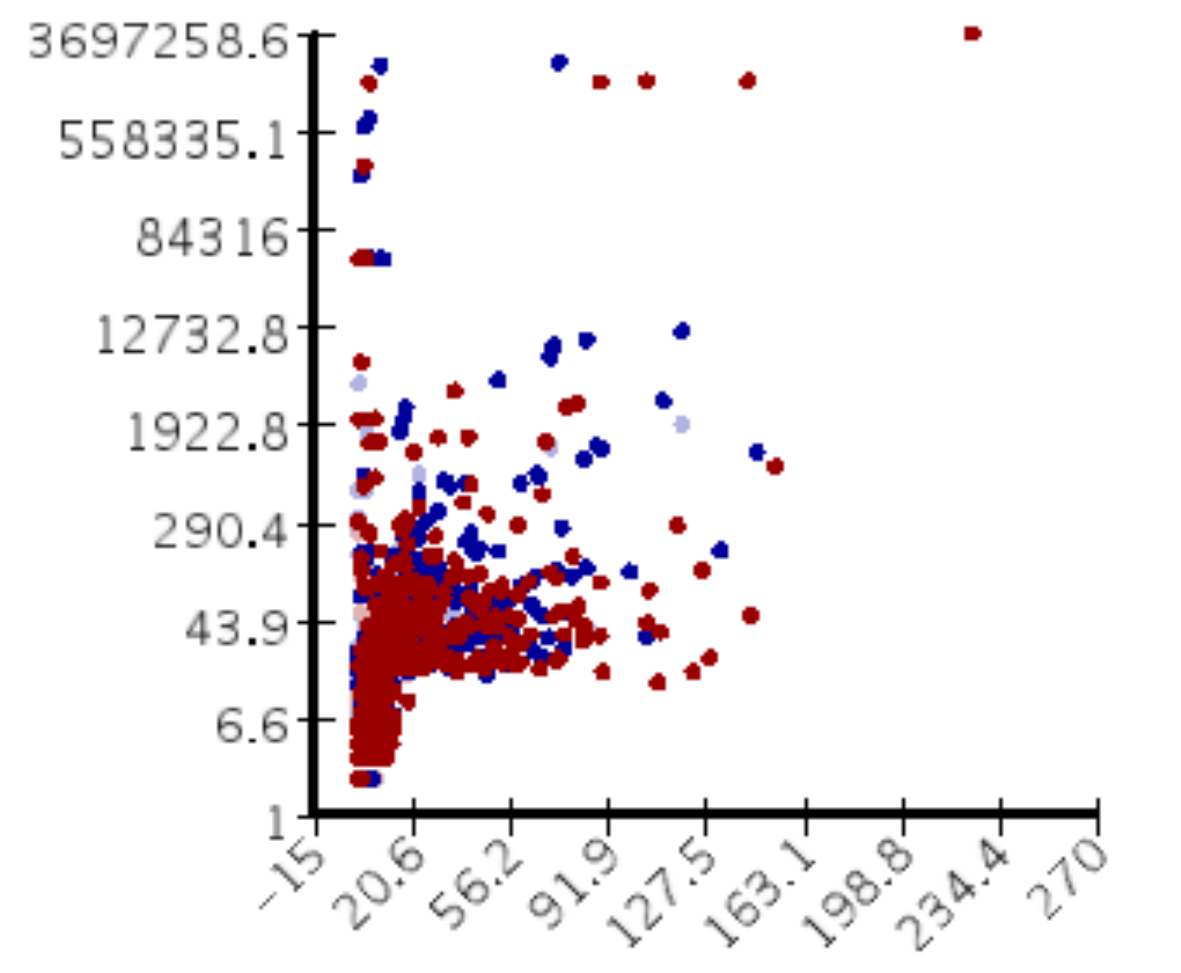}
        \end{minipage}
    }
    \hfill
    \subfigure[Reference Technique~\cite{Kervrann2019}\label{fig.ALflorala.ref}]{
        \begin{minipage}[m]{0.305\linewidth}
        \centering
        \includegraphics[height=1.025in,width=1\linewidth]{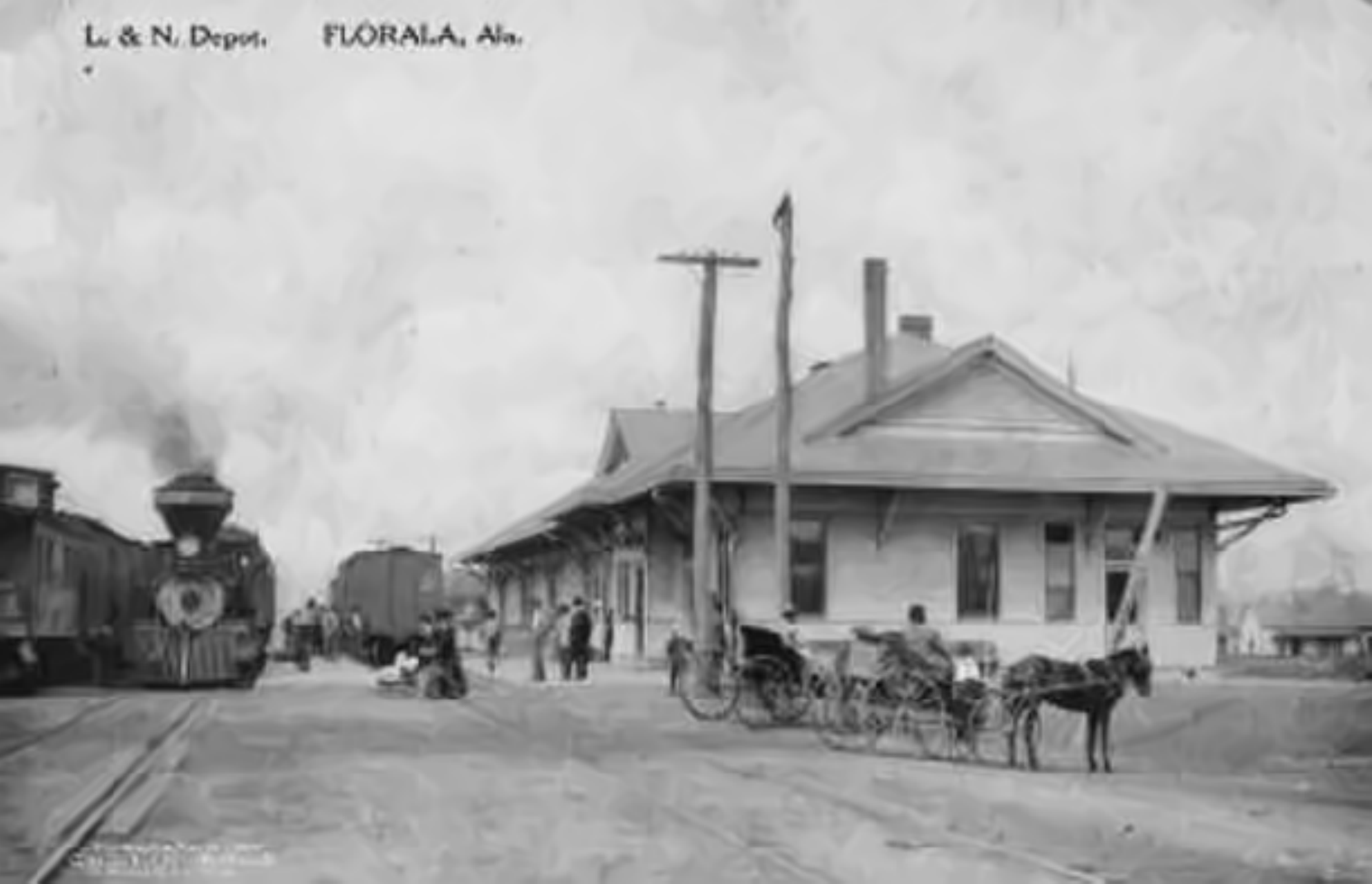} 

        \vspace{5pt}
        \includegraphics[height=0.60in]{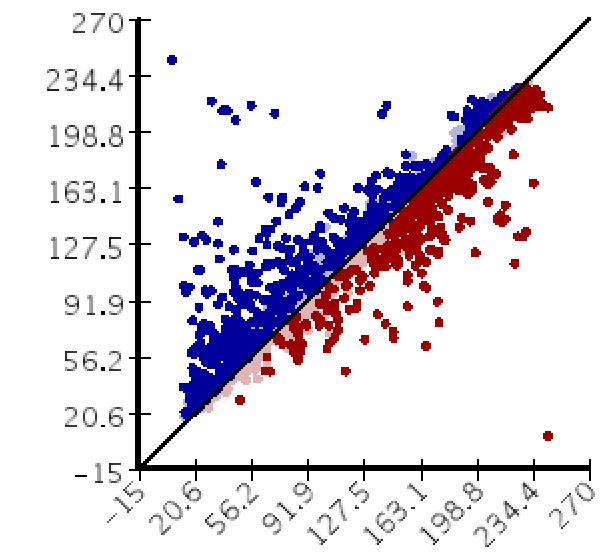}\hfill
        \includegraphics[height=0.60in]{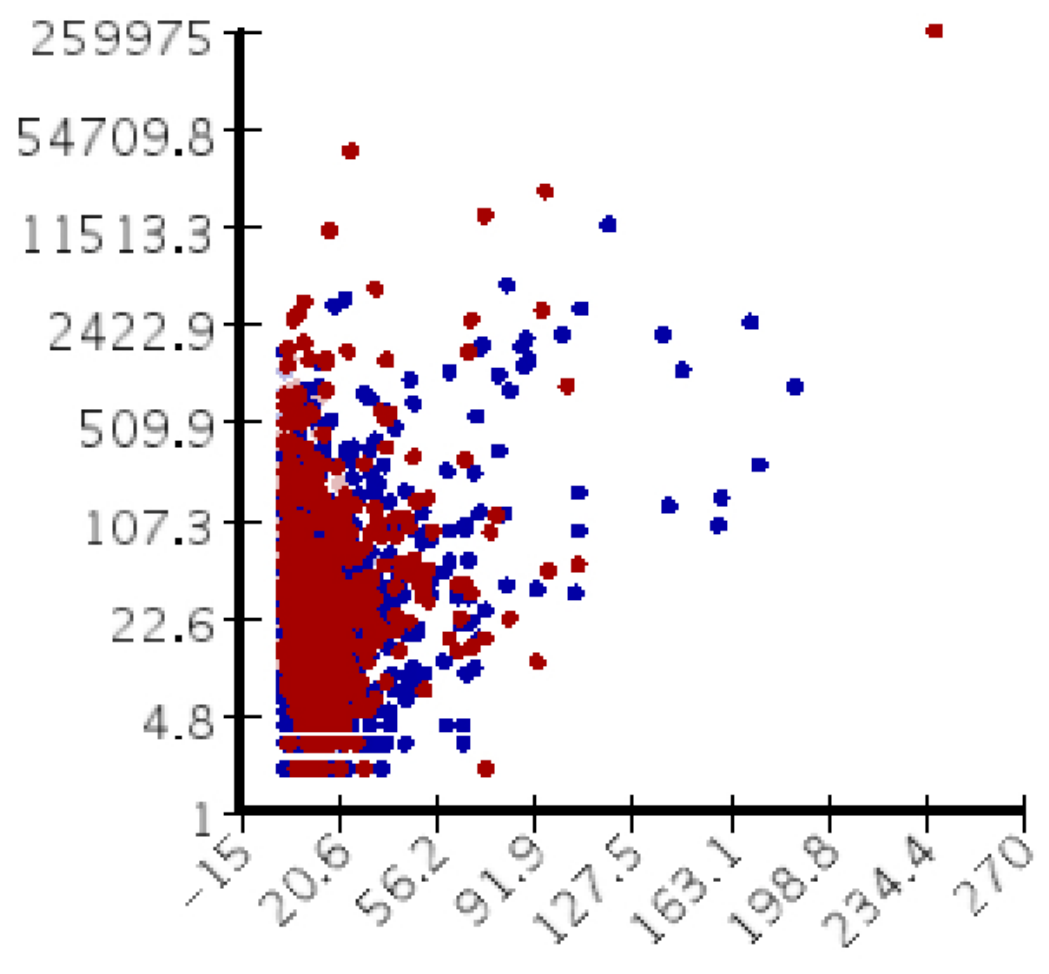}
    \end{minipage}
    }
        
    \caption{\textit{Florala dataset.} The (a) original, (b) cleaned by a combination of denoised, contrast enhancement, and gamma correction, is compared to (c) a reference image from~\cite{Kervrann2019}. A persistence (left) and persistence-volume (right) diagram is shown for each.}
    \label{fig.ALflorala}
\end{figure}

\textbf{Interface.}
The interface for selecting the editing mode and scale is shown in \figref{fig.controls.interface}. The interface designed represents how a chosen function modifies the persistence of the feature (horizontally) and the birth and death of a feature (vertically). For example, contrast enhancement increases persistence, while denoising decreases it. Brightness enhancement modified both the birth and death of a feature, leaving persistence unchanged. Finally, gamma correction changes nothing about the birth, death, or persistence. To use this interface, the user simply selects the button (i.e., the arrow or the circle) for the edit type they prefer. Finally, the level of the transfer function is selected using a slider.

\textbf{Topology Preservation.}
It is important to note that many of these edits will ultimately modify the local topology (i.e., within a subtree) and/or global topology (i.e., entire contour tree) of the image. Contrast enhancement and denoising change the persistence of local topology, and they can change feature pairs in the global context. Brightness enhancement makes no modification of the local topology, but significant changes can occur in the global topology (e.g., creation of new features, changes in persistence, etc.). Finally, gamma correction can modify the persistence of local topology, but it will have no effect on the global topology. To handle these cases, as part of the normal process of image editing, the contour trees are recalculated after an edit is applied to the image.

\begin{figure}[!b]
    \centering

    \begin{minipage}[m]{0.485\linewidth}
    \centering
    \subfigure[Original Image\label{fig.brain.input}]{
        \begin{minipage}[m]{0.475\linewidth}\includegraphics[height=1.30in,width=1\linewidth]{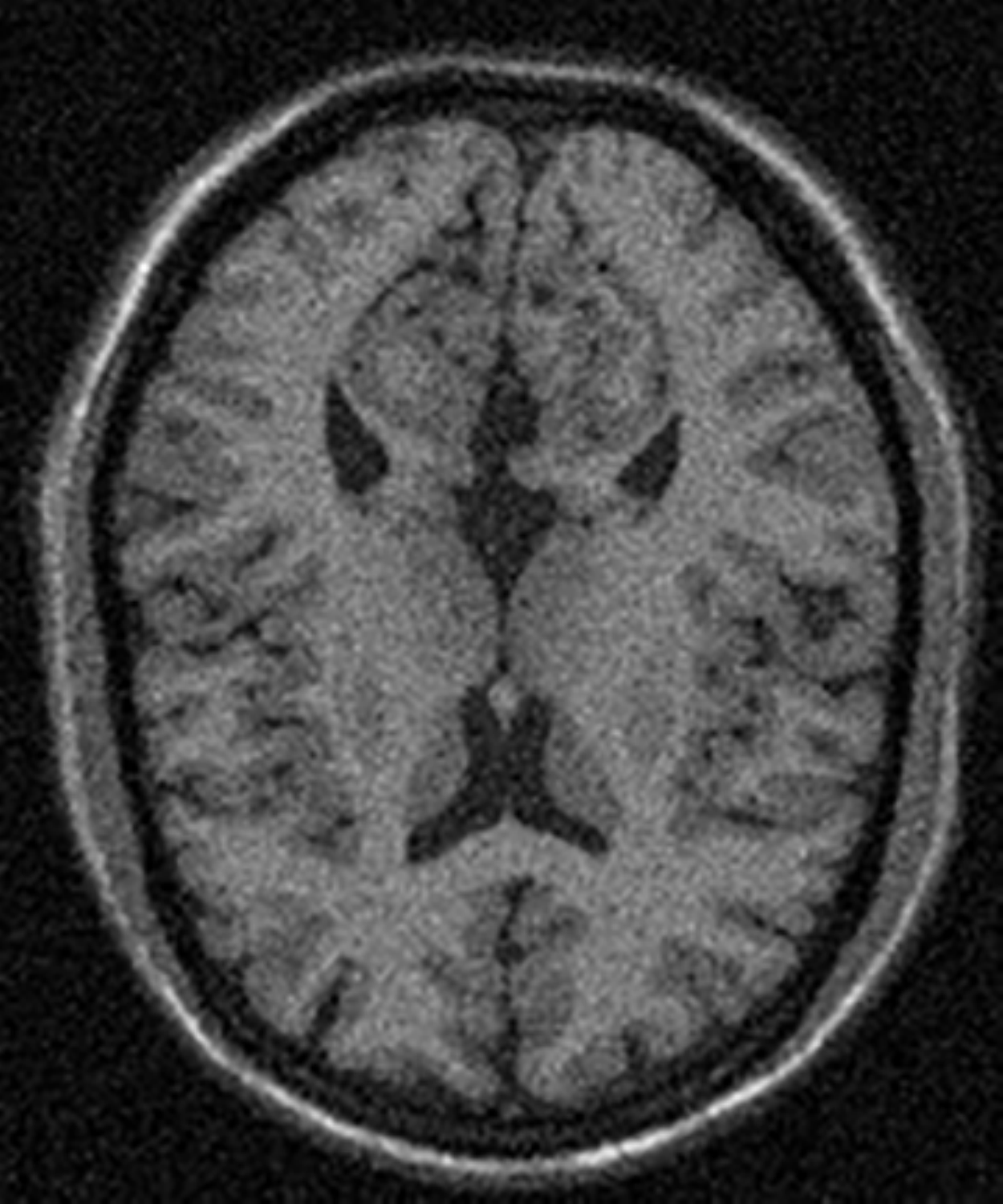}\vspace{5pt}\end{minipage}\hspace{5pt}
        \begin{minipage}[m]{0.33\linewidth}\includegraphics[width=1\linewidth]{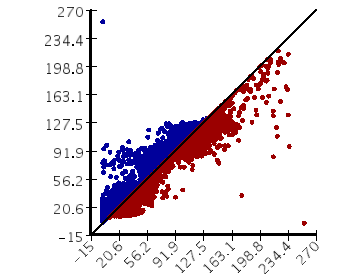}
        \includegraphics[width=1\linewidth]{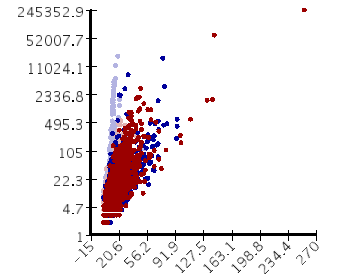}\end{minipage}
    }
    \end{minipage}
    \begin{minipage}[m]{0.485\linewidth}
    \centering
    \subfigure[Reference Image~\cite{Kervrann2019}\label{fig.brain.ref}]{
        \begin{minipage}[m]{0.475\linewidth}\includegraphics[height=1.25in,width=1\linewidth]{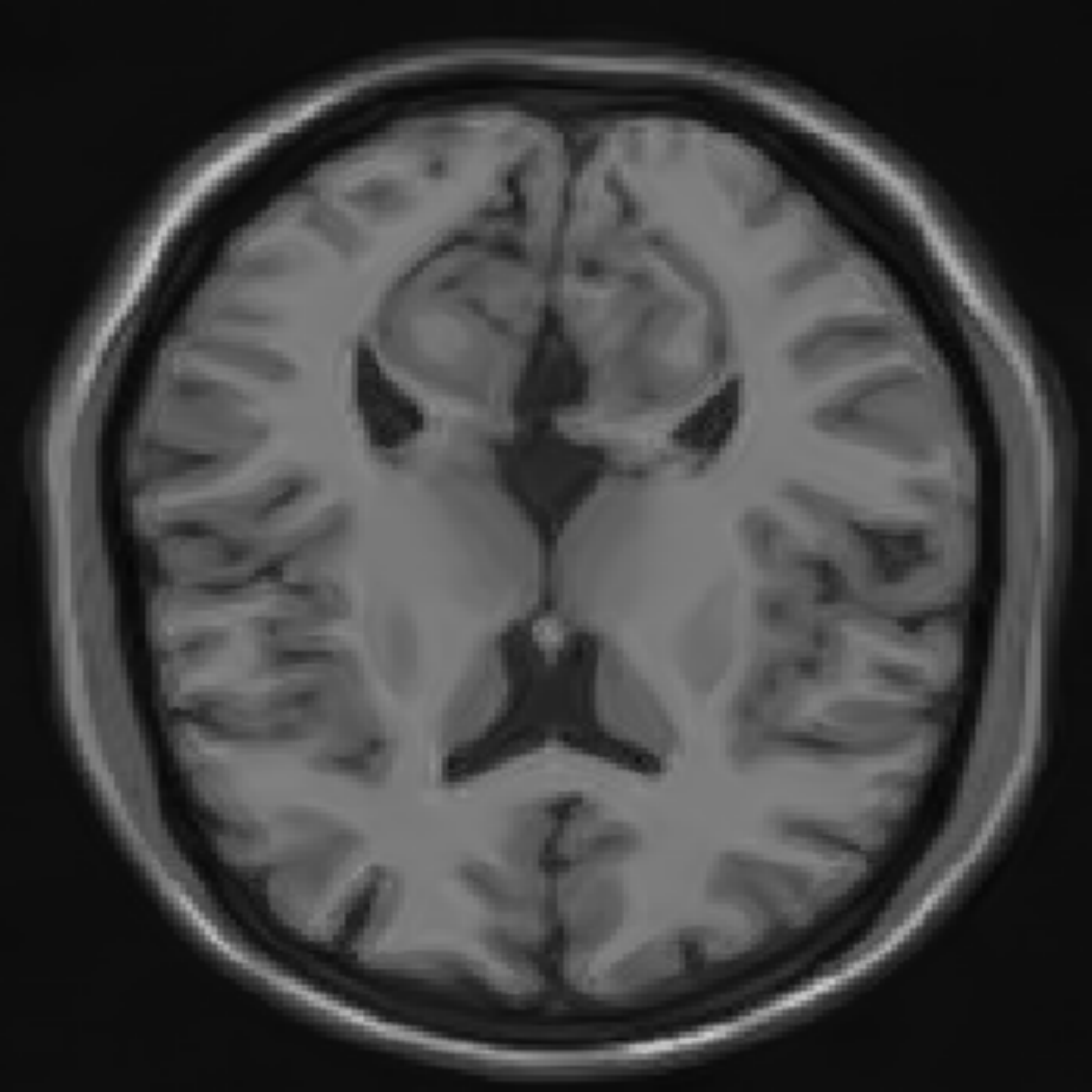}\vspace{5pt}\end{minipage}\hspace{5pt}
        \begin{minipage}[m]{0.315\linewidth}\includegraphics[width=1\linewidth]{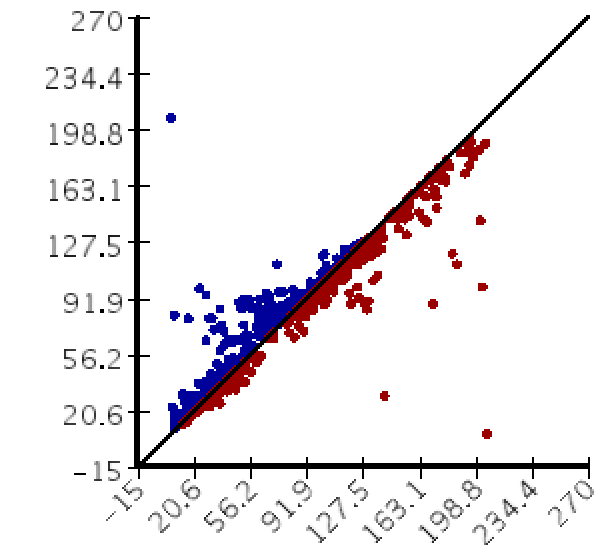}
        \includegraphics[width=1\linewidth]{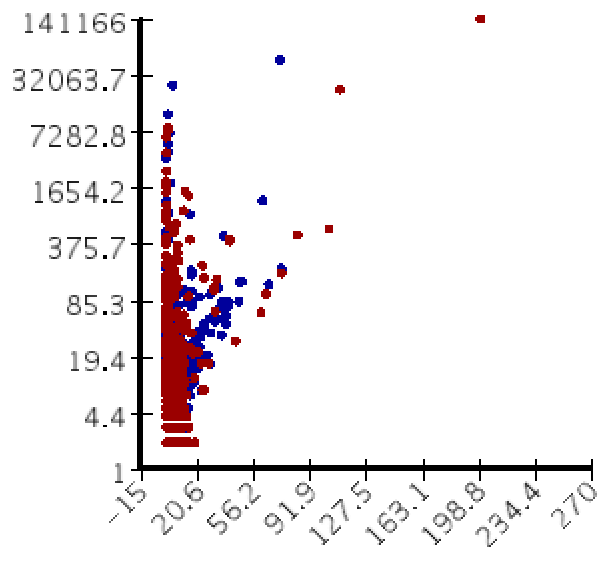}\end{minipage}
    }        
    \end{minipage}
    
    \begin{minipage}[m]{0.485\linewidth}
    \centering
    \subfigure[Midpoint Enhancement\label{fig.brain.output1}]{
        \begin{minipage}[m]{0.475\linewidth}\includegraphics[height=1.25in,width=1\linewidth]{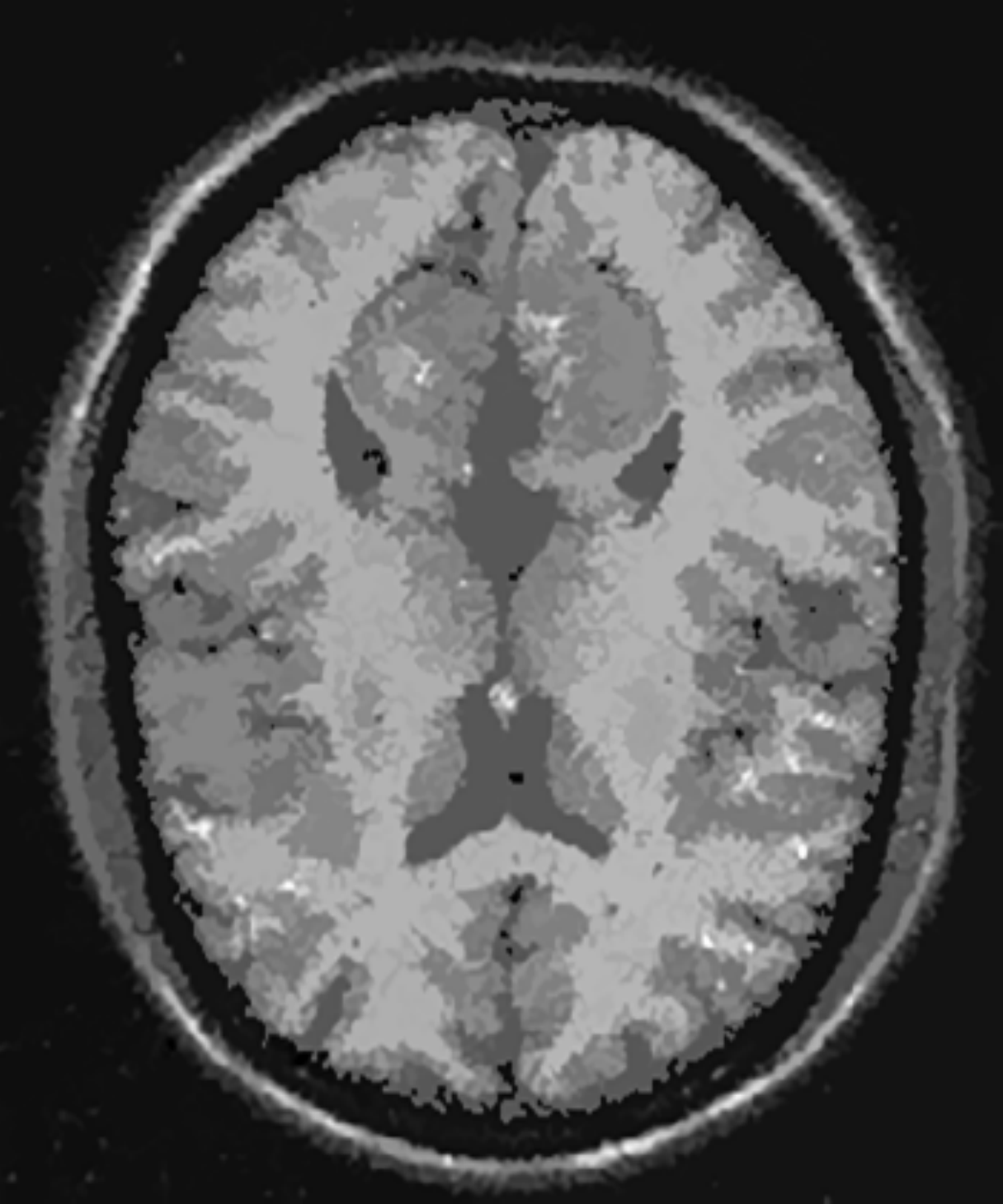}\vspace{5pt}\end{minipage}\hspace{5pt}
        \begin{minipage}[m]{0.33\linewidth}\includegraphics[width=1\linewidth]{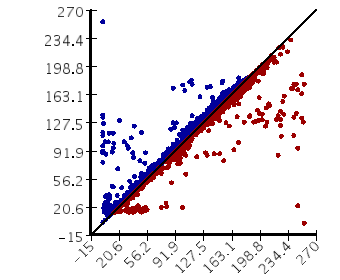}
        \includegraphics[width=1\linewidth]{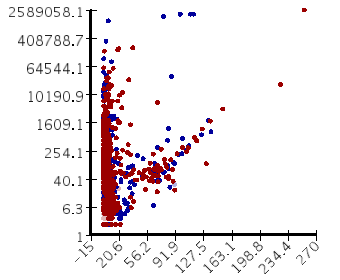}\end{minipage}
    }
    \end{minipage}    
    \begin{minipage}[m]{0.485\linewidth}
    \centering
    \subfigure[Final Enhancement\label{fig.brain.output2}]{
        \begin{minipage}[m]{0.475\linewidth}\includegraphics[height=1.25in,width=1\linewidth]{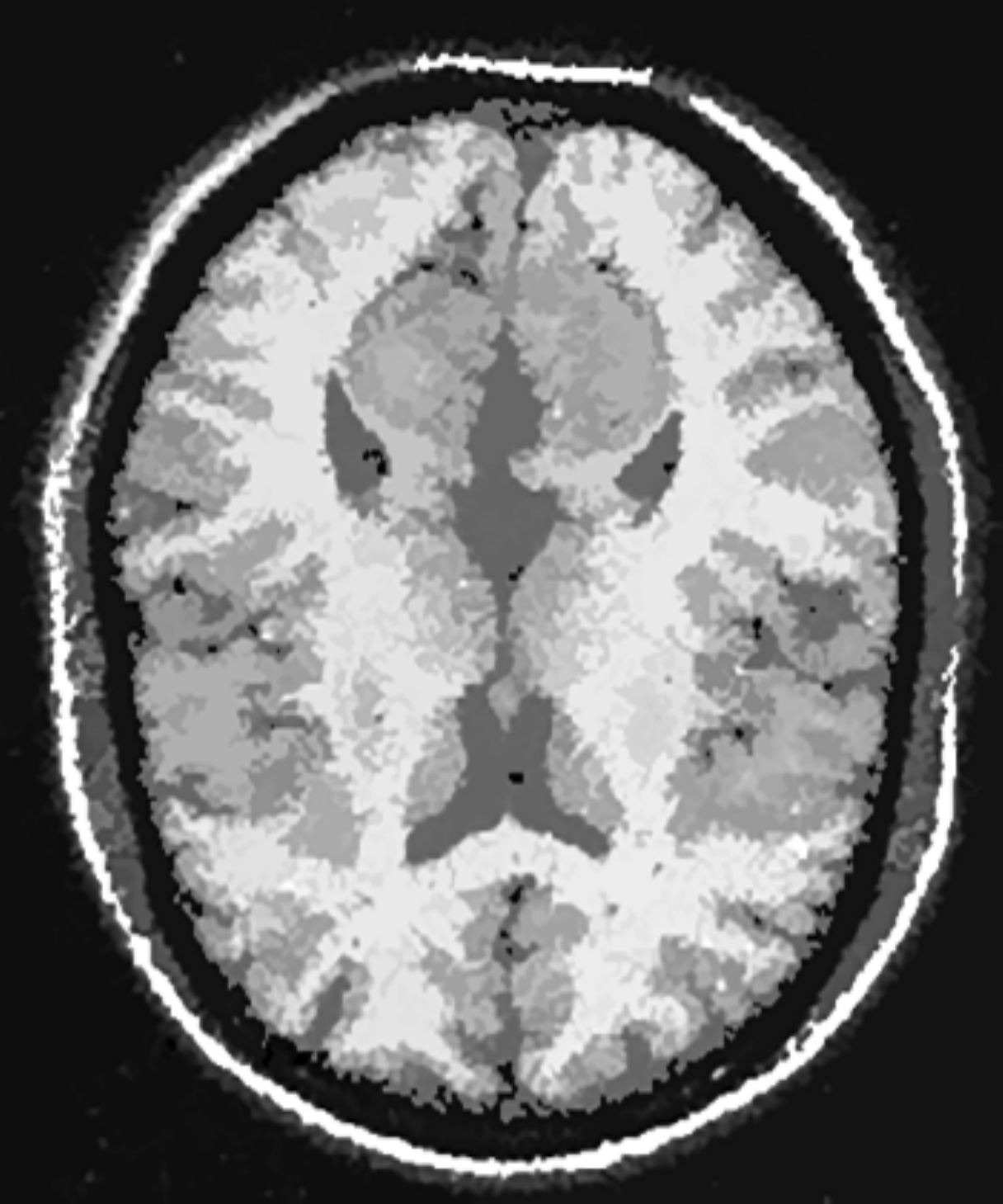}\vspace{5pt}\end{minipage}\hspace{5pt}
        \begin{minipage}[m]{0.33\linewidth}\includegraphics[width=1\linewidth]{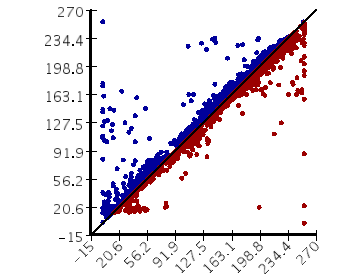}
        \includegraphics[width=1\linewidth]{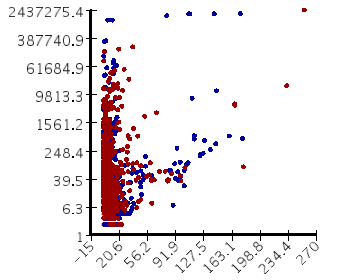}\end{minipage}
    }
    \end{minipage}

    \caption{\textit{Brain dataset} cleaned using a combination of all 4 functions. Each image include its persistence diagram (top) and persistence-volume diagram (bottom).}
    \label{fig.brain}
\end{figure}

\section{Examples}
\label{sec:experiments}

We have implemented a prototypes of our approach using Java. Images were generated using a 2017 MacBook Pro. All aspects of our approach are interactive, except the construction of the contour trees and extraction of feature pair subtrees, which took between 5 and 30 seconds, depending upon the size and complexity of the image. Before considering the quality of an enhancement, context, such as how the image will be utilized, needs to be taken into account. We do not consider this context in our analysis, and instead only provide examples enhancements.

\textbf{Synthetic Example.} Our synthetic example, based upon the selection of the $m/o$ feature from \figref{fig.segementation.selection}, has the operations outlined in \figref{fig.controls} applied: contrast enhancement (\figref{fig.synth.contrast}), denoising (\figref{fig.synth.denoising}), brightness enhancement (\figref{fig.synth.brightness}), and gamma correction (\figref{fig.synth.gamma}).

\subsection{Grayscale Images} 
Since contour trees operate on a single channel, grayscale images are a natural way to demonstrate this functionality. For the following datasets, the brightness channel of the HSB colorspace was used for editing. 

The \textit{Florala dataset}, seen in \figref{fig.ALflorala.input}, is a photograph of Florala Alabama, retrieved from~\cite{Kervrann2019}. The photograph had a series of denoising, contrast enhancement, and gamma correction steps applied to recover the final image in \figref{fig.ALflorala.output}. The persistence and persistence-volume diagrams are shown for comparison.

The \textit{Brain dataset}, seen in \figref{fig.brain.input}, is a noisy and low contrast MRI scan of a brain, retrieved from~\cite{Kervrann2019}. The figure shows a midpoint (\figref{fig.brain.output1}) and the final version (\figref{fig.brain.output2}) after a combination of 12 contrast, brightness, gamma correction, and denoising steps. The final image removes noise and highlights important features, such as the skull, white matter, and grey matter. The persistence and persistence-volume diagrams are shown for comparison.

\begin{figure}[!t]
    \centering
    
    \subfigure[Original]{\hspace{2pt}\includegraphics[width=0.21\linewidth]{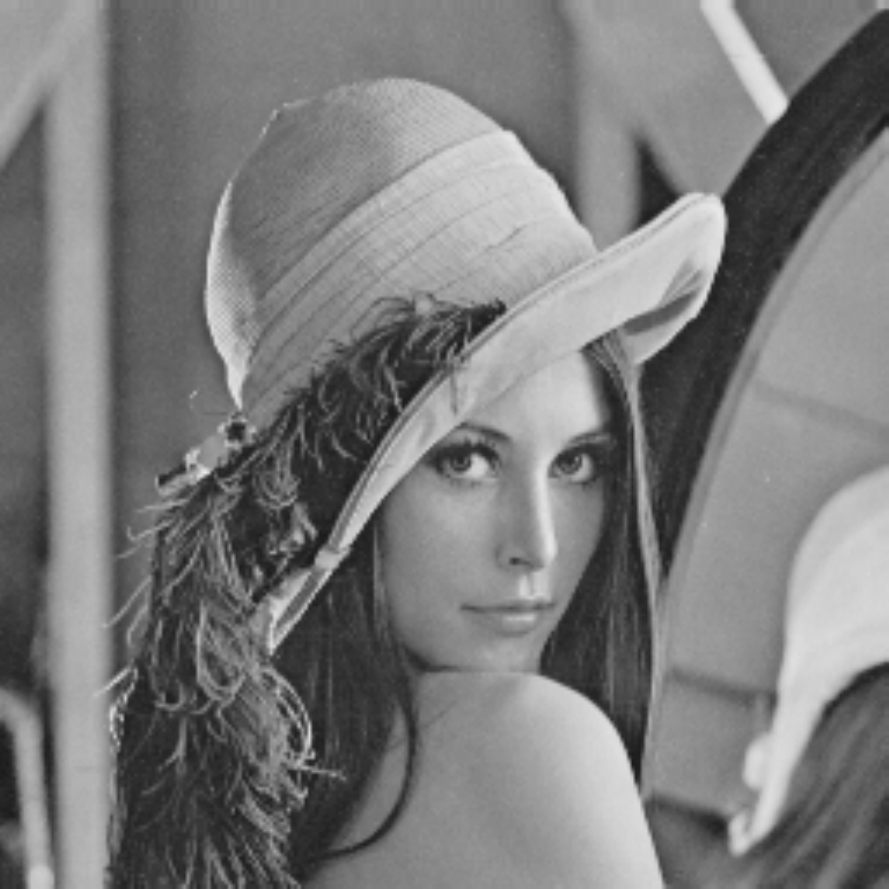}\hspace{2pt}}
    \hfill
    \subfigure[Original w/noise]{\hspace{4pt}\includegraphics[width=0.21\linewidth]{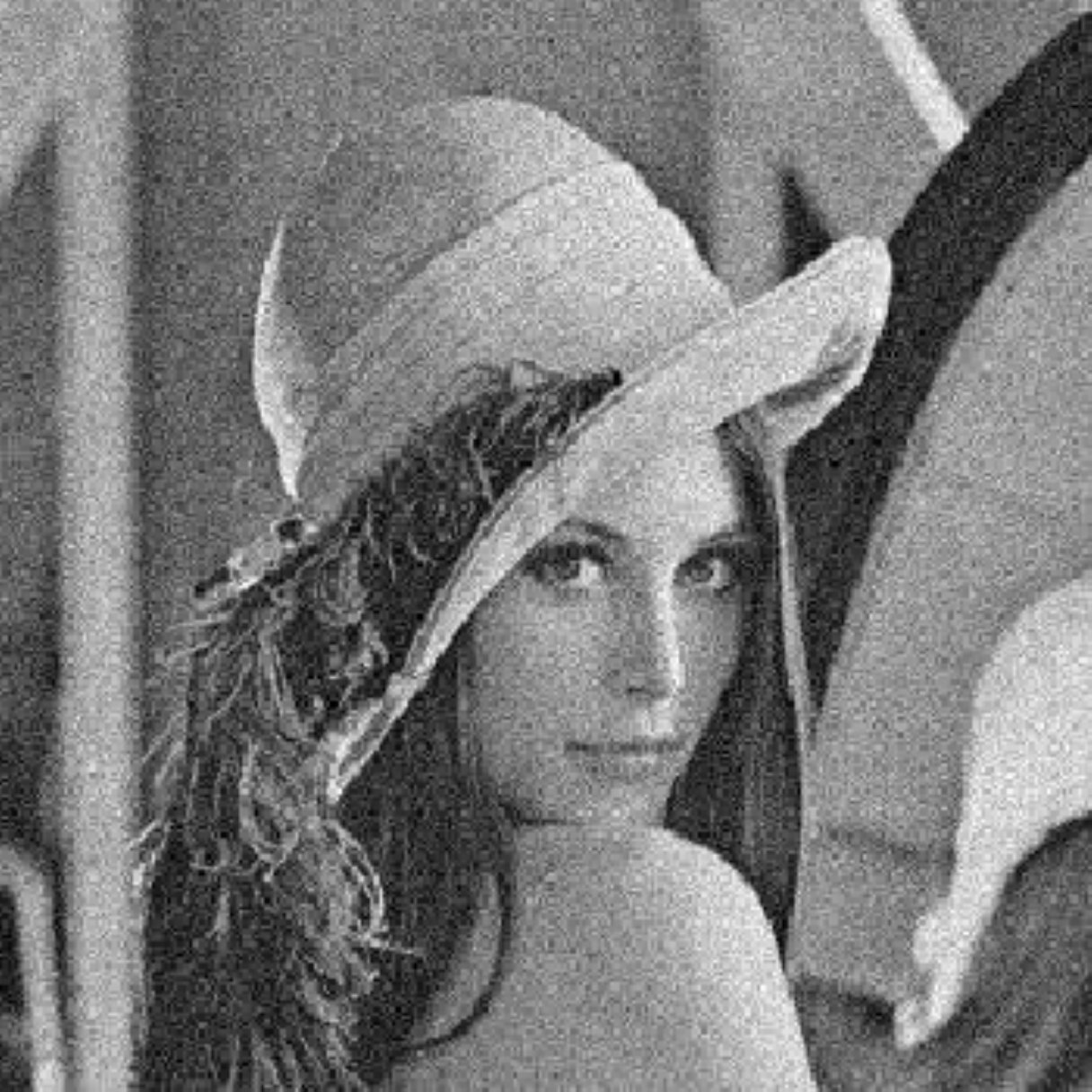}\hspace{4pt}}
    \hfill
    \line(0,1){80}
    \hfill
    \subfigure[Denoising]{\hspace{0pt}\includegraphics[width=0.21\linewidth]{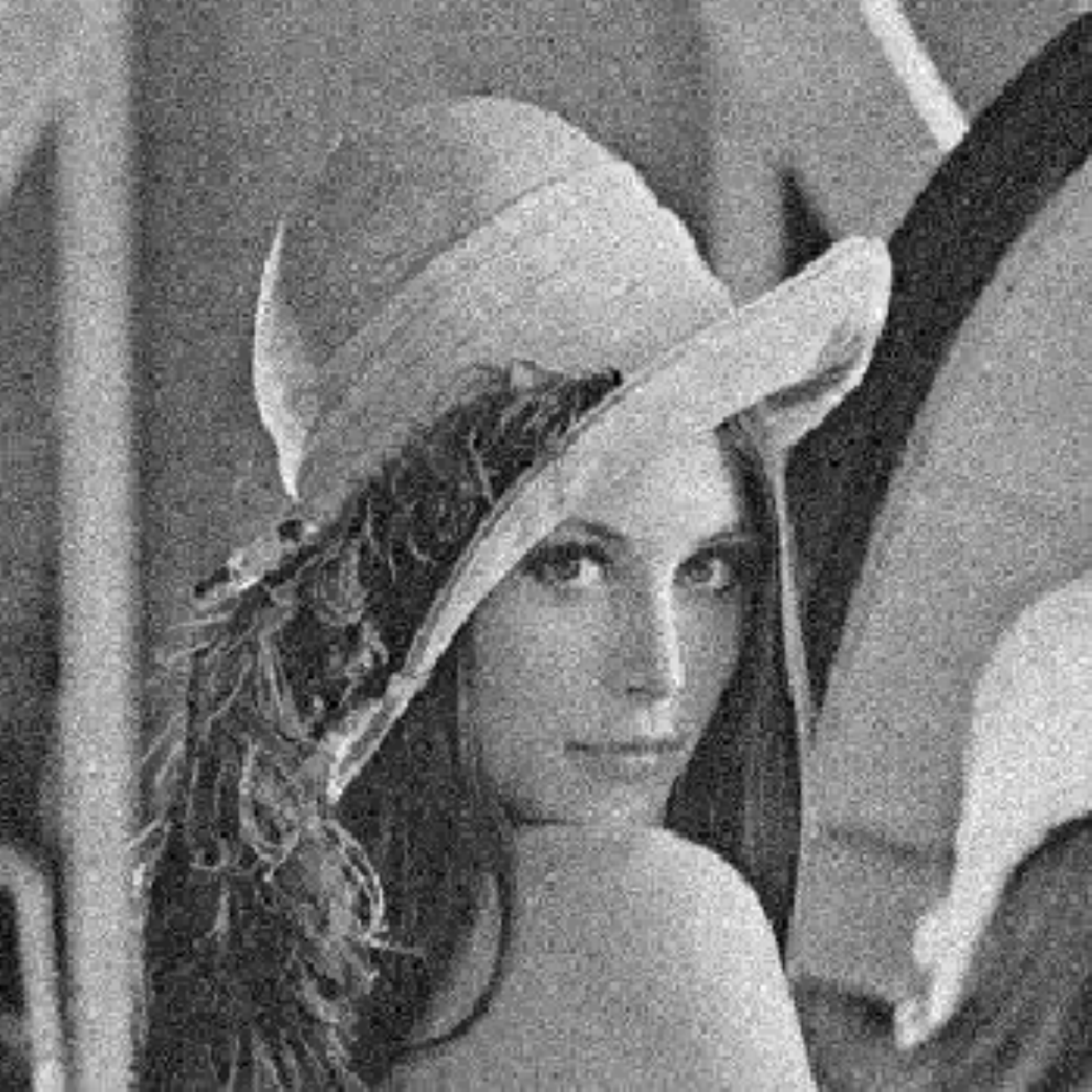}\hspace{0pt}}
    \hfill
    \subfigure[Denoising]{\hspace{0pt}\includegraphics[width=0.21\linewidth]{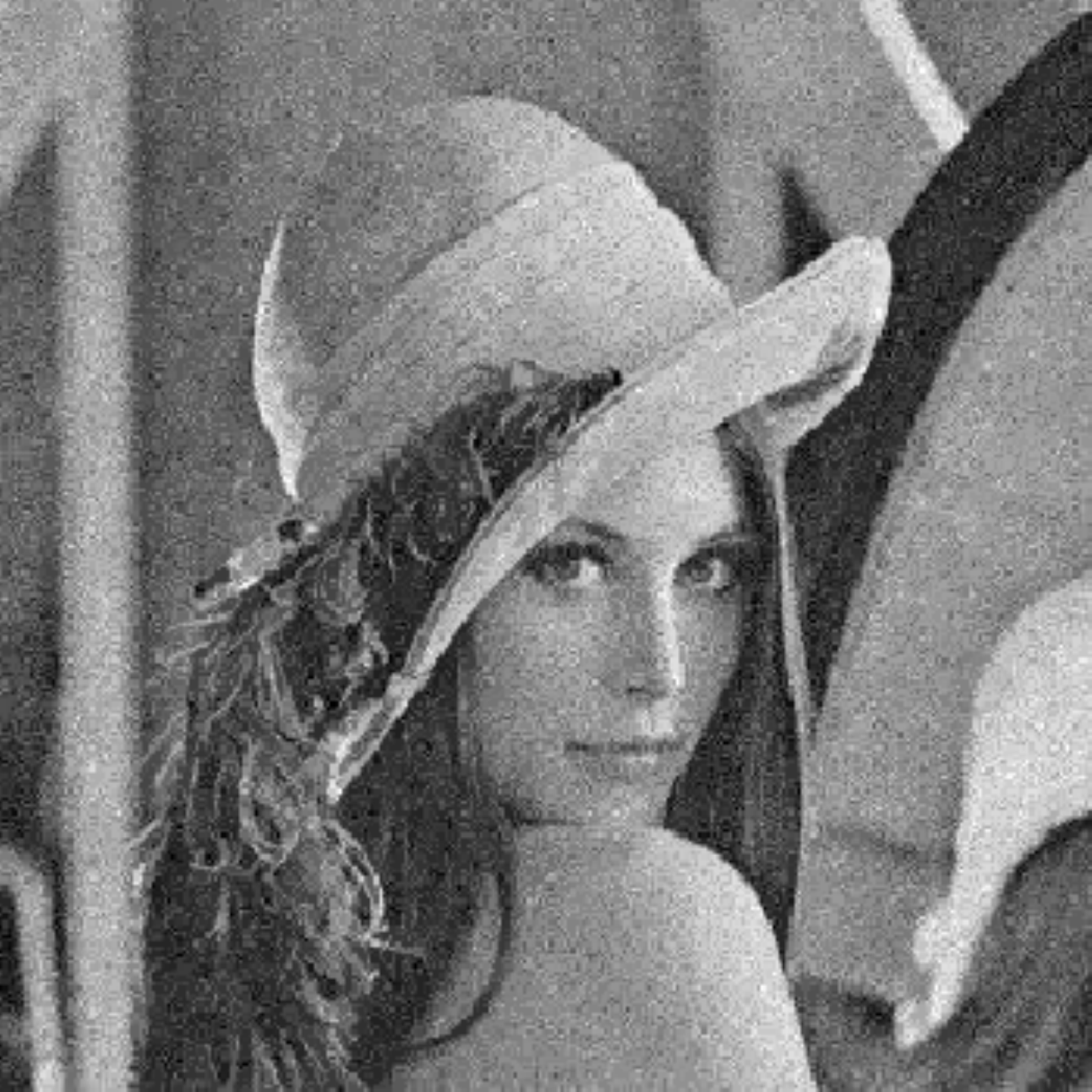}\hspace{0pt}}
    
    \subfigure[Denoising]{\hspace{0pt}\includegraphics[width=0.21\linewidth]{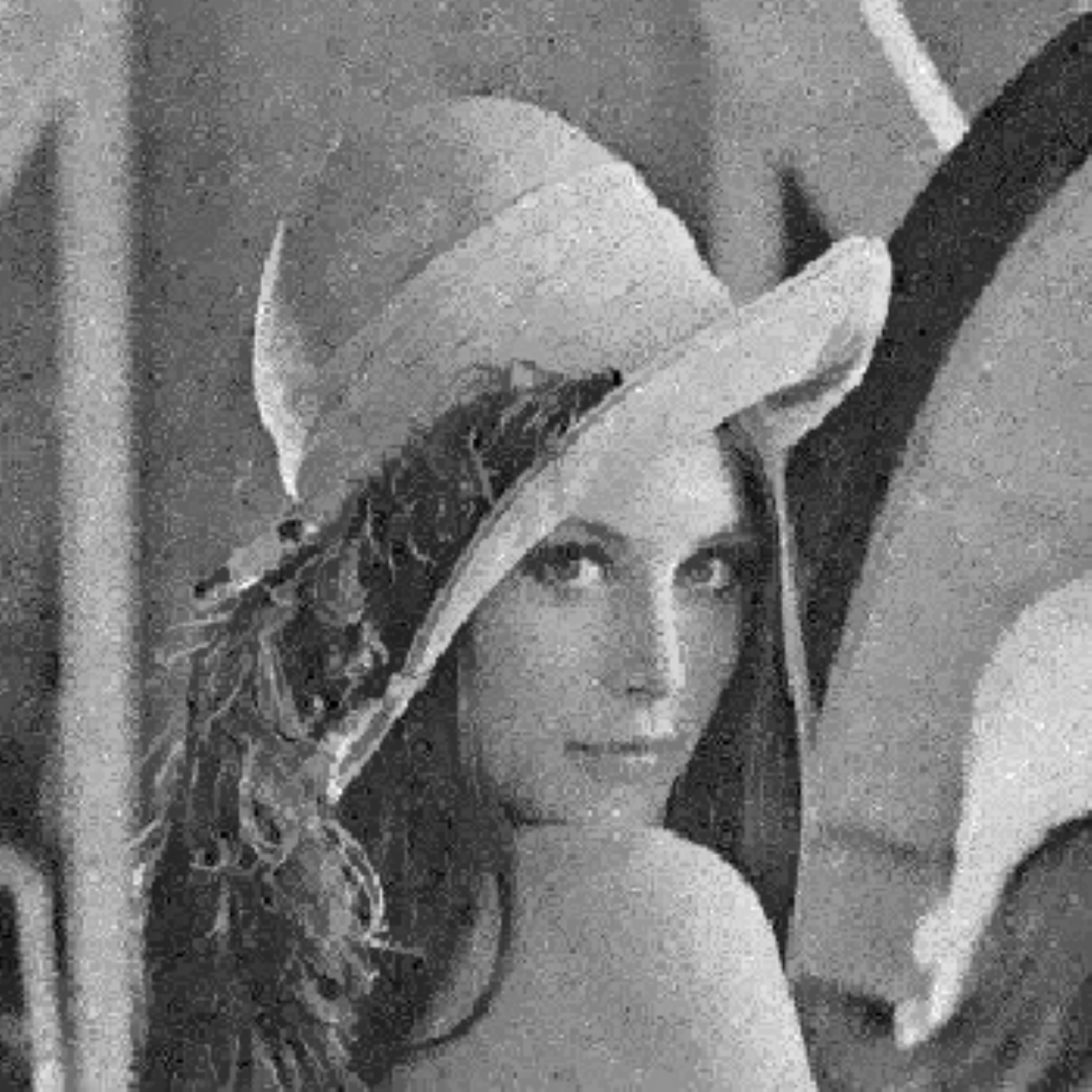}\hspace{0pt}}
    \hfill
    \subfigure[Contrast]{\hspace{0pt}\includegraphics[width=0.21\linewidth]{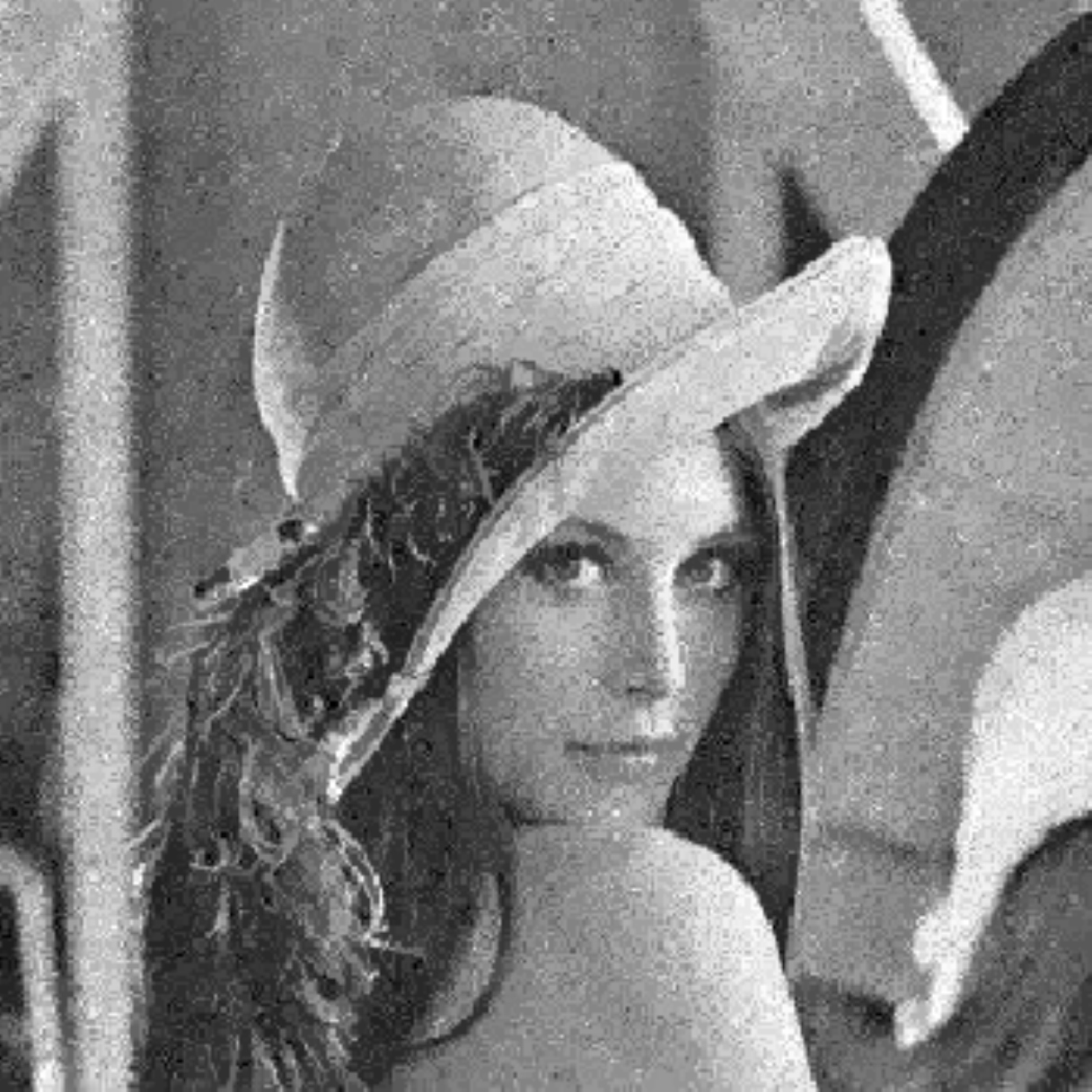}\hspace{0pt}}        
    \hfill
    \subfigure[Contrast]{\hspace{0pt}\includegraphics[width=0.21\linewidth]{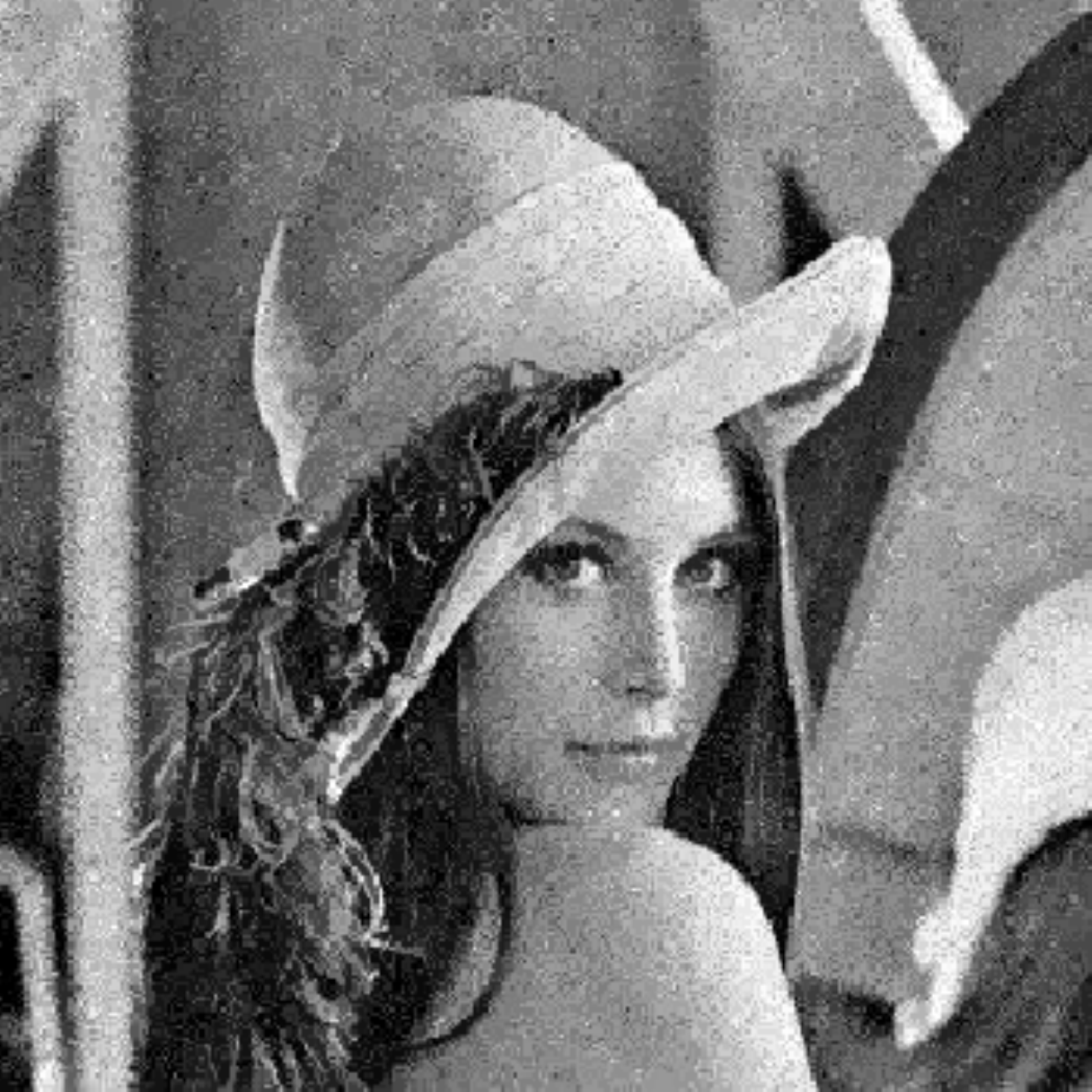}\hspace{0pt}}
    \hfill
    \subfigure[Denoising]{\hspace{0pt}\includegraphics[width=0.21\linewidth]{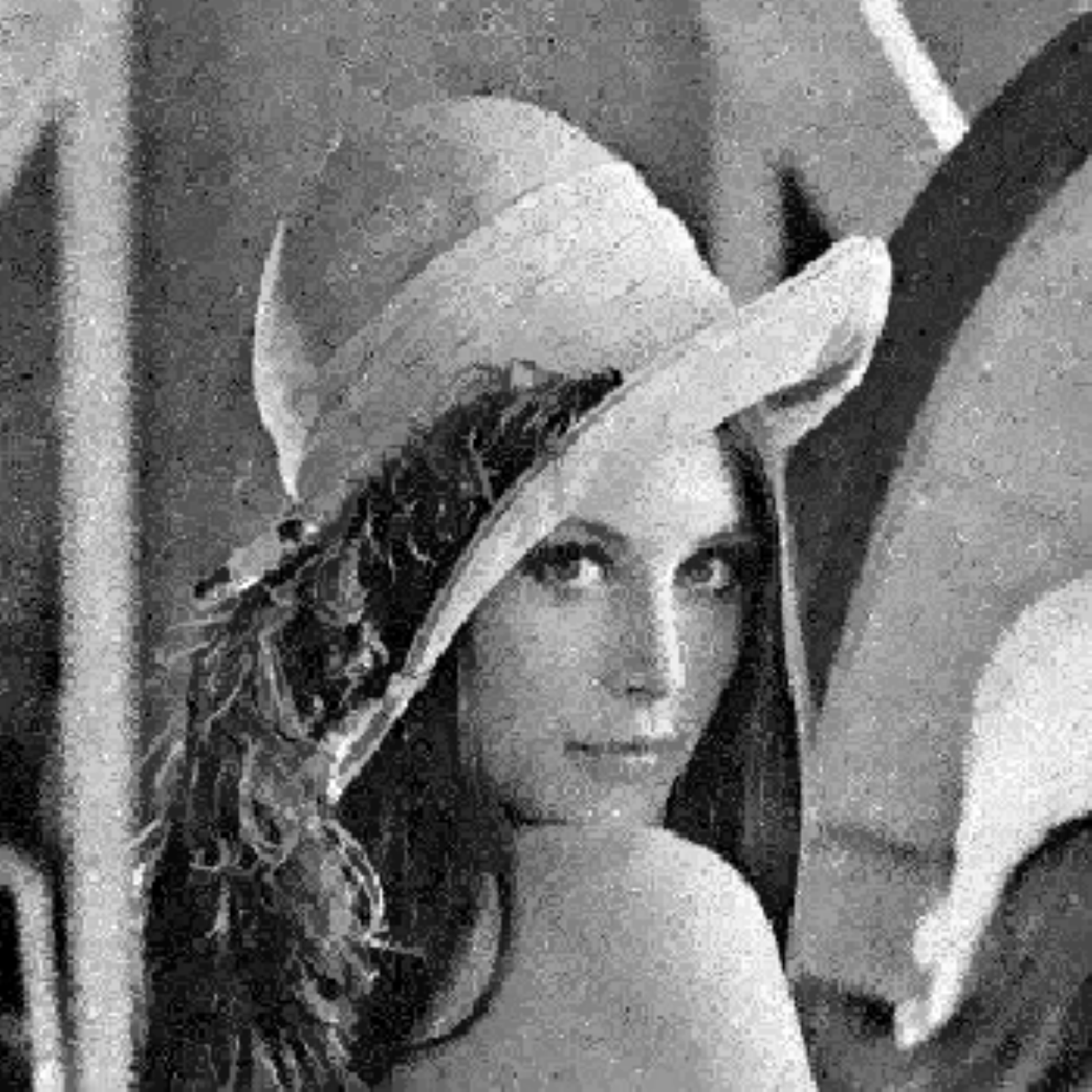}\hspace{0pt}}            

    \caption{(a/b) \textit{Lenna Grayscale}. (c-h) Series of denoising and contrast enhancement.}
    \label{fig:lenna_bw}
\end{figure}

The \textit{Lenna Grayscale Dataset}, seen in \figref{fig:lenna_bw} shows a series of 7 edits to a noisy version of the classic Lenna dataset.

\begin{figure}[!th]
    \centering
    \begin{minipage}{0.575\linewidth}
    \subfigure[Original Data\label{fig.notredam.input}]{\includegraphics[height=0.85in,width=0.485\linewidth]{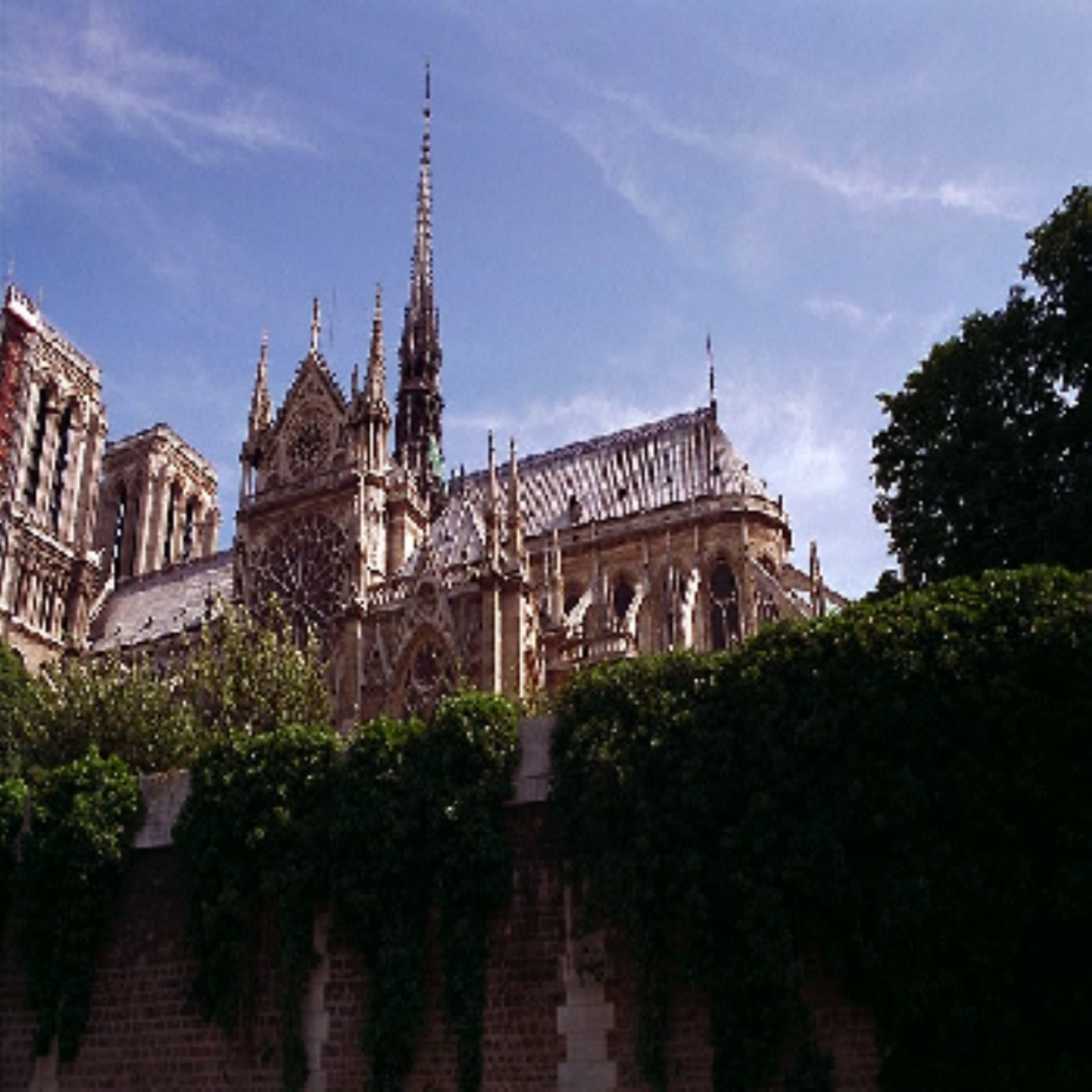}} 
    \hfill
    \subfigure[Reference Method~\cite{fattal2002gradient} \label{fig.notredam.ref}]{\includegraphics[height=0.85in,width=0.485\linewidth]{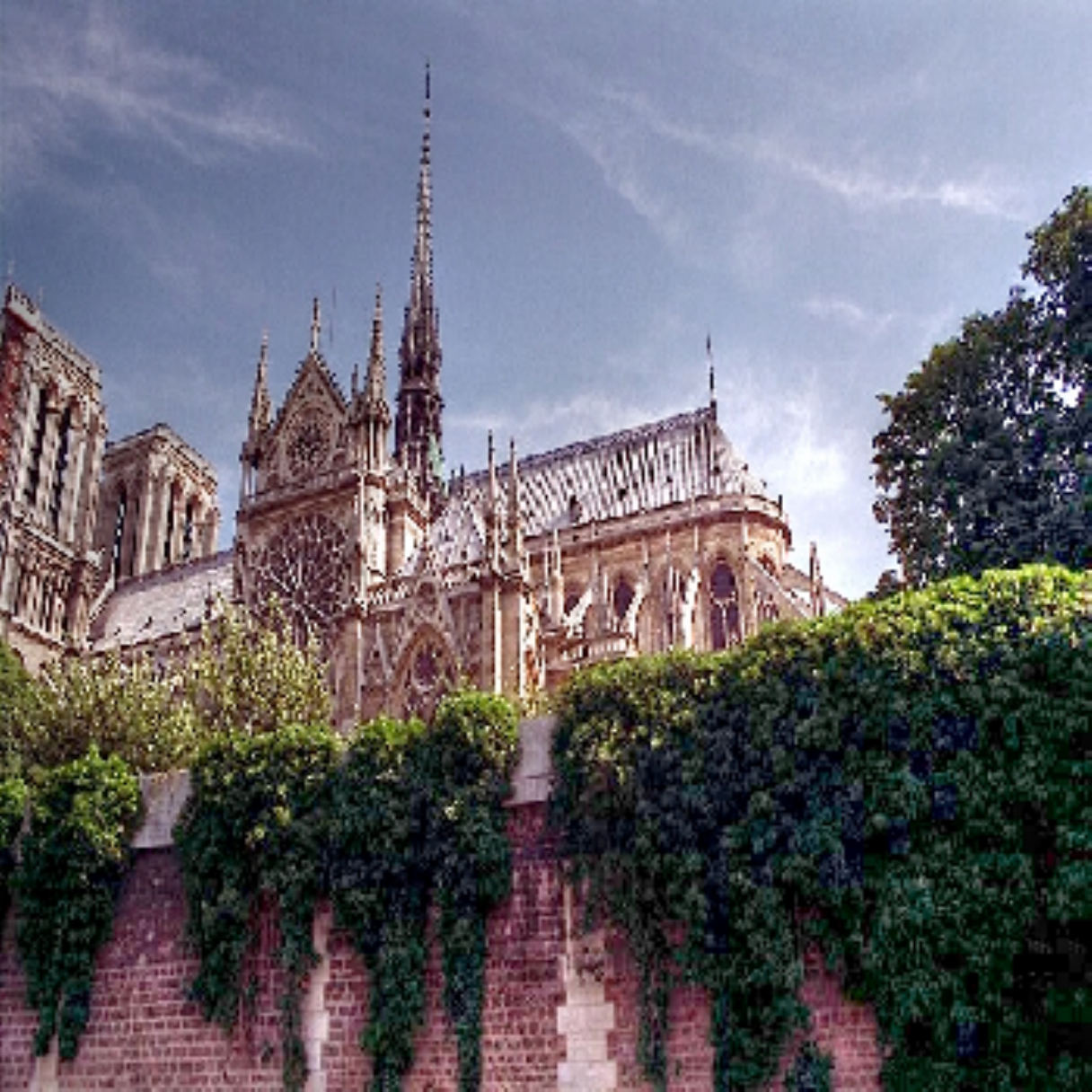}}
    
    \vspace{-9pt}
    \subfigure[Green Channel\label{fig.notredam.output1}]{\includegraphics[height=0.85in,width=0.485\linewidth]{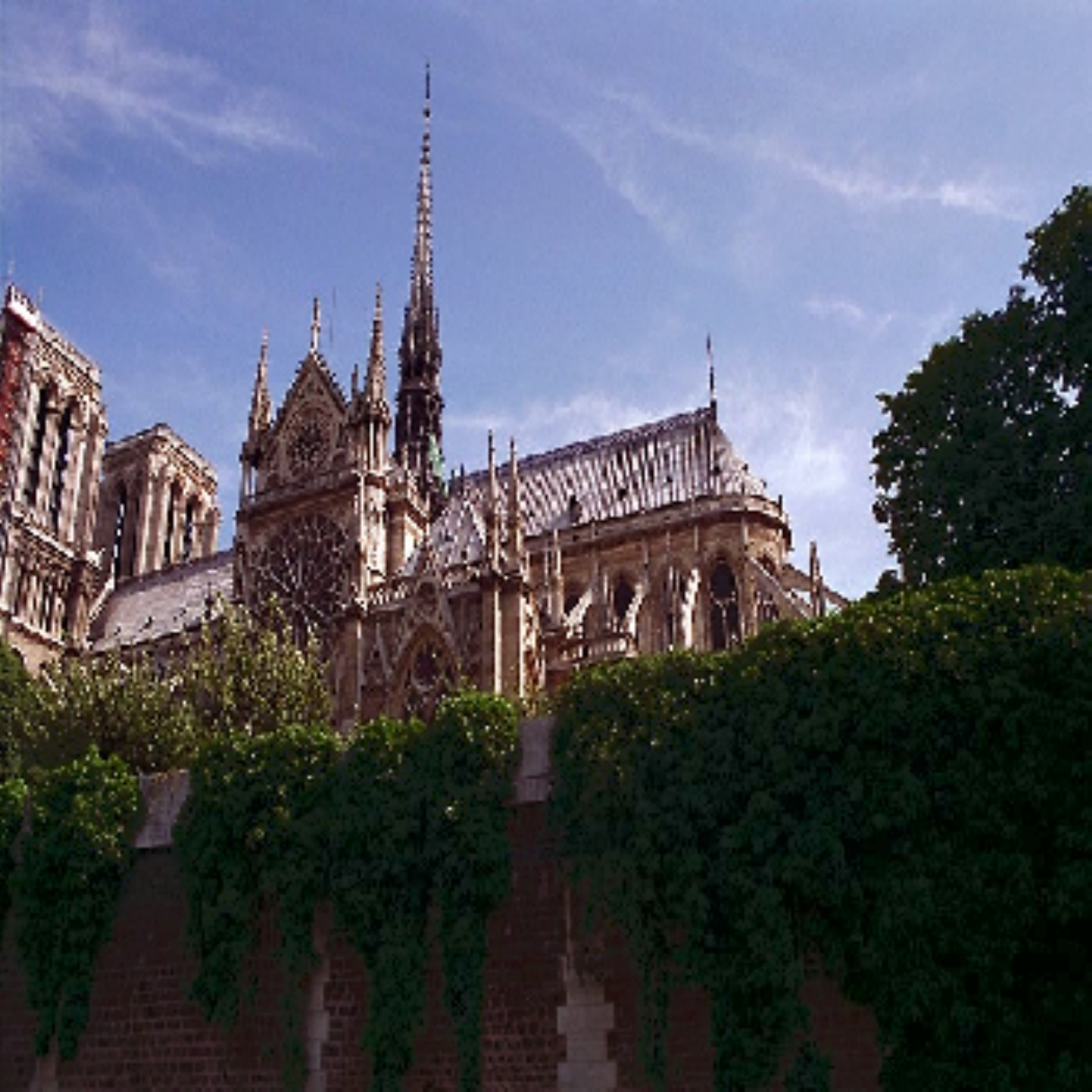}} 
    \hfill
    \subfigure[Saturation Channel\label{fig.notredam.output2}]{\includegraphics[height=0.85in,width=0.485\linewidth]{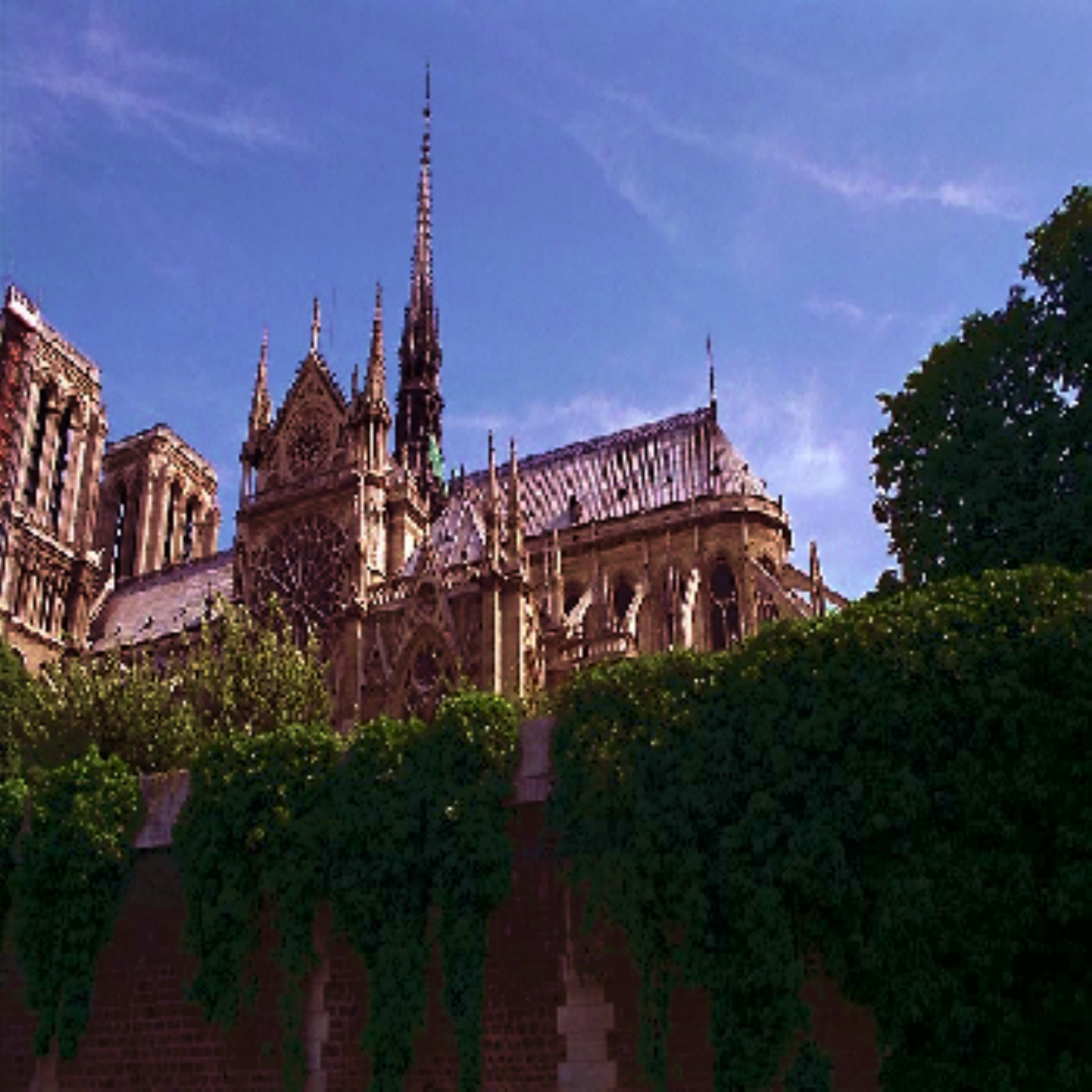}}     
    \end{minipage}
    \hfill
    \begin{minipage}[m]{0.4\linewidth}
    \subfigure[Brightness Channel\label{fig.notredam.output3}]{\includegraphics[height=1.30in,width=1\linewidth]{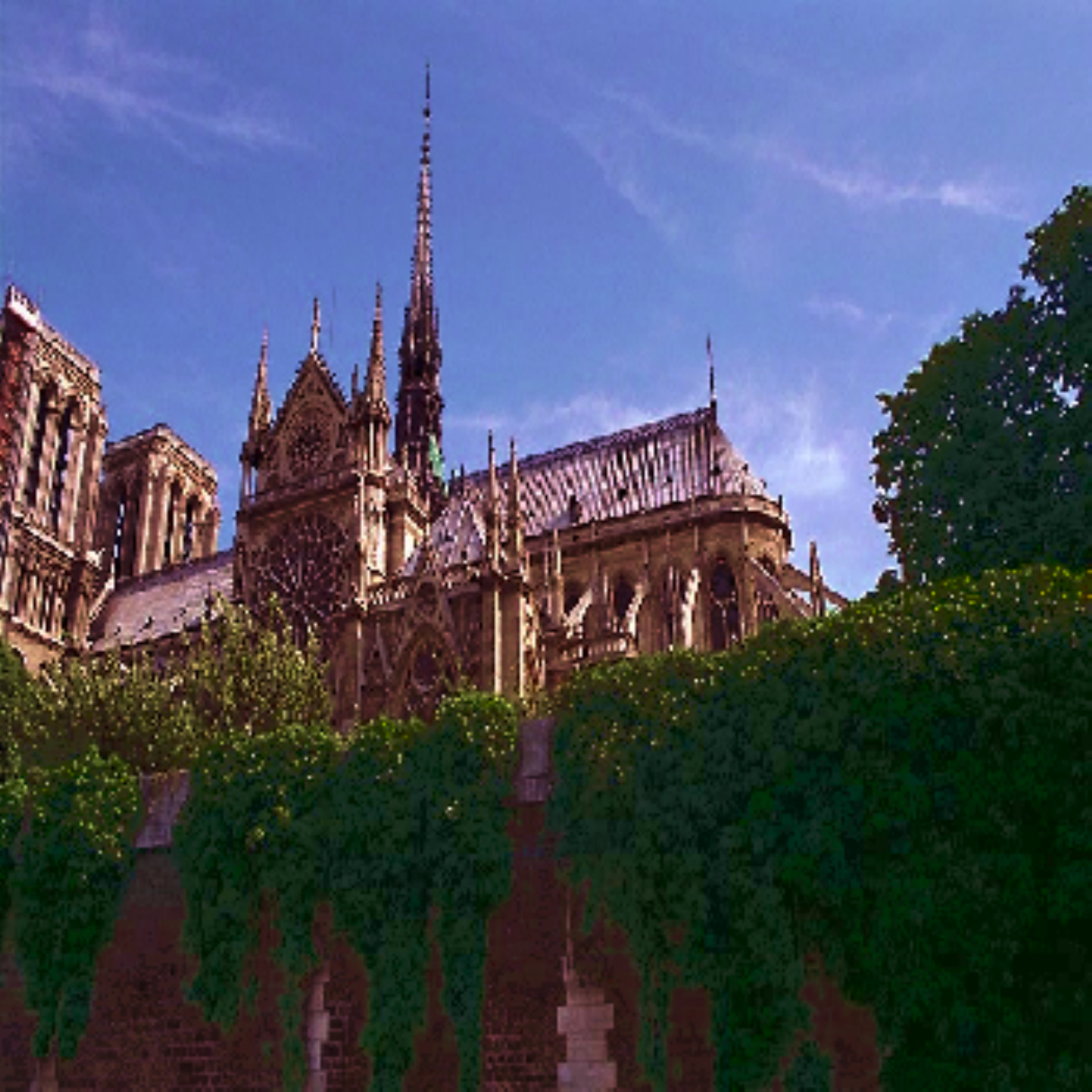}} 
    \end{minipage}

    \caption{(a) \textit{Notre Dame dataset.} (b) Reference method~\cite{fattal2002gradient} is compared to a version modified using (c) the green channel, then (d) saturation, and finally (e) the brightness.}
    \label{fig.notredam}
\end{figure}

\subsection{Color Images} 

The \textit{Notre Dame dataset}, retrieved from~\cite{fattal2002gradient} and shown in \figref{fig.notredam.input}, is a photo with an underexposed foreground. For this dataset, we first performed a (virtually invisible) brightness enhancement to the green color channel, to make the foliage a green hue (see \figref{fig.notredam.output1}). Next, we performed a denoising and contrast enhancement to the saturation channel of the HSB colorspace (see \figref{fig.notredam.output2}). Finally, denoising, gamma correction, and contrast enhancement were applied to the brightness channel of HSB. Persistence and persistence-volume diagram are not shown due to the number of diagrams involved (10 total---2 $\times$ red, green, blue, saturation, and brightness). This is compared to a reference image (\figref{fig.notredam.ref}) generated using a high-dynamic range technique (i.e., using significantly more data than our approach).

\begin{figure*}[!th]
    \centering

    \begin{minipage}[m]{0.575\linewidth}
    \subfigure[Input Image\label{fig.swan.input}]{\includegraphics[height=0.9in,width=0.485\linewidth]{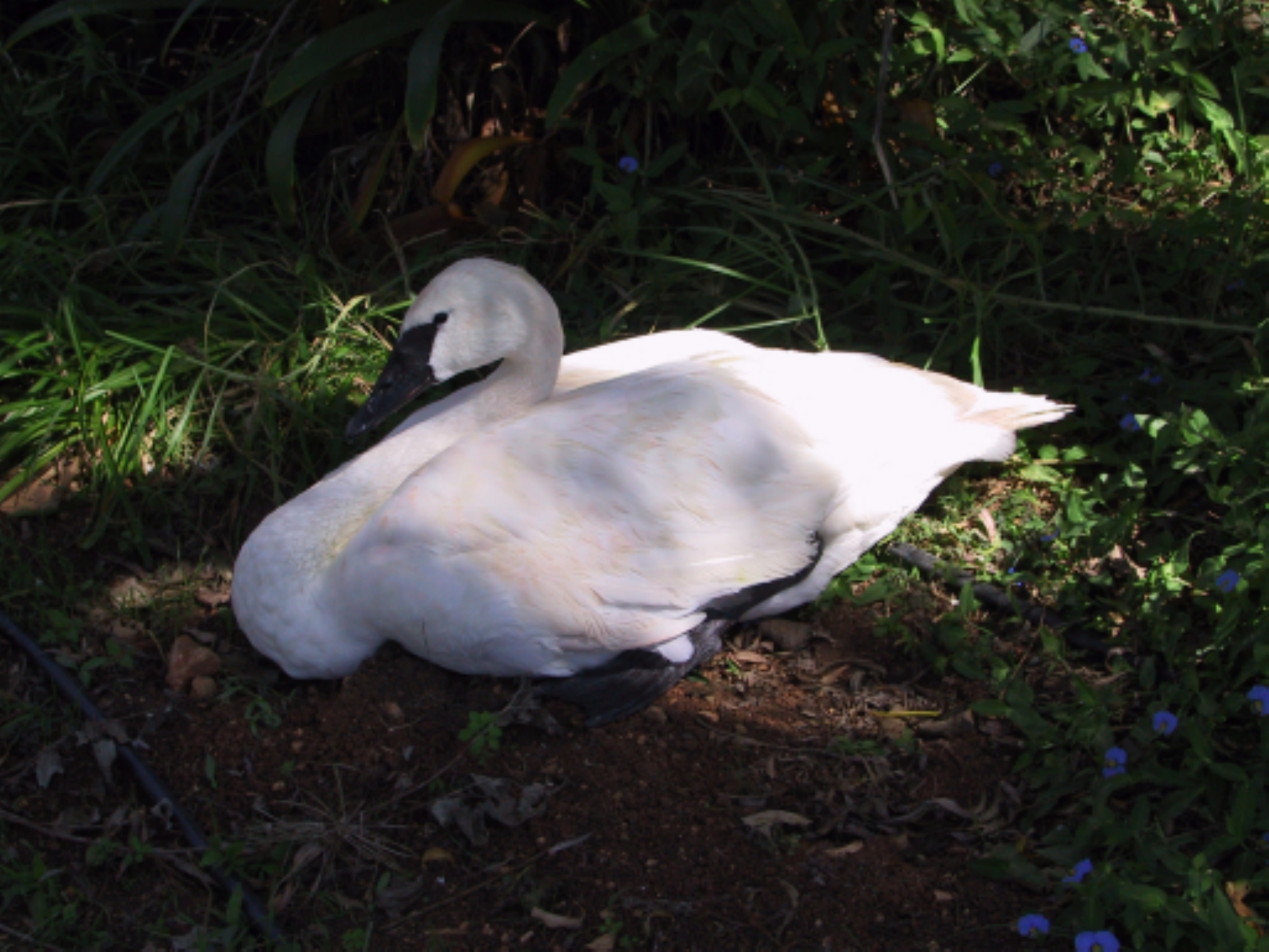}}
    \hfill
    \subfigure[Reference Method~\cite{fattal2002gradient}\label{fig.swan.ref}]{\includegraphics[height=0.9in,width=0.485\linewidth]{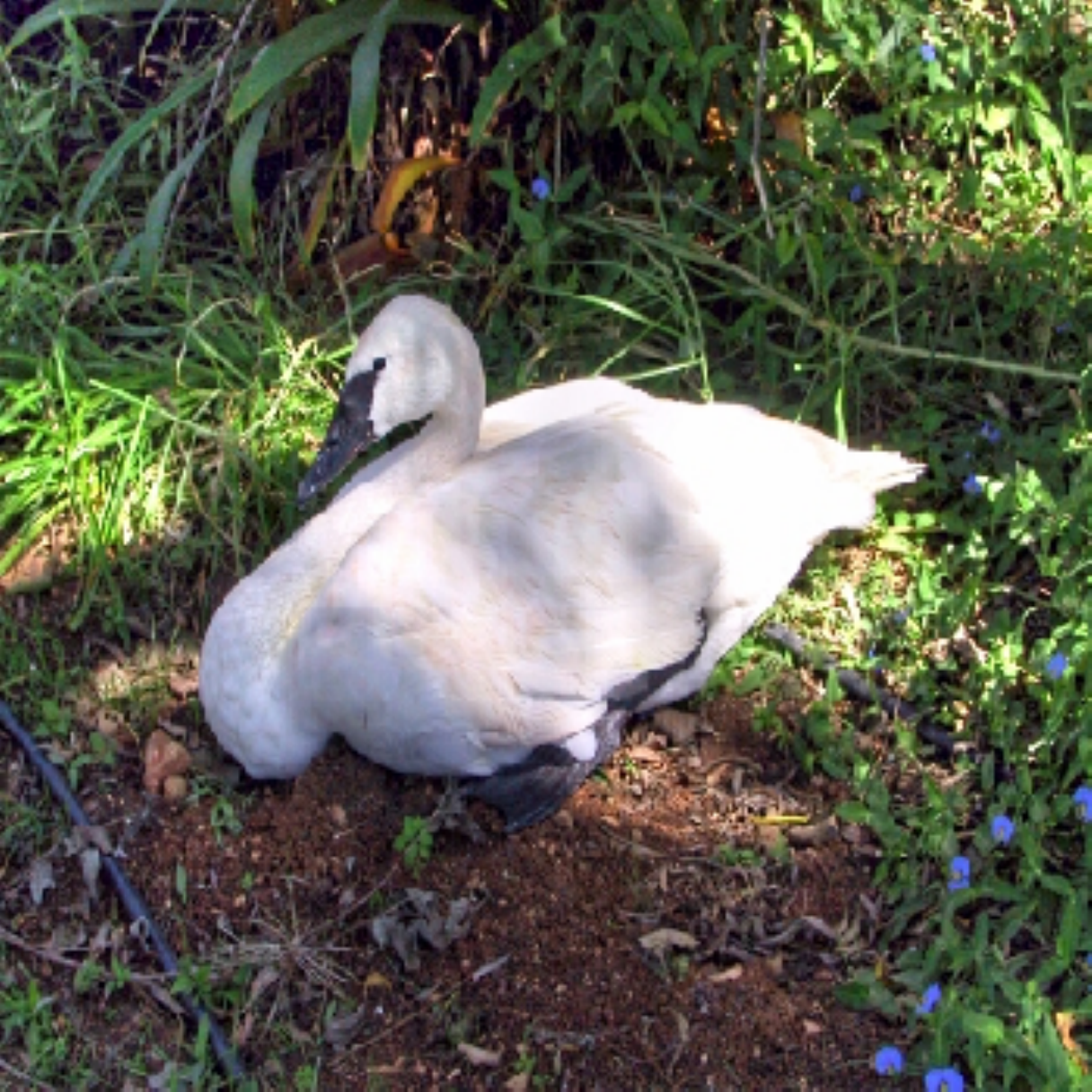}}
    
    \vspace{0pt}
    \subfigure[Brightness Channel\label{fig.swan.output1}]{\includegraphics[height=0.9in,width=0.485\linewidth]{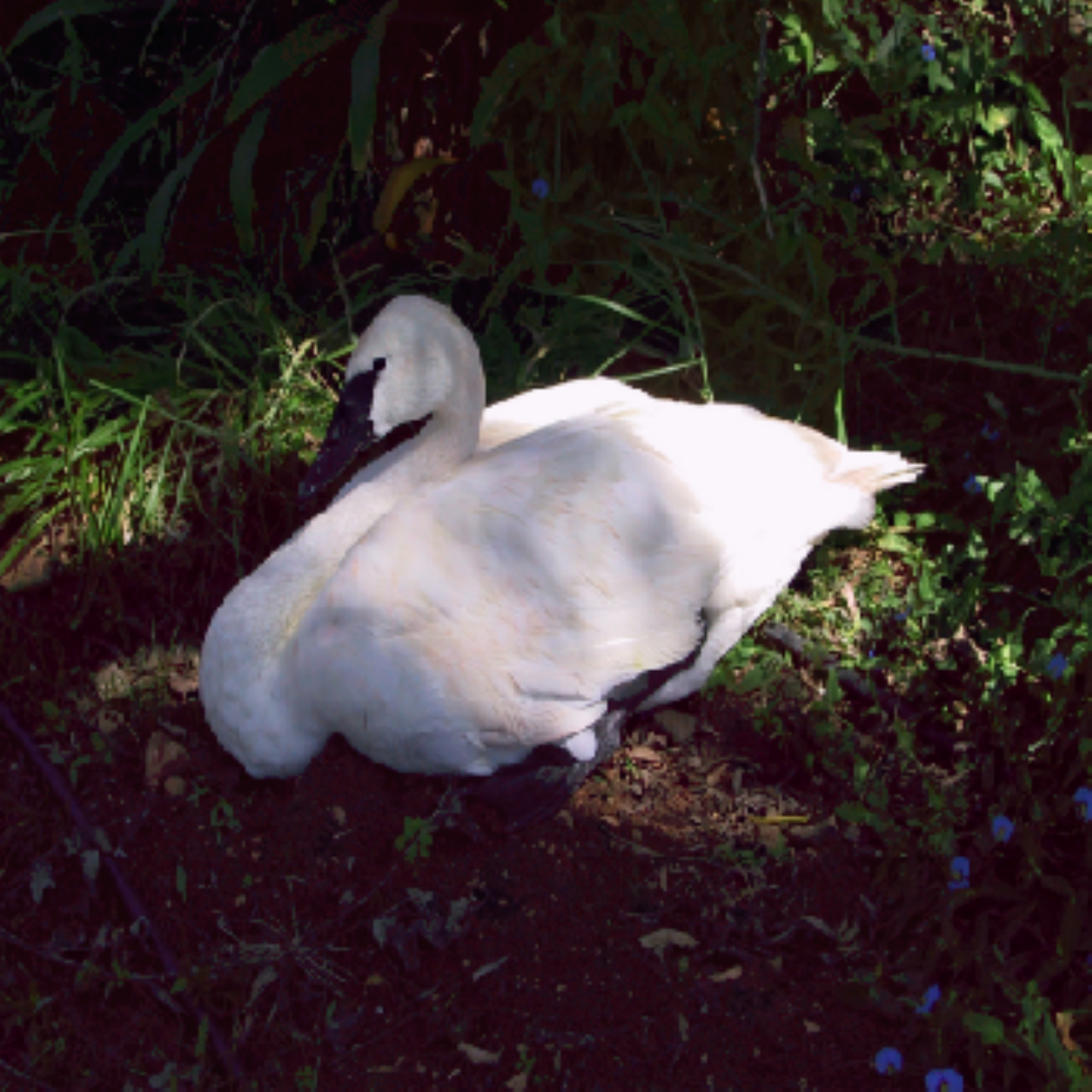}}
    \hfill
    \subfigure[Saturation Channel\label{fig.swan.output2}]{\includegraphics[height=0.9in,width=0.485\linewidth]{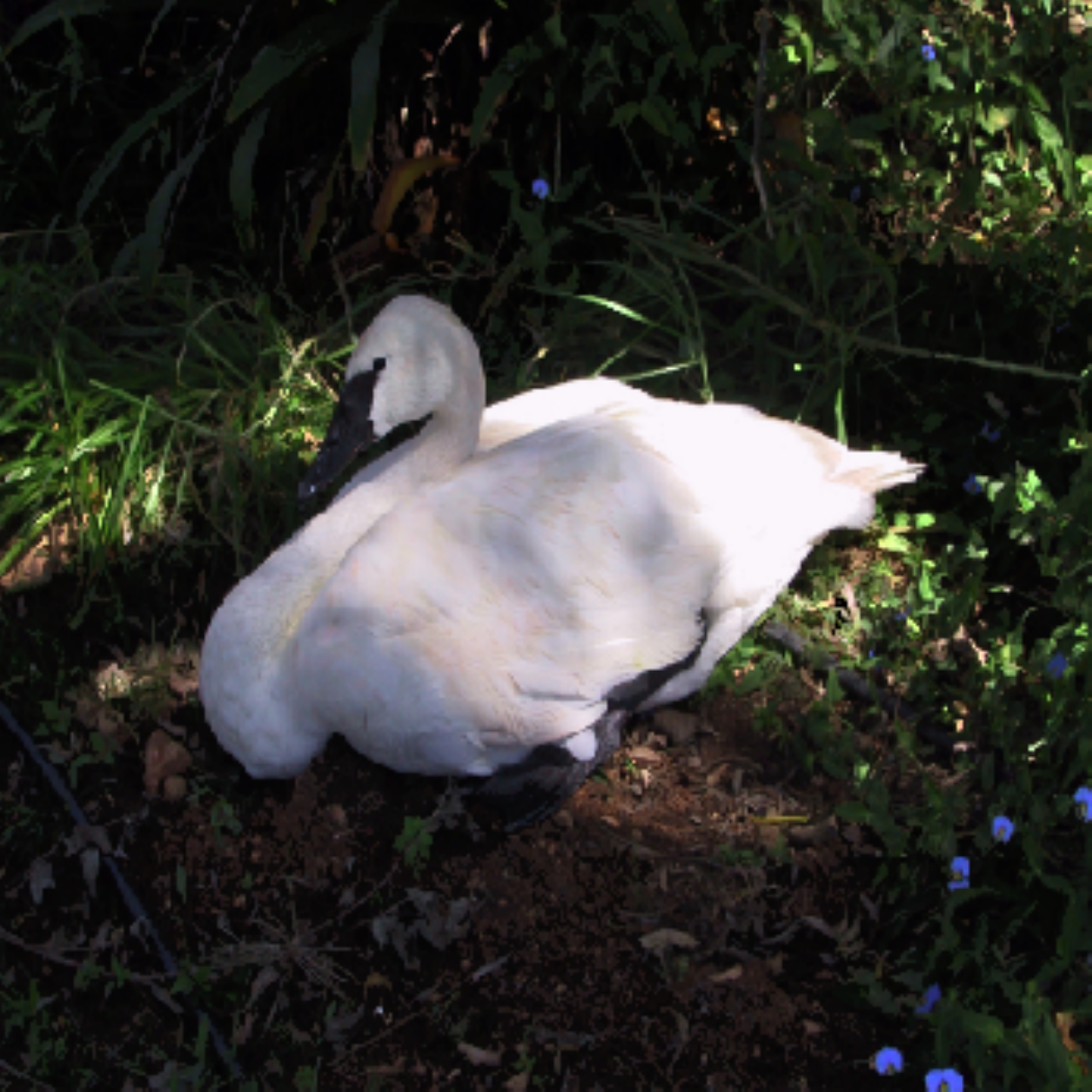}}
    \end{minipage}
    \hfill
    \begin{minipage}[m]{0.4\linewidth}
    \subfigure[Brightness Channel\label{fig.swan.output3}]{\includegraphics[height=1.25in,width=1\linewidth]{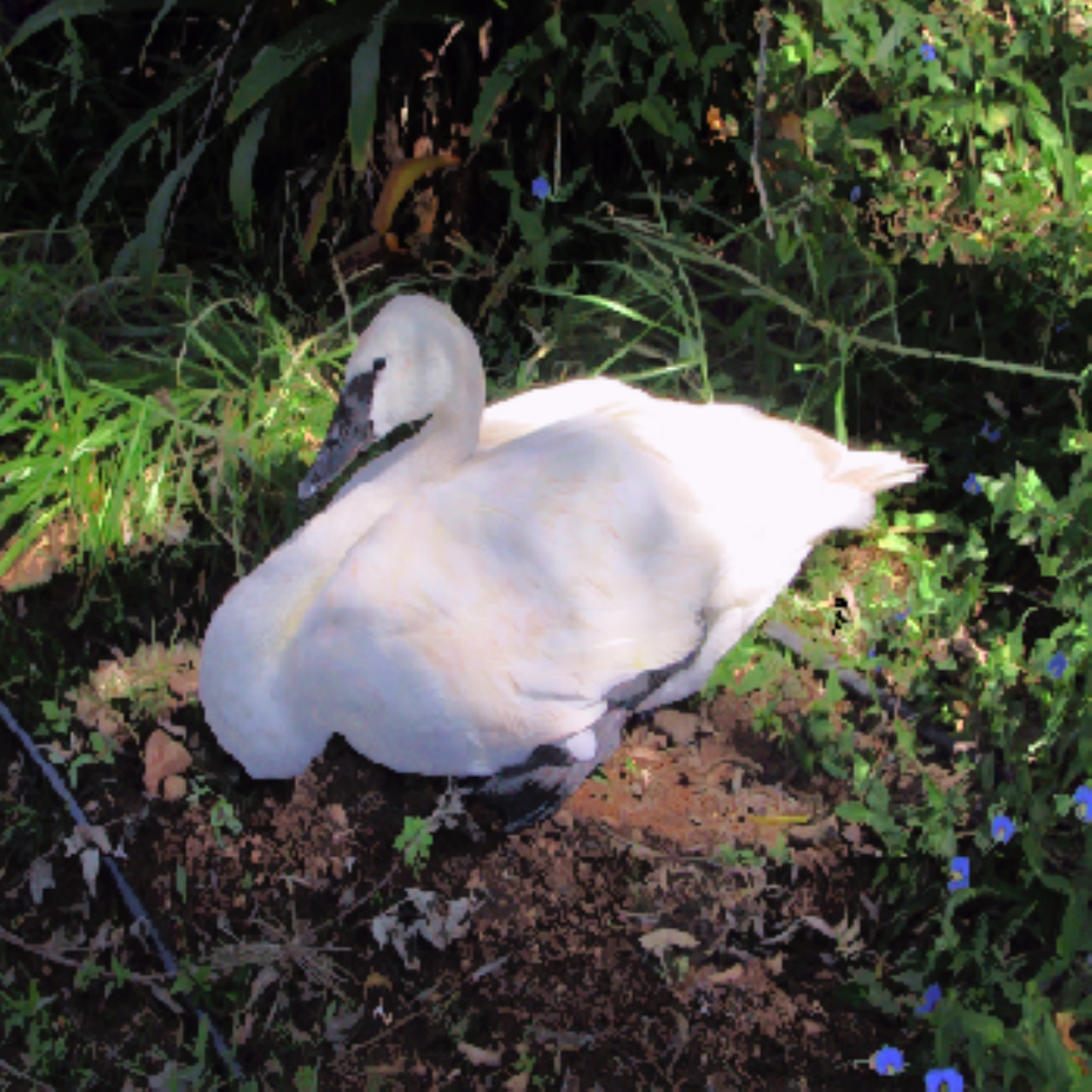}}
    \end{minipage}

    \caption{(a) \textit{Swan dataset.} (b) Reference method~\cite{fattal2002gradient} is compared to image enhanced using (c) denoising of brightness, then (d) brightness enhancement of saturation, and finally (e) contrast enhancement and gamma correction of the brightness.}
    \label{fig.swan}
\end{figure*}

The \textit{Swan dataset}, shown in \figref{fig.swan.input} is a photograph of a swan with a mix of light and shadow, retrieved from~\cite{fattal2002gradient}. We perform a series of enhancements that include denoising of the brightness channel (see \figref{fig.swan.output1}); followed by brightness enhancement in the saturation channel (see \figref{fig.swan.output2}); and contrast enhancement and gamma correction of the brightness channel (see \figref{fig.swan.output3}). This is compared to a reference image (\figref{fig.swan.ref}) generated using a high-dynamic range technique.

\begin{figure}[!t]
    \centering
    \subfigure[Original]{\begin{minipage}[m]{0.2\linewidth}\includegraphics[width=1\linewidth]{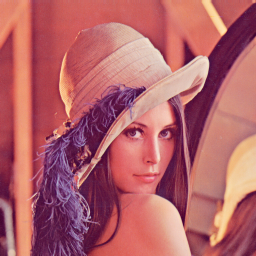}\vspace{5pt}\end{minipage}}\hfill
    \subfigure[Orig.\ w/noise]{\begin{minipage}[m]{0.2\linewidth}\includegraphics[width=1\linewidth]{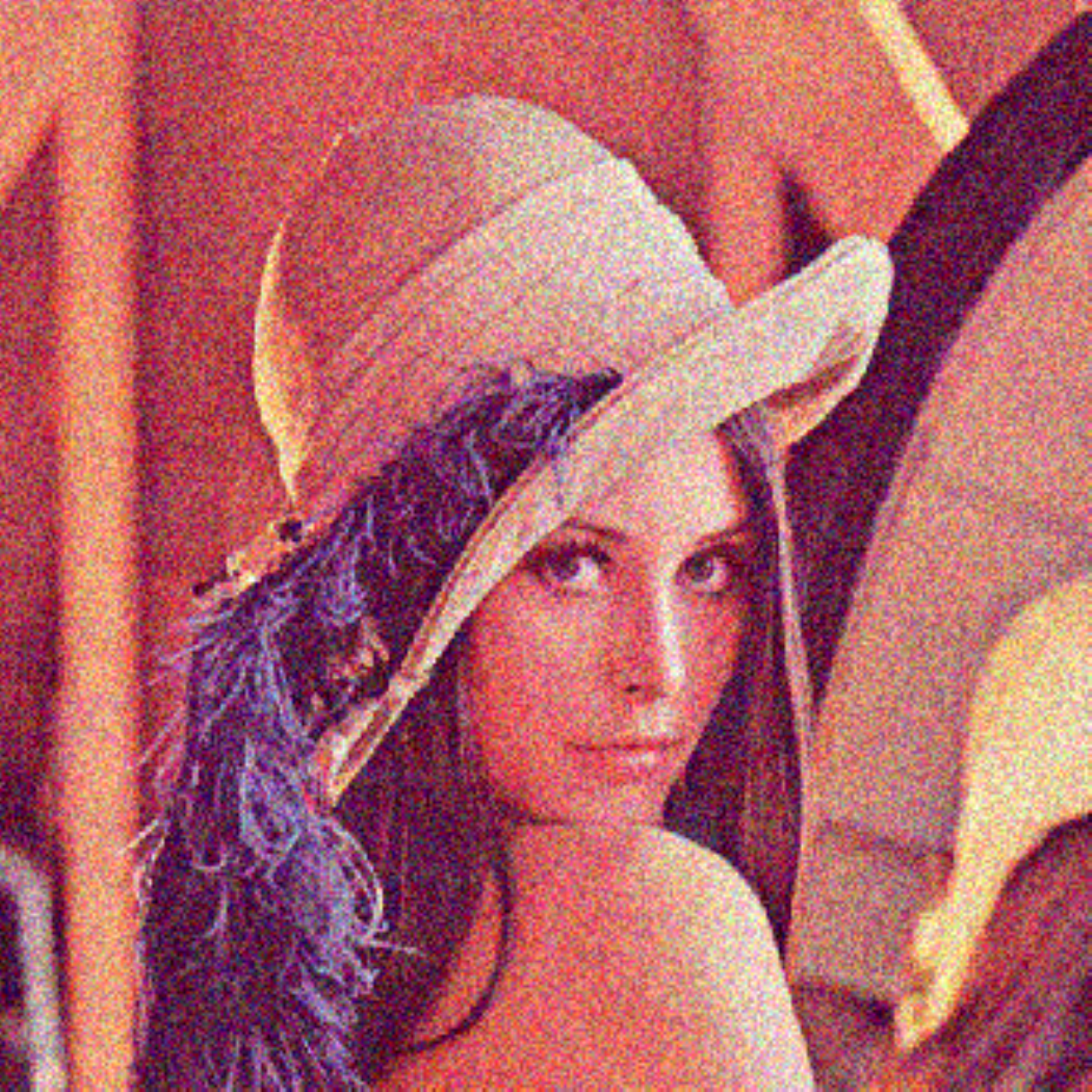}\vspace{5pt}\end{minipage}}
    \hfill
    \begin{minipage}[m]{0.01\linewidth}
        \line(0,1){75}
    \end{minipage}
    \hfill
    \subfigure[Denoising x7]{\begin{minipage}[m]{0.145\linewidth}\includegraphics[width=1\linewidth]{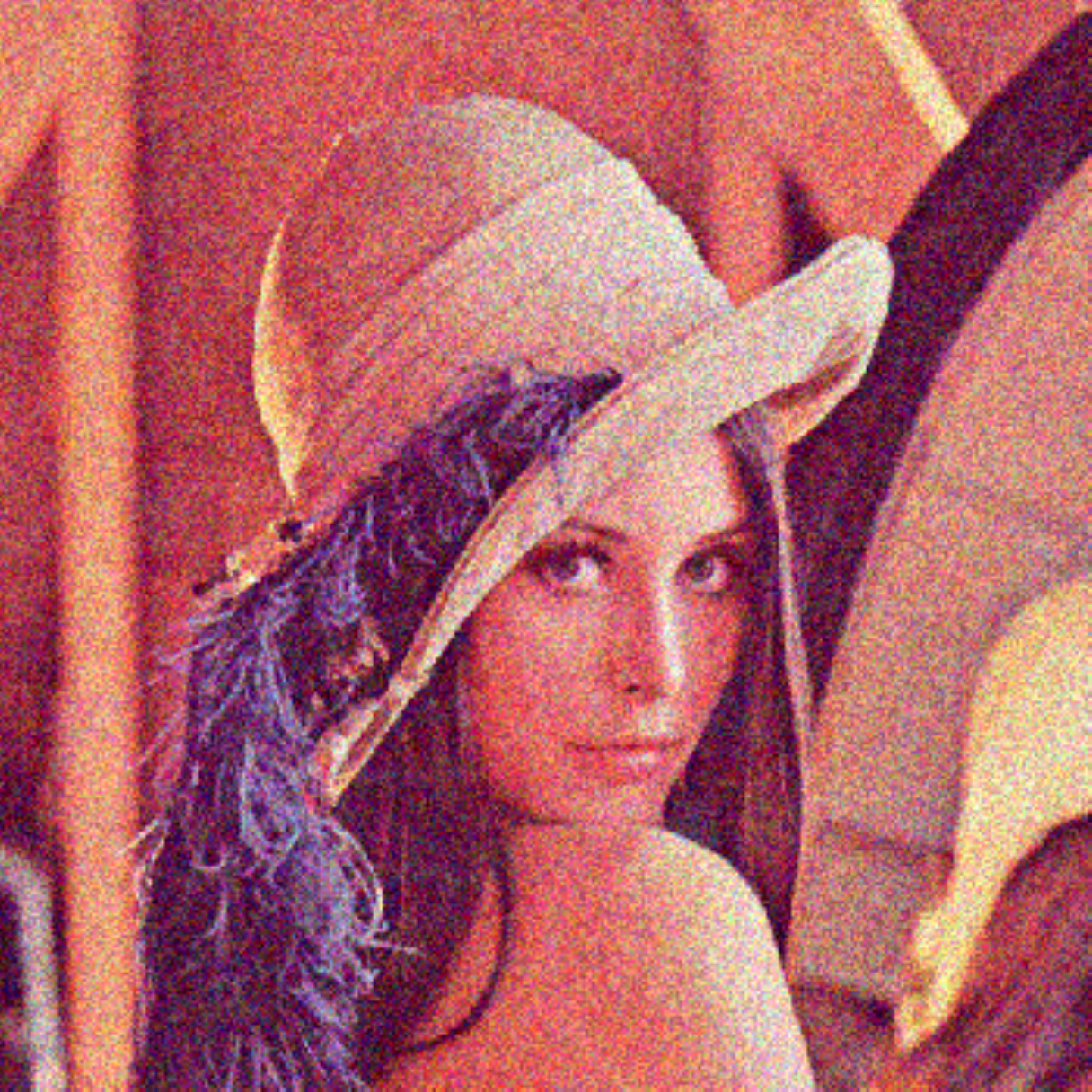}\vspace{5pt}\end{minipage}
    \begin{minipage}[m]{0.03\linewidth}\centering\Large$...$\end{minipage}
    \begin{minipage}[m]{0.105\linewidth}\centering\includegraphics[width=1\linewidth]{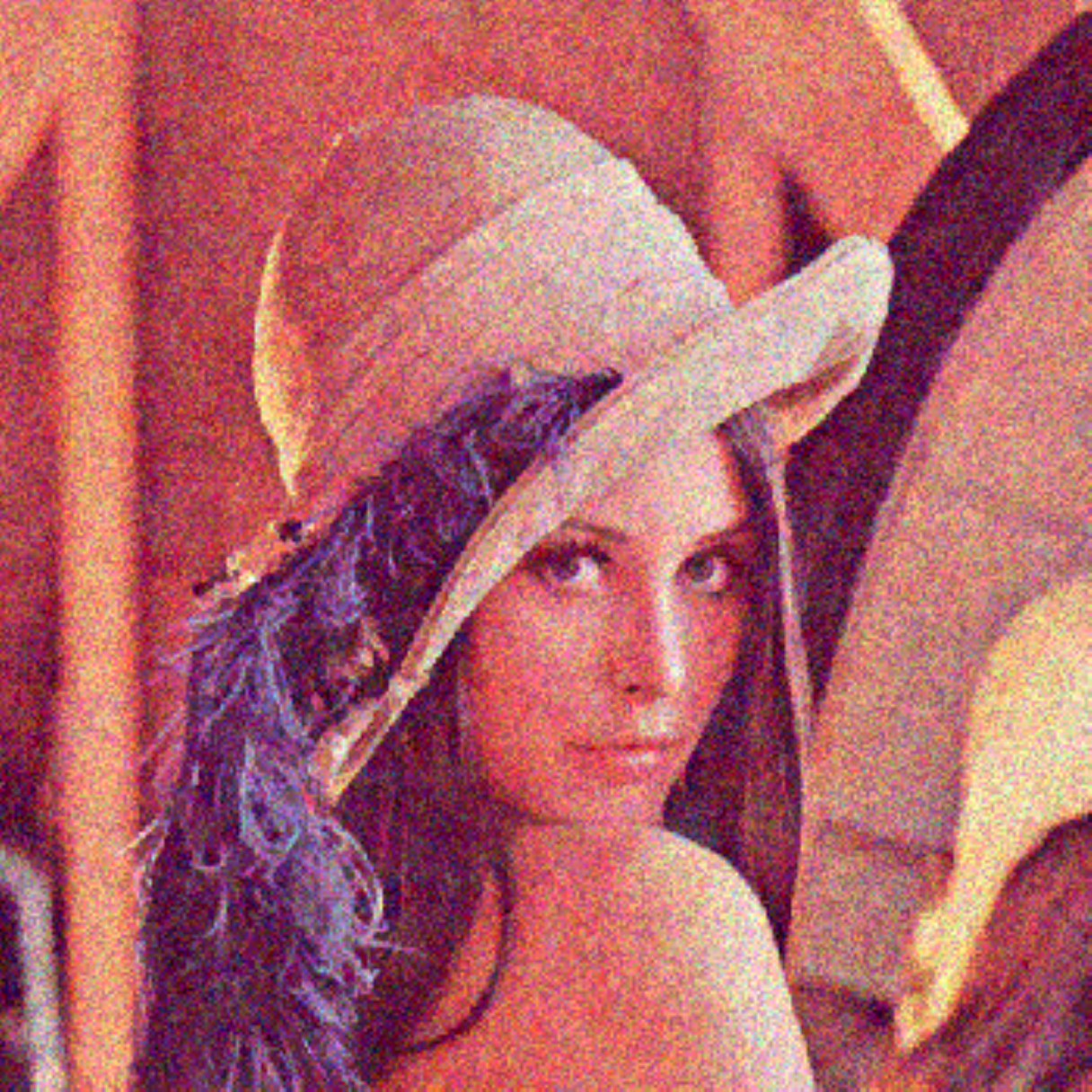}\end{minipage}
    \begin{minipage}[m]{0.03\linewidth}\centering\Large$...$\end{minipage}
    \begin{minipage}[m]{0.145\linewidth}\includegraphics[width=1\linewidth]{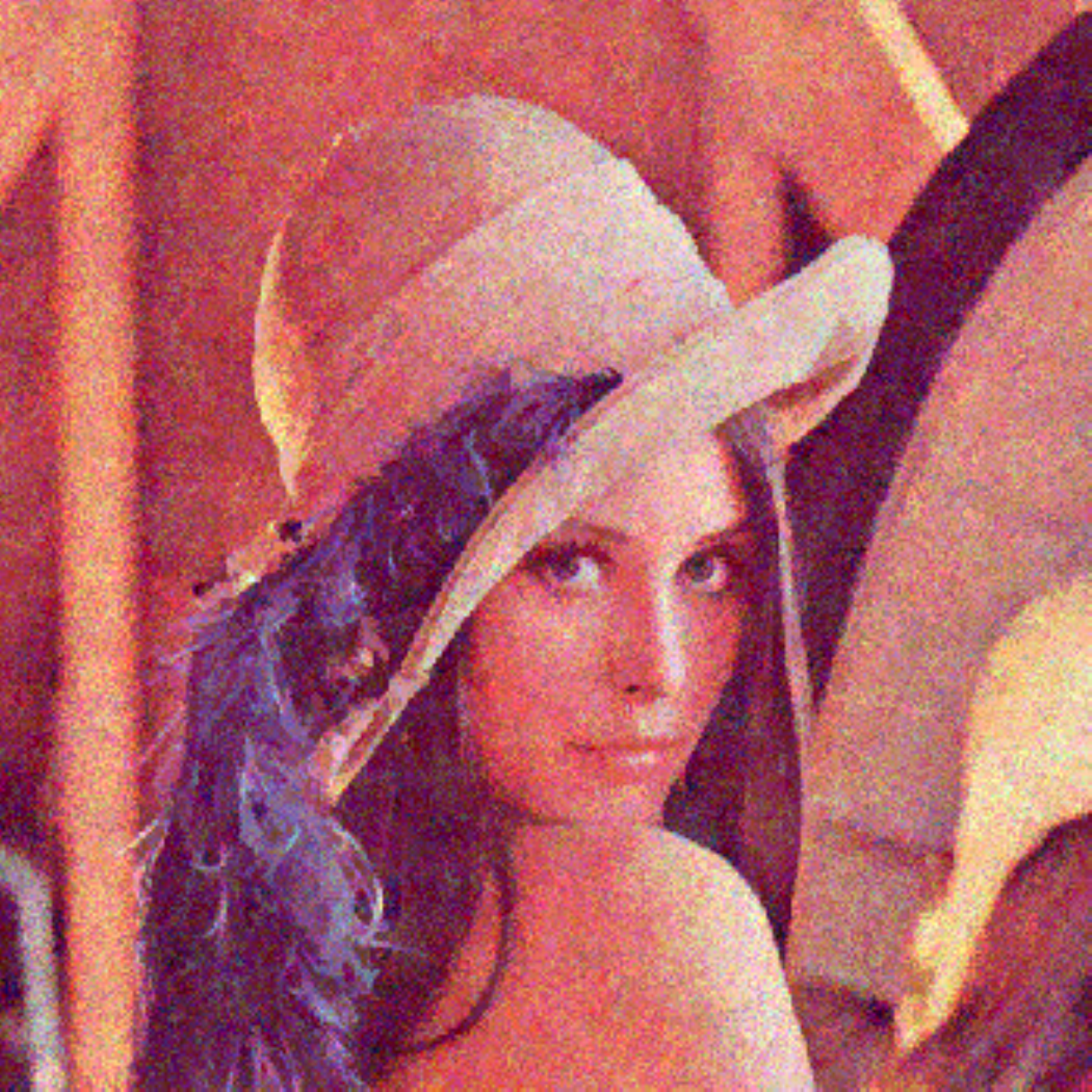}\vspace{5pt}\end{minipage}}            
    
    \subfigure[Denoising]{\includegraphics[width=0.2\linewidth]{figs/lena_color/leaf_outputImage6}}\hfill
    \subfigure[Contrast]{\includegraphics[width=0.2\linewidth]{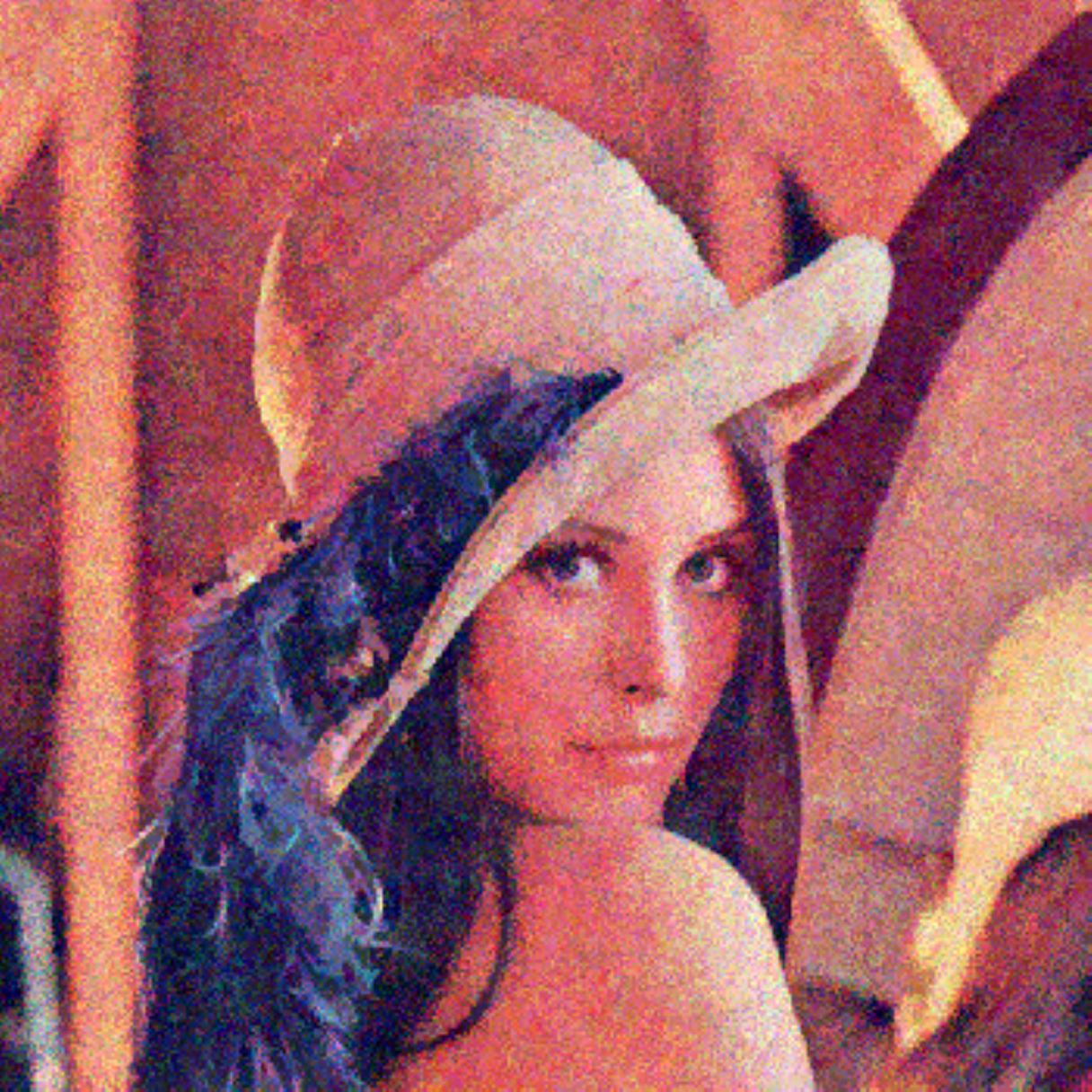}}\hfill
    \subfigure[Denoising]{\includegraphics[width=0.2\linewidth]{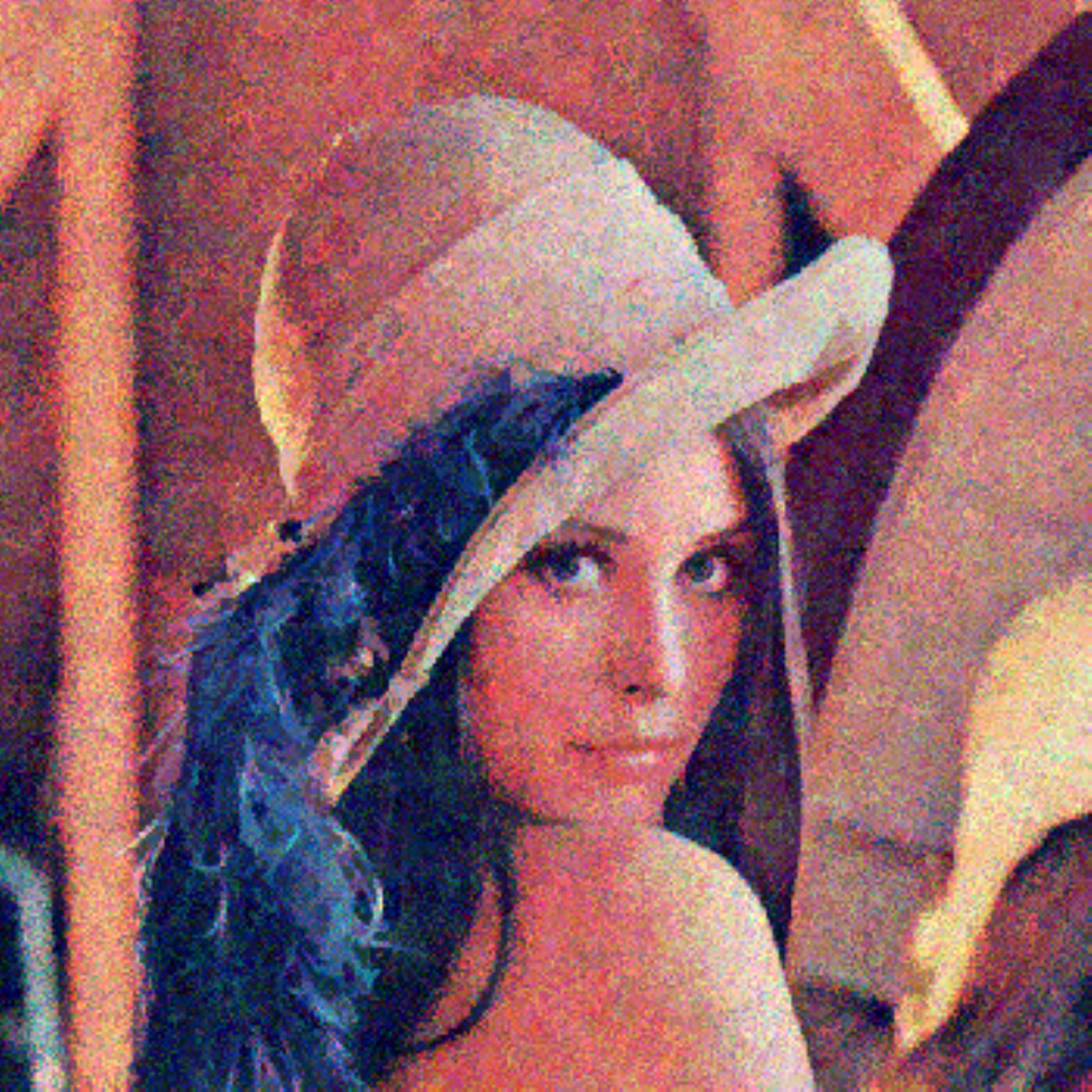}}\hfill
    \subfigure[Brightness]{\includegraphics[width=0.2\linewidth]{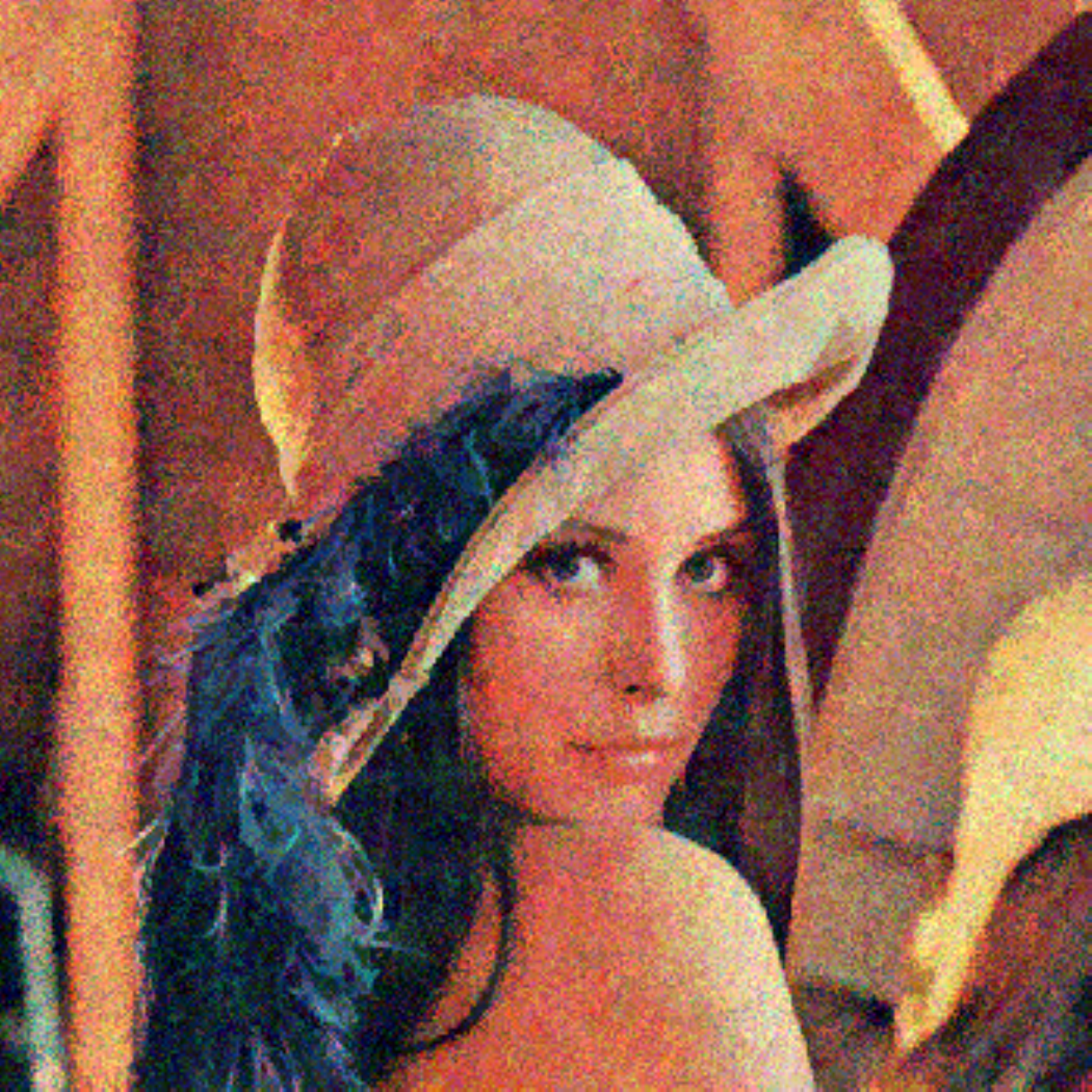}}                
    \caption{(a) \textit{Lenna Color} with noise added. (b-f) A series of denoising, contrast enhancement, and brightness enhancement steps.}
    \label{fig:lenna_color}
\end{figure}

The \textit{Lenna Color Dataset}, seen in \figref{fig:lenna_color} shows a series of 11 edits to a noisy version of the classic color Lenna dataset.

\section{Prior Work \& Conclusions}

We have presented a new approach to image enhancement based upon the topology of an image. Our approach provides a high-level of control to users, while not requiring an extensive number of interactions to achieve desirable results. 

Like with most other image enhancement algorithms, artifacts are an important concern. Our approach does not introduce new artifacts, per se, but instead it may emphasize existing image artifacts. For example in \figref{fig.ALflorala}, blocking artifacts appear due to lack of detail for generating a smooth result. In \figref{fig.brain}, artifacts occur due once again to missing contrast and detail that lead to small differences in intensity ending up emphasized.

Much of the prior work on image enhancement has focused on automated techniques. A number approaches have addressed segmentation. For example, algorithms for edge detection and image segmentation detected ``contours'' (i.e., edges) in a hierarchical manner for segmentation~\cite{amfm_pami2011}. Supervised learning, such as the hierarchical merge trees model in~\cite{LiuTing2016}, are popular in image segmentation. Contrast enhancement has been proposed as an optimization problem that maximizes the average local contrast of an image~\cite{Majumder:2006:CEI:1140491.1140506}. Recently, deep learning has been leveraged to automatically retouch images~\cite{hu2018exposure}. These approaches provide high quality results, but they offer limited opportunity for tuning the output.

TDA has also been used previously in image processing. The first use of the contour tree on a binary image was~\cite{Aydogan2012}. Edge detection and Delaunay triangulation were performed to decompose images into regions indexed by radius of a disk and persistence~\cite{letscher07}. Finally, persistence-based segmentation of noisy 2D point clouds have been studied~\cite{KURLIN20163}. Our approach in this paper is different from these prior works in that it not only captures the structure of images, but it also enables a variety of methods for manipulating the segmented regions of images.

\paragraph{Acknowledgments.} This project was supported in part by the National Science Foundation (IIS-1513616 and IIS-1845204).


\bibliographystyle{splncs04}
\bibliography{main}   

\end{document}